\documentclass[onecolumn,sort&compress,numbers]{els-mrw} 

\usepackage{amsmath,amssymb,amsfonts,amsthm,makeidx,graphicx}
\usepackage{txfonts}
\usepackage{helvet}

\begin{document}


\chapter{Hadronic molecules and multiquark states}\label{chap1}

\author[1]{Christoph Hanhart}%

\address[1]{\orgname{Forschungszentrum J\"ulich}, \orgdiv{Institute for Advanced Simulation}, \orgaddress{D-52425 J\"ulich, Germany}}

\articletag{Chapter Article tagline: update of previous edition, reprint.}

\maketitle

\begin{abstract}[Abstract]
 In this section different theoretical approaches towards multi-quark states are introduced, namely hadrocharmonia, compact tetraquarks and hadronic molecules. The predictions derived from either of them   are contrasted with current and possible future observations.
The focus is on doubly heavy systems, but singly heavy and light systems are mentioned briefly as well.
\end{abstract}

\begin{keywords}
 	Exotic hadrons \sep multi-quarks\sep hadronic molecules
\end{keywords}



\newcommand{\pc}{P_{c\bar c}}
\newcommand{\jp}{J/\psi p}
\newcommand{\sigh}{\Sigma_c^{(*)}\bar{D}^{(*)}}
\newcommand{\sigd}{\Sigma_c\bar{D}}
\newcommand{\sigdstar}{\Sigma_c\bar{D}^*}
\newcommand{\sigstard}{\Sigma_c^*\bar{D}}
\newcommand{\lamh}{\Lambda_c\bar{D}^{(*)}}
\newcommand{\etacp}{\eta_c p}
\newcommand{\etacn}{\eta_c p}
\newcommand{\dsz}{D_{s0}^*(2317)}
\newcommand{\dsone}{D_{s1}(2460)}
\newcommand{\dz}{D_0^*(2300)}
\newcommand{\done}{D_1(2430)}
\newcommand{\bdpp}{B^-\to D^+\pi^-\pi^-}
\newcommand{\cccc}{cc\bar c\bar c}

\begin{figure}[t]
	\centering
	\includegraphics[width=14cm,height=5cm]{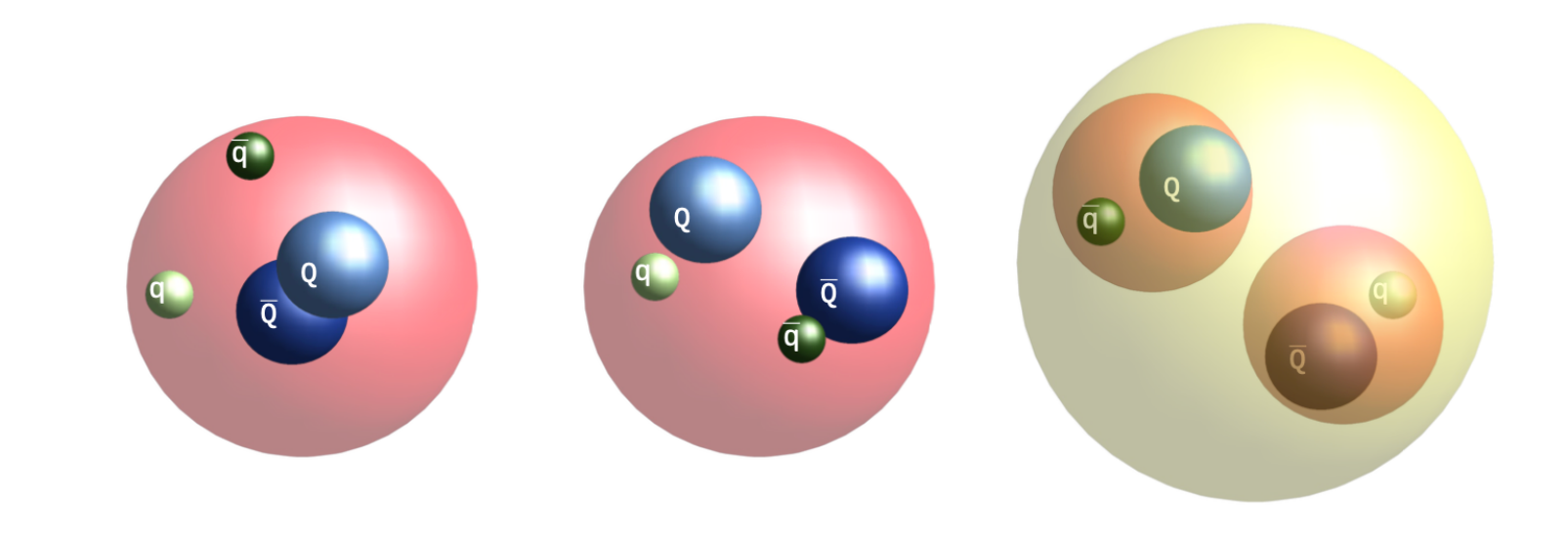}
	\caption{Visualisation of the substructure of possible tetraquark configurations in the doubly heavy sector,
	with $Q$ ($\bar Q$) and $q$ ($\bar q$) denoting heavy and light (anti-)quarks, respectively: (a) hadroquarkonium, (b) compact tetraquark, (c) hadronic molecule.
	While the first two are typically compact with their size dictated by the confinement radius, the last one can be very large if the state is located close
	to the pertinent two-hadron threshold.}
	\label{fig:multiquarks}
	\vspace{-6cm} 
	(a) \hspace{3.6cm} (b) \hspace{3.6cm} (c)\hspace{4cm}

		\vspace{6cm} 

\end{figure}

Since the early days of the quark model, multiquark states are proposed to exist---they are already mentioned in the famous works by Gell-Mann~\cite{Gell-Mann:1964ewy} and Zweig~\cite{Zweig:1964ruk}. 
The first calculation
proposing a multiquark structure for concrete states was performed by Jaffe and Johnson
within the MIT bag model: The nonet of light scalar mesons, nowadays called $f_0(500)$, $f_0(980)$, $a_0(980)$ and
$K^*_0(700)$, were proposed to
be tetraquarks ($\bar q\bar q q q$)~\cite{Jaffe:1975fd}.
Only one year later Voloshin and Okun~\cite{Voloshin:1976ap}
argued for the existence of 
deuteron like states in doubly heavy systems. The idea was picked up later by
T\"ornqvist~\cite{Tornqvist:1993ng}.
In the previous section  strong experimental evidence was presented for the existence of states beyond the most simple
structures allowed by the rules for the formation of hadrons within QCD,  especially in the doubly heavy
sector (for reviews, putting emphasis on different aspects, see Refs.~\cite{Hosaka:2016pey,Esposito:2016noz,Guo:2017jvc,Olsen:2017bmm,Karliner:2017qhf,Brambilla:2019esw, Yang:2020atz,Chen:2022asf,Meng:2022ozq})---this is why we start the discussion for this class of states that we use to introduce the
various structure assumptions currently discussed in the literature.
We will
come back to candidates for multiquark states in other sectors towards the end of this section. 
In the doubly heavy sector there were states found that decay
into a heavy quarkonium state together with a light hadron. Examples are 
$T_{\bar bb}(10610)$ and $T_{\bar bb}(10650)$, also known as $Z_b$ states, observed in the $\Upsilon(nS)\pi$, $n=1,2,3$, and $h_b(mP)\pi$, $m=1,2$, 
final states and $P_{c\bar c}(4312)$, $P_{c\bar c}(4440)$ and $P_{c\bar c}(4457)$ observed in $J/\psi p$ final states. 
Since 
the production of the heavy quarkonium in the course of the decay is heavily suppressed within QCD 
due to the OZI rule, the $\bar QQ$ pair must have preexisted in the wave functions of the observed states.
Since they on the other hand carry charge (or, equivalently, non-vanishing isospin), they are identified as
 tetraquarks and pentaquarks, respectively\footnote{The prefixes tetra and penta, denoting four and five, are of greek origin.}.

In addition to the explicit multiquark states mentioned in the previous paragraph there are also states that qualify for
multiquark states not because of their quantum numbers, but because of their unusual properties, difficult if not impossible
to accommodate within the most simple realisations of the quark model or variants thereof. Prominent examples of this
class are the $\psi(4230)$, also known as $Y(4230)$, a vector state that decays, e.g.,
 into $D\bar D^*\pi$ and $J/\psi\pi\pi$ but
not into $D^{(*)}\bar D^{(*)}$ as is expected for a $\bar cc$ state, and the $\chi_{c1}(3872)$, also known as $X(3872)$,
which decays by far dominantly into $D^0\bar D^{* \, 0}$, although its mass basically coincides with the threshold of this channel,
while the decays into $J/\psi \pi \pi$ and $J/\psi 3\pi$ appear to be heavily suppressed.

In this section the different proposals put forward for the structure those states, sketched for tetraquarks
in Fig.~\ref{fig:multiquarks}, are reviewed. The presentation focusses 
on what imprint in the particle spectrum and observables  the different structures would leave, 
if they were to be the only or the by far dominant
component in a given state. This is also what is mostly discussed in the literature as of today. However, in principle some
mixing between the different structures is possible as well. We come back to this interesting question in section~\ref{sec:multiquarks:subsec:outlook}.
The different structures shown in Fig.~\ref{fig:multiquarks} differ by the assumed sub-structures within the hadron. 
Those are either color neutral (for hadro-quarkonia or hadronic molecules) or carrying a color charge in form of (anti-)diquarks.

The link between QCD and the different assumed structures of the multiquark states is provided by
the approximate symmetries of QCD and their breaking, most notably SU(2)  flavor (also known as isospin
symmetry), embedded in the larger SU(3) flavor symmetry and heavy quark spin symmetry.

In addition, also chiral symmetry plays a crucial role for some systems. SU(2) (SU(3)) flavor symmetry were an exact symmetry of
QCD, if the charges and masses of up and down (up, down and strange) quarks were equal, since
the QCD interaction is flavor blind (up to some quark type dependence entering through the scale
dependence of the QCD coupling constant). Since the up and down quark mass difference is much
smaller than any hadronic scale, isospin symmetry is typically realised with an accuracy of better
than a few percent. As the strange quark is much heavier, SU(3) flavor symmetry is realised only
with some 30\% accuracy, however, SU(3) breaking mass differences within multiplets of compact
states are typically (largely)
explained by the mass added in by the strange quark. For hadronic molecules the situation is 
more complicated as is discussed below.

 Heavy quark spin symmetry is a consequence
of spin dependent interactions scaling as $q_{\rm typ}/M_Q$, where $M_Q$ denotes the mass of the
heavy quark and the typical momentum inside a hadron may be estimated by the non-perturbative
QCD scale, $q_{\rm typ}\sim \Lambda_{\rm QCD}$. In the strict heavy quark limit ($M_Q\to \infty$),
spin multiplets appear formed from the heavy quark spin being coupled differently to the same
light quark cloud. In other words, in this limit a light quark cloud of some $j_\ell$ (this quantum number
should also contain some possible orbital angular momentum among heavy quarks in the system) coupled to some
heavy quark spin $S_h$ will generate a whole set of states with total angular momentum $$|j_\ell - S_h|\le J
\le j_\ell + S_h \ .$$
In systems with more than one heavy quark the multiplets get even larger, since
the properties of the emerging hadrons in leading order of the expansion
do not depend on the value of the total spin $S_h$ the spins of the different 
heavy quarks in the system get coupled to.
Since the symmetry is exact in the heavy quark limit it is realised in the hadron spectrum regardless
the assumed structure of it. However, the symmetry violations turn out to be structure dependent~\cite{Cleven:2015era}
are expected to become an important diagnostic to deduce the structure of exotic hadrons from the spectra.

Theoretical approaches to multiquark states range from phenomenological models,
either on the quark level or of meson exchange type, over effective field
theories, again on the quark level or employing chiral perturbation theory 
on the hadron level, to lattice QCD, each of which are described in other sections
of this encyclopedia. In what follows some account will be given on their 
application to multiquark states.

\section{Hadroquarkonia}

For tetraquarks the assumed structure of a hadroquarkonium is shown in Fig.~\ref{fig:multiquarks}(a). 
The underlying idea is that the multiquark
consists of a compact, color neutral $Q\bar Q$ core, like the $J/\psi$, the $\psi(2S)$ or the $\eta_c$ surrounded by 
a (typically excited) light quark cloud~\cite{Dubynskiy:2008mq}. Since a $Q\bar Q$ state does not contain light quarks,
the interaction of the cloud with the compact doubly heavy core is suppressed (e.g. at leading order chiral perturbation theory
the interaction of pions with quarkonia vanishes), however, the polarisabilities might still be sufficiently large to allow for
some binding of a light quark cloud to e.g. $J/\psi$ or excitations thereof. The same mechanism might also 
provide sufficient binding, to generate states of pairs of $J/\psi$s~\cite{Dong:2021lkh}. We come back to this scenario
in Sec.~\ref{sec:multiquarks:allheavy}.

The proposal for the existence of a hadrocharmonium structure was triggered by the observation of $\psi(4230)$ also
known as $Y(4230)$ (at the time called $Y(4260)$), since this state shows up as a clear peak in the $J/\psi\pi\pi$
spectrum while at the same time being absent in the spectra for $D\bar D$, $D\bar D^{*}$ and $D^{*}\bar D^{*}$.
Such a pattern emerges naturally within the hadrocharmonium picture, since the observed decay is merely 
a fall apart mode of the building blocks, while a transition to the open charm channels requires a break up of
the compact quarkonium as well as some re-arrangement of the light quark cloud.  
Note that recently the $Y(4230)$ was also observed as a clear peak in the reaction $e^+e^-\to\bar D\pi D^{*}$
 with an even higher rate than
in $J/\psi\pi\pi$~\cite{BESIII:2022qal,BESIII:2018iea}, questioning somewhat the logic put forward above~\cite{Wang:2013kra}.

The $Y(4230)$ is also observed in the $h_c\pi\pi$ channel. In contrast to the $J/\psi$ that has a spin 1, the $h_c$ has spin 0.
Thus, if the $Y(4230)$ were to contain a pure $J/\psi$ core, heavy quark spin symmetry would prevent it from decaying
into a final state with a spin 0 quarkonium. To overcome this discrepancy with experiment, in Ref.~\cite{Li:2013ssa}
it was suggested that the $Y(4230)$ is not a pure state but, together with the next heavier state, the $\psi(4360)$ also
known as $Y(4360)$,
emerges from a spin symmetry violating mixing of two states with $h_c$ and $\psi(2S)$ cores, respectively. 
Here it is assumed that the suppression of the spin symmetry violation which is shown to appear
already at order $\Lambda_{\rm QCD}/m_c$, is overcome by the close
proximity of the mixing states leading to a small energy denominator in the mixing amplitude. 
More concretely, starting from the unmixed basis
\begin{equation}
\Psi_3 = (1^{--})_{c\bar c}\otimes (0^{++})_{q\bar q} \qquad \mbox{and} \qquad
\Psi_1 = (1^{+-})_{c\bar c}\otimes (0^{-+})_{q\bar q} \ ,
\end{equation}
where the heavy cores are assumed to be $\psi(2S)$, with a mass of 3686 MeV,  and $h_c(1P)$, 
with a mass of 3525 MeV, one gets for the physical states
\begin{equation}
Y(4230) = \cos(\theta) \Psi_3 - \sin(\theta)\Psi_1  \qquad \mbox{and} \qquad Y(4360) = \sin(\theta) \Psi_3 + \cos(\theta)\Psi_1 \ .
\end{equation}
A fit to data revealed a mixing angle of the order of 40 degrees accompanied by near 
degenerate unmixed states with masses of approximately 4.30 and 4.32 GeV for $\Psi_3$ and $\Psi_1$,
respectively. The apparent close proximity of the mixed states must emerge somewhat accidental from
the interplay of the core states that show a mass difference of the order of 160 MeV with the light
quark clouds. 

\begin{figure}[t]
	\centering
	\includegraphics[width=8cm,height=5cm]{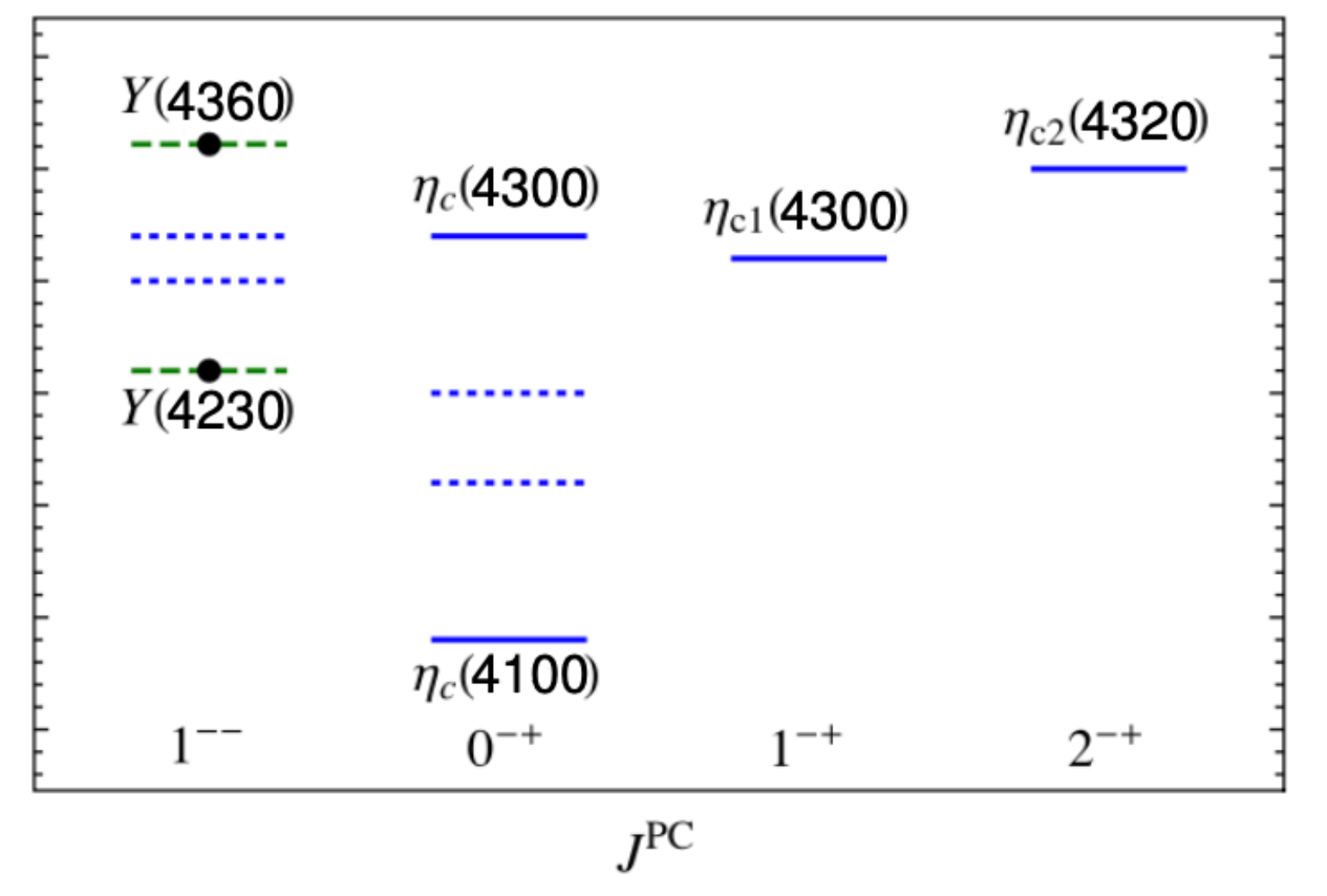}
	\caption{Illustration of the implications of employing spin symmetry to predict the
	spin partners of the $Y(4230)$ and $Y(4360)$ in the hadrocharmonium scenario.
	Solid lines show masses of the physical states (input or predicted), dashed line the
	deduced masses of the unmixed basis states.
	The $1^{--}$ states are used as input --- the masses of the predicted spin partner states are 
	approximate only. Figure adapted from Ref.~\cite{Cleven:2015era}. }
	\label{Fig:hadrocharmoniumspinpartners}
\end{figure}

What testable predictions emerge from this scenario? 
If spin symmetry were exact and a given exotic were a hadroquarkonium, it would imply that replacing
the core of such a state by its spin partner(s) while keeping the light quark cloud the same, must lead to
another hadroquarkonium state, whose mass can be estimated from the mass differences between
the different seed states. While one can expect some spin symmetry violation in systems with
charm, predictions derived from spin symmetry, allowing for the above-mentioned mixing, should
 capture the relevant patterns emerging from the assumed structure. This idea was exploited quantitatively
in Ref.~\cite{Cleven:2015era}---the spin partner states here are found by the replacements
\begin{equation}
\psi(2S)\to \eta_c(2S) \qquad  \mbox{and} \qquad h_c(1P)\to \{\chi_{c0}(1P), \ \chi_{c1}(1P), \ \chi_{c2}(1P)\} \ .
\end{equation}
The emerging spectrum is shown in Fig.~\ref{Fig:hadrocharmoniumspinpartners}.
As a solid prediction of the hadrocharmonium scenario, a relatively light exotic $\eta_c$ state is predicted at
4.1 GeV that should
not decay to $D\bar D^*$ but instead to $\eta_c\pi\pi$. Moreover, at about 4.3 GeV there should be a 
spin exotic state with quantum numbers $1^{-+}$, which cannot be generated for $\bar cc$ states, that
is near degenerate with another $\eta_c$ state. This spectrum nicely shows how using spin symmetry allows
for testable predictions for assumed underlying structures for the exotic states. 
Similar arguments can be applied to decay patterns. For example, in~\cite{Voloshin:2018vym} it is argued that, if
$Z_c(4100)$ and $Z_c(4200)$, also known as $T_{c\bar c1}(4100)^+$ and $T_{c\bar c1}(4200)^+$, respectively,  were hadrocharmonia with seeds $\eta_c(2S)$ and $\psi(2S)$, respectively,
their partial widths to final states containing the core state and a pion should be equal up to spin-symmetry
violating corrections.

The building blocks of hadrocharmonia are color neutral by construction, since the core states
are conventional charmonia. Thus, one might be tempted to put them into one class with hadronic
molecules to be discussed in Sec.~\ref{sec:hadmol}. However, while hadronic molecules typically have masses
close to the thresholds of the channels that form the molecule, which can result in hadrons
of unusually large size, hadroquarkonia can be in mass quite far
above the most pertinent threshold and accordingly show typical hadronic sizes. 
For example, the $Y(4230)$ mentioned above is more than 600~MeV heavier than the
sum of the masses of the $J/\psi$ and the lightest scalar-isoscalar resonance, the $f_0(500)$.
For the $Y(4230)$ the hadrocharmonium and hadronic molecule picture are contrasted in Ref.~\cite{Wang:2013kra}.
The only case discussed so far in the literature that could be viewed as both a hadronic molecule
and a hadrocharmonium is $\psi(4660)$ also known as $Y(4660)$, which is proposed to have
a substructure made of a $\psi(2S)$ and an $f_0(980)$~\cite{Guo:2008zg,Dubynskiy:2008mq}.
A prediction that will allow for a test of this proposal is that, if the assumed structure is correct, there needs to be a pseudoscalar
bound state formed of $\eta_c(2S)$ and $f_0(980)$, located at around 4616~MeV (lighter than 
the $Y(4660)$ by the $\psi(2S)$-$\eta_c(2S)$ mass difference) with a partial width into $\eta_c(2S)\pi\pi$
of $60\pm 30$~MeV~\cite{Guo:2009id}.

Clearly, for doubly heavy states of the type $QQ\bar q\bar q$ like the $T_{cc}(3875)^+$ discovered 
at LHCb, one may expect in the spirit of this section a compact $QQ$  core surrounded by 
a light $\bar q\bar q$ cloud. Thus, here the building blocks are (anti-)diquarks and thus qualify at
the same time as compact tetraquarks. This kind of states is discussed in some detail at the end
of the next section.

\section{Compact Tetraquarks}

The crucial building blocks of compact tetra- and pentaquarks are (anti-)diquarks---for
a  review about diquark properties we refer to Ref.~\cite{Barabanov:2020jvn}. Since quarks (antiquarks) live
in the color $[3]$ ($[\bar 3]$) representation, a quark (antiquark) pair lives either in the color $[\bar 3]$ or the
color $[6]$ ($[3]$ or $[\bar 6]$) representation. In many works as well as here, only the diquarks in the color $[\bar 3]$
representation are being kept, motivated, e.g., by the observation that at least for the one gluon exchange
only in this quark-quark channel the interaction is attractive. 
Thus (anti-)diquarks carry a color charge and are therefore
subject to confinement---the hadrons with these as building blocks are necessarily compact with their size given by the confinement radius. 
Model calculations reveal that the scalar (spin 0) diquarks are 
lighter than the axial-vector diquarks (spin 1)---this is why they were dubbed good and bad diquarks,
respectively. With this being said, one finds for the basis states forming doubly heavy tetraquarks~\cite{Maiani:2004vq}
two states with $S_h^{C}=0^{+}$,
\begin{equation}
|0^{+}\rangle = |0_{cq},0_{\bar c\bar q}; S_h=0\rangle \ , \ |0^{+'}\rangle = |1_{cq},1_{\bar c\bar q}; S_h=0\rangle \ ,
\label{eq:diquarks0++}
\end{equation}
three states with $S_h=1$  
\begin{equation}
|A\rangle = |1_{cq},0_{\bar c\bar q}; S_h=1\rangle \ , \ |B\rangle = |0_{cq},1_{\bar c\bar q}; S_h=1\rangle \ ,
 \ |C\rangle = |1_{cq},1_{\bar c\bar q}; S_h=1\rangle \ ,
\end{equation}
that can be combined into states with defined charge parity $C$ as
\begin{equation}
|1^{+}\rangle = \frac{1}{\sqrt{2}}(|A\rangle + |B\rangle) \ , \ |1^{-}\rangle = \frac{1}{\sqrt{2}}(|A\rangle - |B\rangle) \ , \
|1^{-}\rangle = |C\rangle \ , \
\label{eq:diquarks1}
\end{equation}
and finally one state with $S_h^{C}=2^{+}$, namely
\begin{equation}
|2^{+}\rangle = |1_{cq},1_{\bar c\bar q}; S_h=2\rangle \ .
\label{eq:diquarks2++}
\end{equation}
If the (anti-)diquark structures can be treated as well defined building blocks of the emerging hadrons,
to estimate the spectrum of compact tetraquarks one can
straightforwardly write down the pertinent interactions within the given hadron in analogy to what is known
from atomic physics as
\begin{eqnarray}
H=2m_{[cq]} + 2(\kappa_{cq})_{\bar 3}[(\vec S_c\cdot \vec S_q) + (\vec S_{\bar c}\cdot \vec S_{\bar q}) ]
+ 2(\kappa_{c\bar q})[(\vec S_c\cdot \vec S_{\bar q'}) + (\vec S_{\bar c}\cdot \vec S_{q}) ]
+ 2\kappa_{\bar qq}(\vec S_{q}\cdot \vec S_{\bar q'})+2\kappa_{\bar cc}(\vec S_{c}\cdot \vec S_{\bar c}) 
+\frac{B_{\cal Q}}{2}\vec{L}^2+2a_Y\vec L\cdot \vec S+b_Y S_{12} \ ,
\label{eq:Hdiquark_full}
\end{eqnarray}
where the first term captures the diquark masses and the next four account for 
the spin-spin interactions between the different quarks in the hadron.
The following two terms need to be added to describe states with inter-diquark angular
momenta larger than zero~\cite{Ali:2017wsf}, where $\vec S=\vec S_{\mathcal Q}+\vec S_{\mathcal{\bar Q}}$
with $\vec S_{\mathcal Q}=\vec S_c+\vec S_q$ and analogously for $\vec S_{\mathcal{\bar Q}}$.
Finally, the tensor-operator is defined via 
$$S_{12}=3(\vec S_{\mathcal Q}\cdot \vec n)(\vec S_{\mathcal{\bar Q}}\cdot \vec n)-\vec S_{\mathcal Q}\cdot \vec S_{\mathcal{\bar Q}} \ ,$$
where $\vec n$ is either $\vec r/r$ or $\vec q/q$ for calculations in coordinate space or 
momentum space, respectively, with $\vec r$ the distance between the building blocks and
$\vec q$ the momentum transfer.
The matrix elements of the above operators can be evaluated with standard methods and
the strength parameters need to be determined from experiment. The label on 
the first spin-spin term indicates that in the model sketched here the diquarks are considered in the
color $[\bar 3]$ representation only. Heavy quark spin symmetry
suggests that $\kappa_{cq}/\kappa_{q\bar q}\sim M_q/M_c$ and $\kappa_{c\bar c}/\kappa_{q\bar q}\sim (M_q/M_c)^2$.

We start with a discussion of $S$ wave states ($L=0$).
The Hamiltonian of Eq.~(\ref{eq:Hdiquark_full}) is diagonal in the $J^{PC}=1^{++}$ and $2^{++}$ channels
and Ref.~\cite{Maiani:2004vq} quotes
\begin{equation}
M(2^{++})=M(1^{++})+2[(\kappa_{cq})_{\bar 3}+\kappa_{c\bar q}] = 3952 \ \mbox{MeV} \ ,
\end{equation}
having identified the $1^{++}$ state with the $\chi_{c1}(3872)$ aka $X(3872)$ and estimated the various spin-spin interactions
from the masses of regular charmonia. It is furthermore argued that the $2^{++}$ state can be identified
with the tensor state observed at 3940~MeV. The given interaction predicts a $1^{+-}$ state at
3882~MeV which is nicely consistent with the mass of the isovector state $T_{c\bar c1}(3900)^+$ also
known as $Z_c(3900)$. However, the corresponding spin partner state is predicted at 3754~MeV, significantly
lighter than the $X(3872)$, 
while it is observed experimentally more than 100 MeV above, namely at 4020~MeV. To accommodate
this, in Refs.~\cite{Maiani:2014aja,Giron:2019cfc} it was proposed that contrary to the heavy quark spin symmetry scaling
quoted above, all spin-spin interactions but the first should be negligible. A justification for this unexpected pattern
was provided later in Ref.~\cite{Maiani:2017kyi}, where it was argued that the diquarks are separated by
some potential well, no allowing the short ranged spin-spin interactions to operate between different diquarks---also the dynamical diquark picture proposed in Ref.~\cite{Brodsky:2014xia} leads to a sizeable separation
of diquark and anti-diquark within the exotic hadron.
Then the strength parameter
of the remaining spin-spin interaction is estimated from the mass difference of $Z_c(3900)$ and $Z_c(4020)$ aka
$T_{c\bar c1}(3900)^+$ and $T_{c\bar c1}(4020)^+$, respectively,
 to be $(\kappa_{cq})_{\bar 3}=67$~MeV~\cite{Maiani:2014aja},
about three times larger than the value provided in Ref.~\cite{Maiani:2004vq}, where it was extracted
from the $\Sigma_c$-$\Lambda_c$ mass difference. In this way the $2^{++}$ and the second $1^{+-}$ state are predicted to have
a mass quite close to the $D^*\bar D^*$ threshold (in line with the
prediction in the molecular approach as detailed below).

\begin{figure}[t]
	\centering
	\includegraphics[width=7cm,height=6cm]{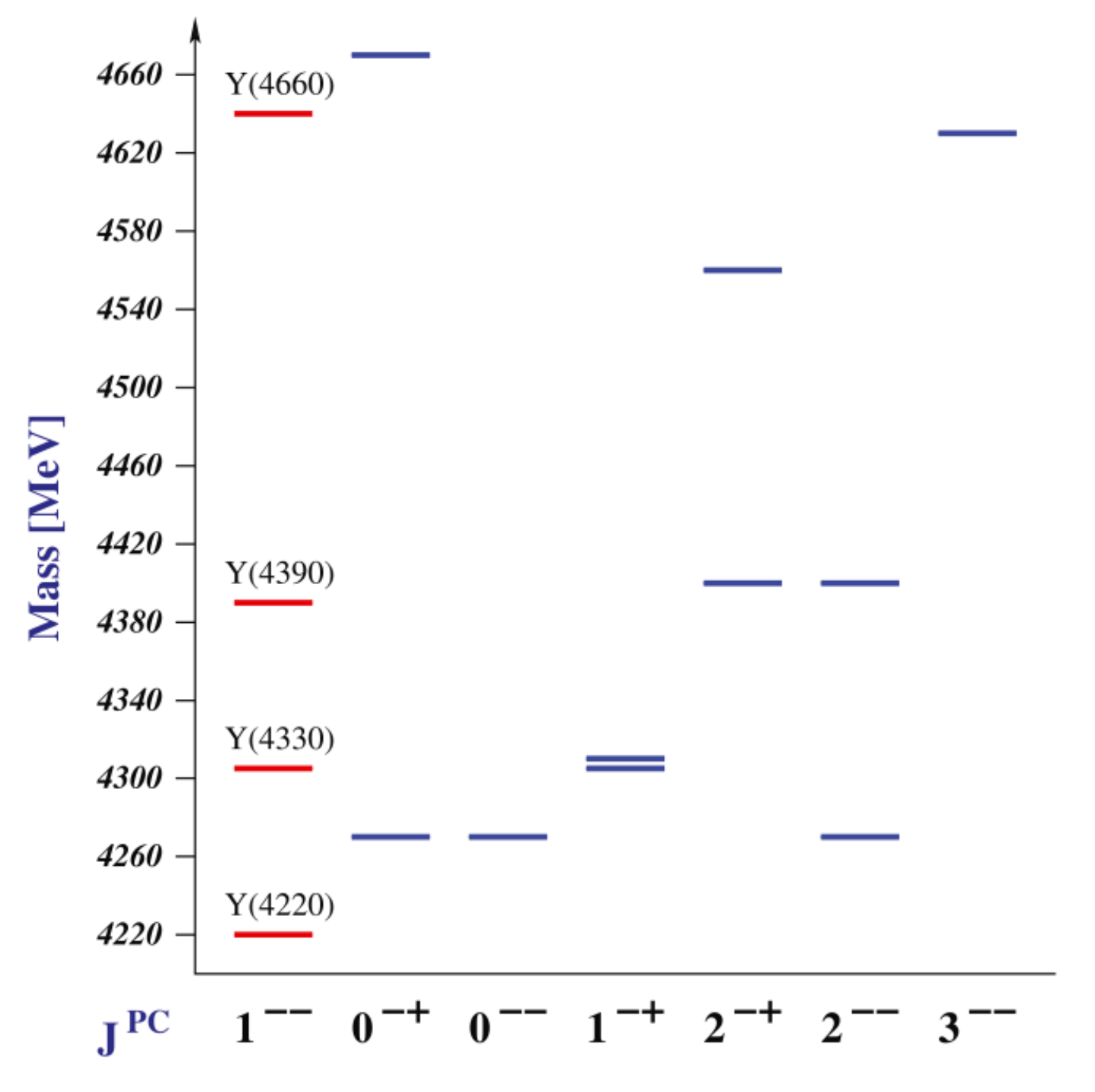}
	\caption{Spectrum of negative parity tetraquarks predicted in Ref.~\cite{Ali:2017wsf}.
	Note that the masses of the $1^{--}$ states used as input are slightly different 
	compared to those employed in the other sections, however, an adaption would not
	change the overall picture. Note that the quantum numbers $0^{--}$ and $1^{-+}$ are
	exotic---they cannot be reached from quark-antiquark states. Fig. from Ref.~\cite{Brambilla:2019esw}.}
	\label{Fig:tetraquarkspectrum}
\end{figure}

The building blocks for tetraquarks containing a heavy quark and its anti-quark with negative parity 
 are still the ones provided in Eqs.~(\ref{eq:diquarks0++}), (\ref{eq:diquarks1}), and (\ref{eq:diquarks2++}).
The negative parity is then provided by an angular momentum between the diquark and the anti-diquark.
For charge and flavor neutral states an odd angular momentum changes also the charge parity. Thus,
one get e.g. five states with $J^{PC}=1^{--}$: Two by adding one unit of angular momentum to
the two states provided in Eq.~(\ref{eq:diquarks0++}) or by coupling the total $S_h^+$ states
 of Eqs.~(\ref{eq:diquarks1}) and (\ref{eq:diquarks2++}) with an angular momentum
of one to a total angular momentum of one. Finally the state given in  Eq.~(\ref{eq:diquarks2++}) can be
combined with $L=3$ to a total angular momentum of one. The  $\vec L^2$-term pushes the vector state with $L=3$ 
so far up, that it does not need to be considered any further. One thus gets three additional parameters
that can be fixed from the vector states. Alternatively one can include the parameter $(\kappa_{cq})_{\bar 3}$
in the fit and see if a value consistent with that found for the $S$-wave states is found. This is the strategy
followed in Ref.~\cite{Ali:2017wsf} and indeed, one of the fits is consistent with the existing data as well
as the spin coupling term within 2$\sigma$~\footnote{In this work two fitting schemes were applied, however,
scheme I included the $Y(4008)$ which seems not to be confirmed by the data. We thus here
only quote the results for scheme II.}. The four vector states included in the fit were
$Y(4230)$, $Y(4320)$, $Y(4390)$ and $Y(4660)$, see Fig.~\ref{Fig:tetraquarkspectrum}. Once the parameters are fixed, the masses of other
exotics with different quantum numbers can be predicted, most strikingly the authors find two pseudoscalar
states at about 4700 and 4270 MeV, respectively, and even three states with quantum numbers forbidden in
the naive quark model, namely a $0^{--}$ state at a similar mass as the $Y(4230)$ and two near 
degenerate $1^{-+}$ states
at about 4310 MeV. 

As in the regular quark model, also in the compact tetraquark model  radial excitations
of the ground states should appear. For example a natural candidate for a radial excitation of $Z_c(3900)$
is the $Z_c(4430)$~\cite{Maiani:2014aja} aka $T_{c\bar c1}(3900)^+$  and $T_{c\bar c1}(4430)^+$, respectively, a state difficult to understand in other approaches. Moreover,
as in the standard quark model compact tetraquarks should fill complete SU(3) flavor multiplets, with
the amount of SU(3) breaking being driven by the strange quark mass 
$M_s-M_u=120-150$~MeV~\footnote{Note that quark masses in QCD depend on the renormalisation
scale and scheme---what is meant here are effective mass parameters employed in the model.}. 
In particular, as long as no quark type dependence is included in the interaction hamiltonian, each isoscalar state
is necessarily accompanied by a triplet of isovector states, with very similar mass---in analogy of the
near degeneracy of the isovector $\rho$ and the isoscalar $\omega$ in the light quark sector.
While this nicely explains the large isospin violation in the decays of $X(3872)$, it in effect leads to
a larger number of states than observed in experiment so far, although there are indications
that the SU(3) multiplets start to fill up~\cite{Maiani:2021tri}---note, however, that the experimental evidence as of today for the states that contain strangeness is rather weak.
As a further refinement of the diquark--anti-diquark interaction in Ref.~\cite{Giron:2019cfc} an isospin dependent contribution
is discussed. 

Recently LHCb discovered a very intriguing state in the $D^0D^0\pi^+$ mass distribution, the $T_{cc}(3875)^+$
with a minimal quark content of $cc\bar u\bar d$. In the tetraquark model such states were discussed 
qualitatively in Ref.~\cite{Esposito:2013fma}---more quantitative investigations can be found in Refs.~\cite{Karliner:2017qjm,Eichten:2017ffp}.
Contrary to the heavy-light diquarks discussed above, for this state the constituents need to be heavy-heavy
and light-light diquarks. Now the diquarks are subject to the Pauli principle. Therefore the charmed diquark needs to be
in the spin one state. The light anti-diquark appears as the good combination, spin 0, in the anti-symmetric flavor $[3]$, while the bad, spin 1, anti-diquarks are in the symmetric flavor
$[\bar 6]$, since still only the totally antisymmetric color $[3]$ configuration is kept. Accordingly, the lightest
tetraquark with the mentioned quark content needs to have the quantum numbers $J^P=1^+$ consistent with
current experimental observations, although an amplitude analysis is still missing.
The mass of the $T_{cc}(3875)^+$ basically coincides with the $D^{*+}D^0$ threshold. However, if indeed
the $cc$-diquark forms an essential ingredient in this state one should expect the flavor partner of the
$T_{cc}(3875)^+$ with $cc$ being replaced by $bb$ to quite deeply bound, simply since the heavier
quark pair should sit closer together feeling the much stronger attraction provided by the one gluon exchange (the increase in strength driven by the
$1/r$ scaling of the potential vastly overcomes the decrease in the
strong coupling $\alpha_s$ which comes in since the pertinent mass 
scale rises). 
Already in the 80ties it was stressed that as soon as the ratio of heavy to light quark masses  is well beyond
10, binding should occur for states of the kind discussed here~\cite{Ballot:1983iv,Zouzou:1986qh}. 
Note that these studies were performed within
constituent quark models and accordingly we here talk about ratios of constituent masses, which for
light quarks are of the order of 300 MeV not the elementary
current quark masses of the order of a few MeV. Thus, based on those studies one should expect sizeable
binding for the $bb\bar q\bar q$ systems only.

\begin{figure}[t]
	\centering
	\includegraphics[width=7cm,height=6cm]{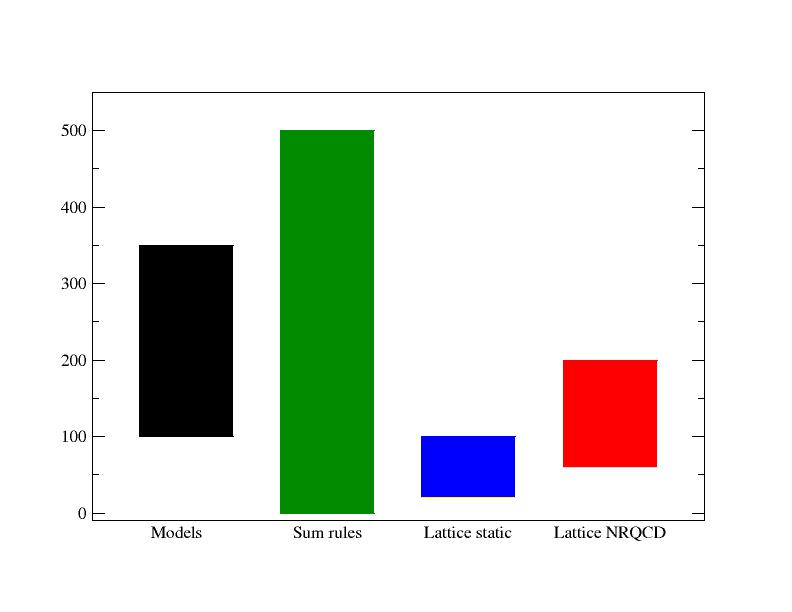}
	\caption{Comparison of the binding energies in MeV relative to the $BB^*$ threshold
	 predicted for $bb\bar u\bar d$ tetraquarks
	with quantum numbers $I(J^P)=0(1^+)$ from different approaches. The columns show in order the
	spread in the predictions from different works, typically with much smaller individual
	uncertainties, for phenomenological models, QCD sum rules, 
	lattice QCD employing static $QQ$ potentials and lattice QCD employing NRQCD for the heavy quarks. }
	\label{Fig:bbtetraquarks}
\end{figure}

By now there is a 
large number of studies for doubly heavy tetraquarks including phenomenological models~\cite{Ballot:1983iv,Zouzou:1986qh,Brink:1998as,Karliner:2017qjm,Eichten:2017ffp,
Park:2018wjk,Noh:2021lqs,Ma:2023int} lattice QCD either employing
static potentials~\cite{Bicudo:2012qt,Brown:2012tm,Bicudo:2015kna,Bicudo:2016ooe} or NRQCD for
the $b$-quarks~\cite{Francis:2016hui,Junnarkar:2018twb,Meinel:2022lzo,Mohanta:2020eed,Hudspith:2023loy,Alexandrou:2024iwi}
 as well as QCD sum rules~\cite{Navarra:2007yw,Du:2012wp,Wang:2017uld} .
They all agree that a $T_{bb}$ should
be bound deeply in the $I=0$ and $J^P=1^+$ channel, although there is still a considerable spread in the predicted binding energies relative to the lowest open charm threshold ($BB^*$), which
range from near zero to 500 MeV---see Fig.~\ref{Fig:bbtetraquarks}.

\section{Hadronic Molecules}
\label{sec:hadmol}

Atomic nuclei are well understood as bound systems of nucleons, protons and neutrons. The lightest non-trivial
nucleus is the deuteron, a bound state of a single proton with a single neutron and a binding energy of only
2.2 MeV. A relatively strongly bound nucleus is the $\alpha$ particle, made of two protons and two neutrons --- is has a binding
energy of 7 MeV per nucleon which translates into a separation energy into two deuterons of 12 MeV. 
In addition there are bound systems
of hyperons and nucleons, the so-called hypernuclei. Thus also there exists some SU(3) structure of hadron-hadron
bound systems. The hypertriton, which may be viewed as a bound system of a $\Lambda$ baryon and a deuteron, 
has a $\Lambda$ separation energy of only 0.13 keV. We thus observe that nuclear binding energies span over 
two orders of magnitude, however, always stay well below 100 MeV. We may use this as some guidance for what to
expect for hadronic molecules in general. 
Hadronic molecules are multi quark states whose substructure is given by color neutral hadrons. Thus they are the
analogue of atomic nuclei, 
only that their constituents are different hadrons than ground state baryons.

Hadronic molecules are a special realisation of states one may want to classify as two-hadron states: The corresponding
poles are generated through the hadron-hadron dynamics and not from e.g. gluon exchanges between quarks. The 
corresponding states may not only appear as near threshold bound or virtual states that qualify as hadronic molecules,
but even wide resonances such as the $f_0(500)$ which has a mass of the same order of magnitude as its width 
can fall into this class (see the discussion in~\ref{sec:multiquarks:light}).

To provide some general understanding
 under what conditions the scattering of two particles of mass $m_1$ and $m_2$ that feel
an attractive force develops a pole near threshold, we may study a zero range interaction---this is justified
as soon as the de Broglie wave length connected to the bound system is significantly larger than the 
range of forces.  Then the scattering amplitude at tree level in momentum space is given by a constant 
and we get for the loop expansion of the scattering amplitude in non-relativistic kinematics~\cite{Guo:2017jvc}
\begin{equation}
T_{\rm NR}=C \sum_{n=0}^{\infty}(C \Pi(E))^n =  \frac{C}{1+C \Pi(E)} \qquad \mbox{with}
\qquad \Pi(E)= -\int \frac{d^3 q}{(2\pi)^3}\frac{F^2_\Lambda(q)}{E-q^2/(2\mu)+i\epsilon} = \frac{\mu}{2\pi}
\left(\Lambda+i\sqrt{2\mu E+i\epsilon}\right) + {\mathcal O}(\Lambda^{-1}) \ ,
\label{eq:TNR}
\end{equation}
where $F_\Lambda(q)$ is some regularisation function and $\Lambda$ the regularisation scale.
For  $F_\Lambda$  one may either chose a sharp cut off, $F_\Lambda(q)=\theta(\pi\Lambda/2 - q)$, or
a Gaussian cut-off or employ more sophisticated methods like the
 so-called power divergence subtraction~\cite{Kaplan:1998tg} (for a detailed discussion see Ref.~\cite{Epelbaum:2008ga}). Here 
 $\mu=m_1m_2/(m_1+m_2)$ denotes the reduced
mass of the external particles. The condition for the appearance of a near threshold pole
in the amplitude may thus be expressed as
\begin{equation}
C^{-1} = -\Pi(-E_B)  = \frac{\mu}{2\pi}
\left(-\Lambda + \gamma\right) + {\mathcal O}(\Lambda^{-1}) \ ,
\end{equation}
with the binding momentum $\gamma = \sqrt{2\mu E_B}$, where $E_B=m_1+m_2-m_B$ with
$m_B$ for the mass of the studied bound state.
Formally the scale $\Lambda$ is not an observable and needs to be absorbed into the coupling strength
$C$. However, as soon as one would introduce more dynamics into the potential, $\Lambda$ would acquire 
a physical meaning like, e.g., the mass of the lightest exchange particle. Thus the appearance of a near threshold
pole is controlled by some interplay of the binding potential, the masses of the external particles and the binding
momentum. For $m_1\approx m_2$ we have $\mu \approx m_2/2$ and for $m_1\ll m_2$ we have $\mu\approx
m_1$. Thus we deduce that it is more natural for doubly heavy systems to generate bound states than for 
singly heavy ones, since in the former case already a rather weak attraction ($|C|$ small) leads to a bound state. 

The renormalised scattering amplitude, where $\Lambda$ is absorbed into
$C$, takes the simple form
\begin{equation}
T_{\rm NR}=\left( \frac{2\pi}{\mu}\right)\frac{1}{\gamma+i\sqrt{2\mu E+i\epsilon}} \ ,
\label{eq:mol_boundren}
\end{equation}
which depends, besides on the energy, only on the binding energy and the masses of the external
particles. In particular one gets for the residue at the pole
\begin{equation}
g^2_{\rm NR}=\lim_{E\to -E_B}(E+E_B)T_{\rm NR}=\frac{1}{d\Pi(E)/dE|_{E=-E_B}}=\frac{2\pi\gamma}{\mu^2} \ ,
\label{eq:molres}
\end{equation}
also independent of the detailed dynamics that lead to the appearance of the pole. In fact, shallow bound states develop
some universal properties---for a review see Ref.~\cite{Braaten:2004rn}.

The discussion above applies to purely dynamically generated states. The equations were generalised by Weinberg~\cite{Weinberg:1965zz} (inelastic channels were later included in Ref.~\cite{Baru:2003qq}). 
The study starts from a two component wave function of a bound state interacting via some hamiltonian
$\mathcal H$:
\begin{equation}
|\Psi \rangle = \left(\lambda|\psi_0\rangle\atop 
\chi (\mathbf{p}) |h_1h_2 \rangle \right) 
\qquad \mbox{solving} \qquad \hat {\mathcal H} |\Psi\rangle = E |\Psi\rangle,~~
\hat {\mathcal H} = \left(\begin{array}{cc}
\hat{H}_c&\hat{V}\\
\hat{V}&\hat{H}_{hh}^0
\end{array}
\right) \ .
\end{equation}
Here $\psi_0$ ($h_1h_2$) is the compact (two hadron) component of the wave function. 
The terms in the hamiltonian include originally the interquark potential which also contains the mechanism
of confinement, $\hat{H}_c$, the transition potential between the compact component and the continuum,
$\hat{V}$, and the hamiltonian for the two-hadron system.
The quantity of interest here is 
\begin{equation}
\lambda^2 = \langle \psi_0|\Psi\rangle ^2  \ ,
\end{equation}
which is the probability to find the compact component of the wave function in the full wave function
and is thus a direct measure of the composition of the wave function.
In earlier works Weinberg
demonstrated that the non-perturbative parts of the hadron-hadron interaction can be absorbed into the
effective parameters of the formalism by a proper field redefinition~\footnote{This field redefinition 
is possible only, if there is only one bound-state pole in the system. Otherwise a more complicated
treatment becomes necessary~\cite{Baru:2010ww,Hanhart:2011jz,Guo:2016bjq}.} such that to leading order in a momentum
expansion $\hat{H}_{hh}^0=p^2/(2\mu)$~\cite{Weinberg:1962hj,Weinberg:1963zza}. With this one finds for
the two-hadron wave function
\begin{equation}
\chi(p)=\lambda \, \frac{f(p^2)}{E-p^2/(2\mu)} \qquad \mbox{with} \qquad f(p^2)=\langle h_1h_2|\hat V|\psi\rangle \ .
\end{equation}
The normalisation condition for physical bound states then can be expressed as
\begin{equation}
1=\langle \Psi|\Psi\rangle = \lambda^2 \left(1
+\int \frac{d^3 p}{(2\pi)^3} \frac{f^2(p^2)}{(E_B+p^2/(2\mu))^2}\right)
= \lambda^2 \left(1
+g_0^2\int \frac{d^3 p}{(2\pi)^3} \frac{1}{(E_B+p^2/(2\mu))^2}+\mathcal{O}\left(\frac{\gamma}{\beta}\right)\right)
=\lambda^2 \left(1
+\frac{g_0^2\mu^2}{2\pi\gamma}+\mathcal{O}\left(\frac{\gamma}{\beta}\right)\right)
\ ,
\end{equation}
where $g_0=f(0)$ and $\beta$ denotes the intrinsic momentum scale in the transition form factor
$f(p^2)$. This allows one to relate the effective coupling $g_0^2$ to the quantity of interest, $\lambda^2$,
\begin{equation}
g_0^2 = \frac{2\pi\gamma}{\mu^2}\left(\frac{1}{\lambda^2}-1+\mathcal{O}\left(\frac{\gamma}{\beta}\right)
\right) \ .
\label{eq:g0oflambda}
\end{equation}
It is straightforward to calculate the scattering amplitude within the same formalism under the assumption
that the scattering is dominated by a single pole. All it takes is to replace in Eq.~(\ref{eq:TNR}) $C$ by
$g_0^2/(E-M_0)$. Now the scale dependence is to be absorbed into the bare mass $M_0$ via that
renormalisation condition $E_B=-M_0+g_0^2\mu/(2\pi)(\Lambda-\gamma)$, where we used that 
the analytic continuation of the square root of the energy on the first sheet (here the focus is on bound
states) gives the positive imaginary momentum.
Then the scattering matrix reads~\cite{Guo:2017jvc}
\begin{equation}
T_{\rm NR}=\frac{g_0^2}{E+E_B+g_0^2\mu/(2\pi)(ip+\gamma)} \ .
\label{eq:TNRpole}
\end{equation}
For a pure molecule we have $\lambda\to 0$ and accordingly $g_0^2\to\infty$. In that limit  
Eq.~(\ref{eq:TNRpole}) agrees with Eq.~(\ref{eq:mol_boundren}).
Moreover, to get from the bare coupling provided in Eq.~(\ref{eq:g0oflambda}) to the residue at
the pole, it needs to be multiplied with the $Z$-factor that happens to agree with $\lambda^2$. This
gives
\begin{equation}
g_{\rm NR}(\lambda^2)^2 = \lambda^2 g_0^2 = \frac{2\pi\gamma}{\mu^2}\left(1-{\lambda^2}+\mathcal{O}\left(\frac{\gamma}{\beta}\right)\right) \ ,
\end{equation}
which agrees to Eq.~(\ref{eq:molres}) for $\lambda^2=0$ when the range corrections are omitted. Eqs.~(\ref{eq:g0oflambda})
and~(\ref{eq:TNRpole}) nicely illustrate what the Weinberg criterion really does: It measures the importance of the
term non-analytic in the energy, $p=\sqrt{2\mu E}$, which appears since the two-hadron intermediate state can go
on shell and is thus specific for the molecular component, relative to the terms analytic in energy that also appear
for compact structures.

The effective range expansion (ERE) reads
\begin{equation}
T_{\rm NR}(E)=-\frac{2\pi}{\mu}\frac{1}{1/a+(r/2)p^2-ip} \ ,
\label{eq:ERE}
\end{equation}
with $a$ and $r$ for the scattering length and the effective range, respectively, and we used the particle
physics sign convention for the scattering length. Matching Eq.~(\ref{eq:ERE})
and Eq.~(\ref{eq:TNRpole}) gives
\begin{equation}
\frac{1}{a} =-\frac{2\pi E_B}{\mu g_0^2}-\gamma \ , \quad r = -\frac{2\pi}{g_0^2\mu^2}\qquad \Longrightarrow  \qquad {a=-2\frac{1-\lambda^2}{2-\lambda^2}\left(\frac{1}{\gamma}\right)  
+{\cal O}\left(\frac{1}{\beta}\right)} \ , \quad  {r= -\frac{\lambda^2}{1-\lambda^2}\left(\frac{1}{\gamma}\right)  +{\cal O}\left(\frac{1}{\beta}\right)} \ ,
\label{eq:arrelations}
\end{equation}
where Eq.~(\ref{eq:g0oflambda}) was used to provide relations in terms of the probability $\lambda^2$. 
One thus finds that  for a pure molecule, $\lambda^2=0$, $a=-1/\gamma+{\cal O}\left({1}/{\beta}\right)$ and
$r={\cal O}\left({1}/{\beta}\right)$ and for a purely compact structure, $\lambda^2=1$,  
$a={\cal O}\left({1}/{\beta}\right)$ and $r\to -\infty$. While the scattering length is very sensitive to the
actual binding energy, it is the effective range that is most sensitive to the binding mechanism and
thus to the structure of the state as will be discussed below. Moreover, because of the range corrections
indicated in Eqs.~(\ref{eq:arrelations}) one should not use $a$ and $r$ as input to extract $\lambda^2$~\cite{Song:2022yvz}, but
more use the relations to check if, within errors, the properties of a given system are consistent with
e.g. a pure molecule ($\lambda^2=0$).

\begin{figure}[t]
	\centering
	\includegraphics[width=7cm,height=7cm]{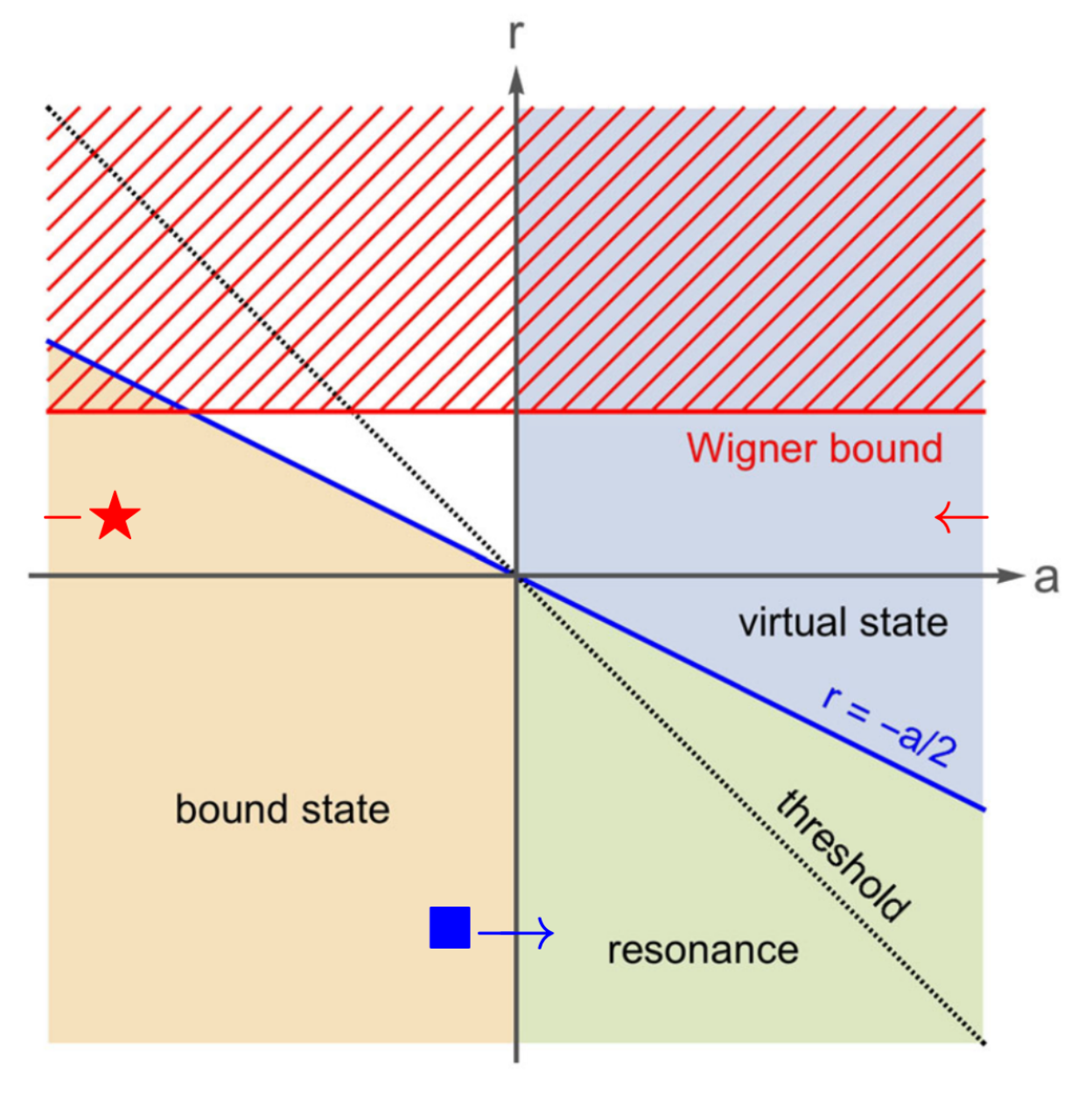}
	\caption{The scatteringlength-effective range plane: Shown is for which
	pairs of values what kind of pole emerges. The area left white and the hatched
one are forbidden 
	from causality---the former because the related poles are located in the complex plane
	of the first sheet, the latter because of the Wigner bound. The dotted line ($r=-a$) refers to a pole with a real part
	exactly at the threshold. The red star (blue square) shows a typical location of the pole of a 
	bound state with a molecular structure (compact structure). The respective arrows show the characteristic
	trajectory of the poles as the interaction gets changed slightly. 
	The figure is adapted from Ref.~\cite{Matuschek:2020gqe}.}
	\label{Fig:raplane}
\end{figure}

In Fig.~\ref{Fig:raplane} the red star and the blue square show the pole locations for bound states 
with a molecular and compact structure in a plane defined by scattering length and effective range, respectively. The figure also assigns regions in the $(r-a)$ plane
to the types of poles, namely bound states, real valued poles on the first sheet of the complex energy plane, virtual states, real values poles
on the second sheet, and resonances, complex valued poles in the complex plane of the second sheet. 
The probabilistic interpretation of $\lambda^2$ was derived from the normalisation condition of the bound
state. Clearly it cannot be copied straightforwardly to virtual states and resonances, since their wave functions
cannot be normalised. However, it is instructive to ask what happens, if the interaction strength that lead
to the pertinent pole gets slightly weakened (e.g. by changing the quark mass). Since the effective range is
controlled by the type of the binding interaction it will change little, however, the scattering length will
change its sign (see, e.g., Refs.~\cite{Hanhart:2008mx,Hanhart:2014ssa} and~\cite{Guo:2009ct} for detailed studies of light and singly heavy systems;
the situation is potentially more complicated for doubly heavy systems as described below): For molecular
structures it is the inverse scattering length that changes smoothly (a molecular bound state exactly at threshold is
characterised by an infinite scattering length), while for compact structures it is the scattering length itself.
As indicated in Fig.~\ref{Fig:raplane}, the weakening of the interaction thus transforms a molecular bound state
into a virtual state, but a compact state into a resonance. One thus needs to conclude that a virtual pole necessarily
is generated from non-perturbative two-hadron interactions and cannot be generated from a compact state.
There are a large number of works available in the literature that discuss generalizations of Weinberg's
criterion also to resonances~\cite{Baru:2003qq,Matuschek:2020gqe,Bruns:2019xgo,Oller:2017alp,Kang:2016jxw,Sekihara:2016xnq,Kamiya:2016oao,Guo:2015daa,Sekihara:2014kya,Aceti:2012dd,Gamermann:2009uq,Kinugawa:2024kwb}, which, however, we cannot discuss in detail here.

The 
derivation provided so far was based on single channel scattering with zero range interactions only, however, 
crucial information is encoded
in the effective range. In fact, it appears that a positive
effective range is an unambiguous signature of a purely molecular state~\cite{Weinberg:1965zz,Esposito:2021vhu,Baru:2021ldu,Li:2021cue,Albaladejo:2022sux},
fully in line with the general theorem that potential scattering in a single channel with purely attractive
interactions necessarily
has a positive effective range~\cite{Bethe:1949yr}. Unfortunately reversing this statement does not
work: In general not even a sizeable   negative effective range
provides a unique signature of a compact structure, since coupled channel effects also induce 
a negative contribution to the effective range. For illustration we may start from generalising
Eq.~(\ref{eq:TNRpole}) to two channels to find~\cite{Baru:2021ldu}\footnote{In
Ref.~\cite{Hanhart:2007yq} an analogous formula was proposed, however, without the
subtraction terms $\gamma_i$. Then, however, $E_B$ is not the binding energy and in fitting
data potentially huge correlations between the parameters appear.}
\begin{equation}
T_{\rm NR}^{(ij)}=\frac{g_0^{(i)}g_0^{(j)}}{E+E_B+g_0^{(1)\, 2}\mu_1/(2\pi)(ip_1+\gamma_1)+g_0^{(2)\, 2}\mu_2/(2\pi)(ip_2+\gamma_2)} \ ,
\label{eq:TNRpole_cc}
\end{equation}
where the on-shell momenta of the particles in channel $i$ are 
 $$p_i(E)=\sqrt{2\mu_i(E-\delta_i)}\Theta(E-\delta_i) + i\sqrt{2\mu_i(\delta_i-E)}\Theta(\delta_i-E)$$
 using the channel specific reduced masses $\mu_i$
and as before $\gamma_i=p_i(-E_B)$. If the total energy is measured relative to the lowest threshold, which
we assign to channel 1, one gets $\delta_1=0$ and $\delta_2=m_1^{(2)}+m_2^{(2)}-m_1^{(1)}-m_2^{(1)}$.
In principle the denominator of Eq.~(\ref{eq:TNRpole_cc}) could also contain additional inelasticities that we
neglect here, for simplicity.  Clearly, the higher channel provides a contribution to the effective range,
which now reads~\cite{Matuschek:2020gqe,Esposito:2021vhu,Baru:2021ldu}
\begin{equation}
r = -\frac{2\pi}{g_0^{(1)\, 2}\mu_1^2} - \frac{g_0^{(2)\, 2}\mu_2}{g_0^{(1)\, 2}\mu_1^2}\sqrt{\frac{\mu_2}{2\delta_2}}
+   {\cal O}\left(\frac{1}{\beta}\right)  \ .
\label{eq:rcorr}
\end{equation}
The second term on the right hand side follows straightforwardly from expanding $p_2$ around $E=0$. 
It should be stressed that the sign of this term is fixed from unitarity. In other words,
the effect of heavier channels on the effective range is necessarily negative and can thus mimic the 
presence of a compact component of a studied shallow resonance.
In case of the $X(3872)$ or the $T_{cc}(3875)$
the distances between the two pertinent thresholds are with 
$\delta_2{=}m_{D^+}{+}m_{\bar D^{*\, -}}{-}m_{D^0}{-}m_{\bar D^{*\, 0}}{=}8$~MeV and 
$\delta_2{=}m_{D^+}{+}m_{D^{*\, 0}}{-}m_{D^0}{-}m_{D^{*\, +}}{=}1.4$~MeV, respectively, very
small, since both emerge from the isospin violating mass differences between the charged and neutral $D^{(*)}$-mesons. 
Accordingly, the second term above is e.g. with $-1.4$~fm for the $X(3872)$ quite sizeable and should be subtracted before the
Weinberg criterion is applied. In Ref.~\cite{Baru:2021ldu} it is argued that the recent analysis
of the line shape of $X(3872)$ by LHCb~\cite{LHCb:2020xds} 
only allows for an extraction of a lower bound for $g_0^{(1)\, 2}$
and accordingly one is to conclude from current line shape studies that the parameters of the $X(3872)$ are
fully consistent with a pure molecule. 

The underlying physics of the above mentioned range corrections is the exchange of mesons
of a finite mass. Mathematically spoken these exchanges introduce a left-hand (energies smaller than 0) 
branch-point and with it a left-hand cut (lhc) into the  amplitude, located at~\cite{Du:2023hlu}
\begin{equation}
E_{\rm lhc}^{x} = \frac{1}{8\mu}\left[(\Delta M)^2-m_x^2\right] 
\end{equation}
for the exchange of a particle of mass $m_x$ and a mass difference of the 
external particles in the process of the emission of $\Delta M$. This branch-point introduces a non-analyticity
into the amplitude that invalidates the ERE in the simple form provided
in Eq.~(\ref{eq:ERE}).
The left-hand branch point is closest to
the physical axis for light exchange particles. Accordingly, the leading lhc in case of nucleon-nucleon
scattering (as long as we neglect the exchange of photons), where $\Delta M=0$, comes from the one-pion exchange and is located at
$E_{\rm lhc}^{\pi}[NN] = -5$ MeV---in this case the ERE can be used to extract the
pole of the deuteron located at $-2.2$~MeV. In case of $BB^*$ scattering, where $\Delta M{=}M_B^*{-}M_B{=}45$~MeV (in
the process of a pion emission a $B$ gets converted into a $B^*$),
we get $E_{\rm lhc}^{\pi}[BB^*] = -2$ MeV. For $DD^*$ scattering there is no lhc, 
since here $\Delta M{=}M_D^*{-}M_D{=}140.6$~MeV${>m_\pi}$ and a positive value for $E_{\rm lhc}^{x}$
indicates the presence of a three-body cut instead of a lhc. However, for slightly larger than physical 
pion masses, as is studied in lattice QCD, the left-hand cut is located very close to the physical axis.
For examply, for $m_\pi=280$~MeV and 1927 and 2049~MeV for the masses of the $D$ and the $D^*$
meson as used in Ref.~\cite{Padmanath:2022cvl}, the left-hand branch point is located at $-8$~MeV while the 
pole extracted from the lattice data using the ERE is located at $-10$~MeV. Thus, in this case
the ERE in its original formulation should not be employed---see Ref.~\cite{Du:2024snq}. Moreover, to extract phase shifts from the lattice energy
levels the L\"uscher method was employed that also calls for a modification in the presence
of left-hand cuts~\cite{Meng:2023bmz,Hansen:2024ffk,Bubna:2024izx}.
Moreover, the pole trajectories in this case look a lot more complicated than what is discussed in Refs.~\cite{Hanhart:2008mx,Hanhart:2014ssa,Guo:2009ct} as detailed in Refs.~\cite{Abolnikov:2024key,Collins:2024sfi}.

\begin{figure}[t]
	\centering
	\includegraphics[width=7cm,height=1.7cm]{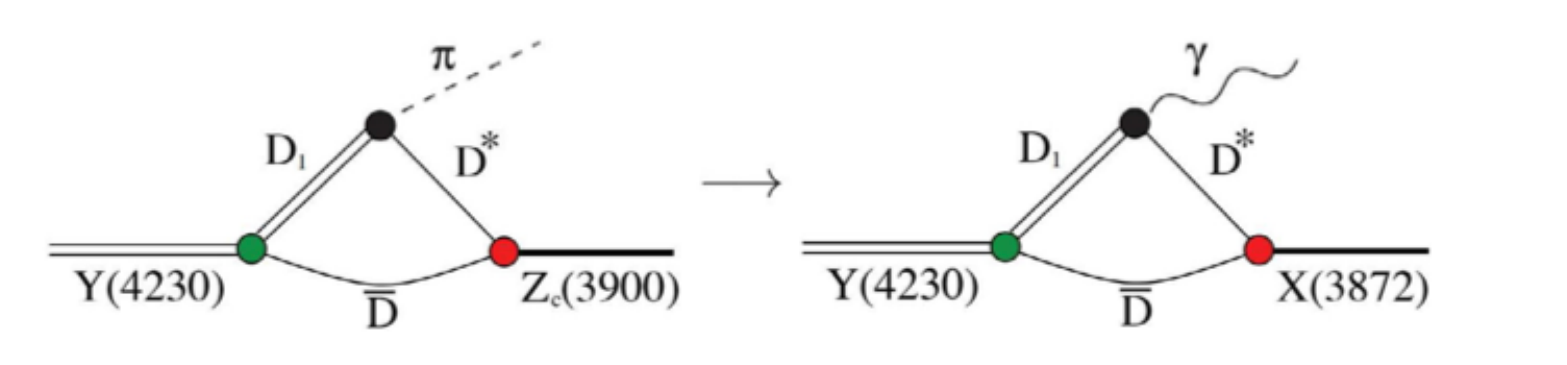}
	\caption{The relation between $Y(4230)\to \pi Z_c(3900)$ and $Y(4230)\to \gamma X(3872)$ in leading
	order EFT und der assumption that all external charmonium like states are hadronic molecules. Fig. from Ref.~\cite{Brambilla:2019esw}.}
	\label{Fig:Y2ZandX}
\end{figure}

As demonstrated above, important information on the
nature of the state is encoded in the effective coupling $g_0^2$ or, more concretely, the related residue,
$g_{\rm NR}(\lambda^2)^2$, which gets maximal for $\lambda^2=0$. Having that said it becomes clear
that only those observables where one is sensitive to the mentioned residue can be sensitive to the molecular component.
However, this is not always given. For example, the decays $X(3872)\to \gamma \psi$, where $\psi$ is either $\psi(2S)$
or $J/\psi$, can either go via a $D^*\bar D\to \gamma D^{(*)}\bar D ^{(*)}\to \gamma J/\psi$ triangle diagram, which
scales with $g_{\rm NR}(\lambda^2)$,
or a contact transition, which does not. The latter type of diagram is necessary to absorb the divergence of the former. Thus, 
the total rate of this transition cannot be sensitive to the molecular component of the $X(3872)$~\cite{Guo:2014taa}
in a model independent way, contrary to what is claimed,
e.g.,
in Refs.~\cite{Swanson:2004pp,Grinstein:2024rcu}. However, clearly these radiative decays are excellent observables
to test model predictions~\cite{Barnes:2003vb,Swanson:2004pp,Dong:2008gb,Dong:2009uf,Giacosa:2019zxw,Lebed:2022vks} (see also the compilation provided
in Ref.~\cite{LHCb:2024tpv}). On the other hand for some other decays non-trivial predictions are possible. For example,
 if $Y(4230)$ and $Z_c(3900)$ are both
molecular made of $D_1\bar D$ and $D^*\bar D$, respectively, there is no leading order counter term
for the transition $Y(4230)\to \pi Z_c(3900)$, which starts at one loop level~\cite{Wang:2013cya}---see left diagram
in Fig.~\ref{Fig:Y2ZandX}. Moreover,
this transition is enhanced by a nearby triangle singularity~\cite{Wang:2013hga}. If, in addition, also the
$X(3872)$ is a $D^*\bar D$ molecular state, then the same topology contributes to $Y(4230)\to \gamma X(3872)$
(see right diagram of Fig.~\ref{Fig:Y2ZandX}) such that its rate could be predicted~\cite{Guo:2013zbw}. The prediction
was confirmed shortly after by BESIII~\cite{BESIII:2013fnz}.

\begin{figure}[t]
	\centering
	\includegraphics[width=16cm,height=12cm]{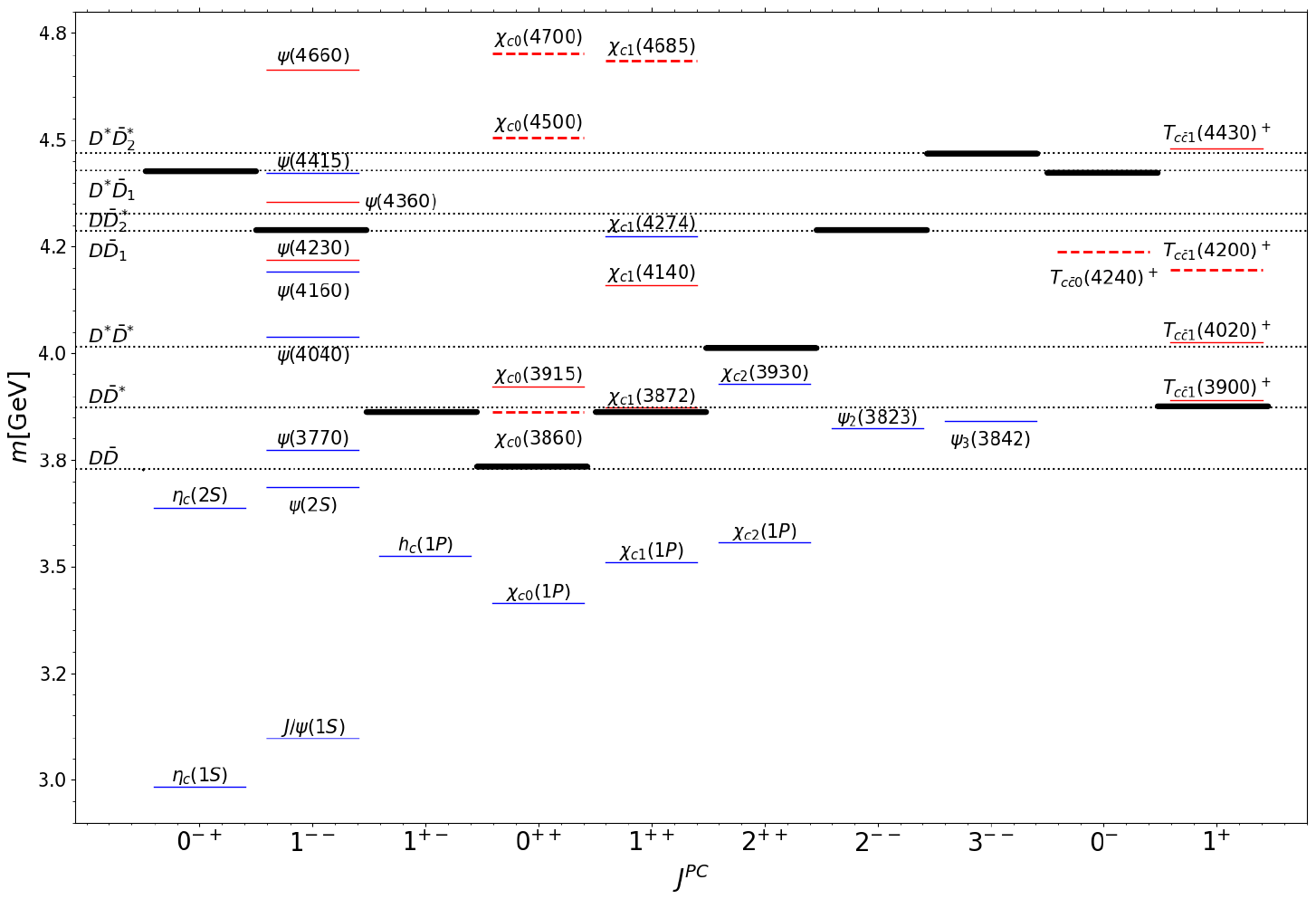}
	\caption{Current spectrum of charmonium and charmonium like states. Solid lines show states 
	well established while dashed lines those where confirmation is still to happen. Blue lines
	relate to states showing properties in line with the quark model (using the results of Ref.~\cite{Godfrey:1985xj} as a guidance), red lines those with unconventional
	properties. To distinguish isovector states a $+$ is 
    attached to the name. Thin dashed lines indicate the thresholds for the particle pairs put on the left
	(charge conjugation is understood to be included). The thick, solid lines show the lowest thresholds,
	where two narrow open charm states can combine to the given quantum numbers.}
	\label{Fig:molspecgeneral}
\end{figure}

After the quite general discussion about hadronic molecules we now come to the direct implications
of a molecular structure for doubly heavy systems. For the rest of this chapter we focus on the hypothesis that the states studied are in fact purely molecular and ask what imprint that has on observables. 
We start with the spectrum. First of all it is important to note that only narrow hadrons can form hadronic
molecules, since typically the widths of the constituents set a lower limit to the width of the molecule~\cite{Filin:2010se}---the width can get smaller
than that of the constituents only, if the reduction in phase space
provided by the bound state mass being smaller that the nominal threshold
of the constituents is relevant as is the case for the $T_{cc}$. Another
view on this situation is to acknowledge that the time scale of the formation of the molecule needs to be
much shorter than the life-time of the constituents~\cite{Guo:2011dd}. Thus we have at our disposal as building
blocks for hadronic molecules the spin doublets $\{D,D^*\}$, with negative parity and negligible widths, and $\{D_1(2420),D_2(2460)\}$
with positive parity and widths below 50~MeV. All 4 states fill a flavor $[\bar 3]$ multiplet.
 In contrast to the listed narrow states, the doublet $\{D_0(2300),D_1(2430)\}$ having widths of the order
of 300~MeV, is not expected to form hadronic molecules. 

Since the centrifugal barrier acts like a repulsive force, in general hadronic molecules should typically appear
in $S$-waves. Accordingly, the quantum numbers of the constituents dictate those of the composite state.
In Fig.~\ref{Fig:molspecgeneral} both is shown the thresholds for the various hidden charm channels (as the
thin dashed lines)
as well as the lowest threshold where the given quantum numbers can be reached in an $S$-wave by combining
the quantum numbers of the constituents~\cite{Cleven:2015era}. It is interesting to observe that with one exception---the $T_{c\bar c0}(4240)^+$, that still awaits confirmation---all states 
with unconventional properties observed so far appear either just below or above the thick black line, as
expected for hadronic molecules. Moreover, from these considerations it follows that, if the $Y(4230)$ 
is a $D_1\bar D$ molecular state~\cite{Ding:2008gr,Wang:2013cya,Cleven:2013mka}, its lightest spin partner state with $J^P=0^-$ must
be located near the $D_1\bar D^*$ threshold and thus be 140 MeV  heavier than the $Y(4230)$. This prediction
was confirmed in a microscopic calculation~\cite{Ji:2022blw} that
even puts a state with the exotic quantum numbers\footnote{Those
are quantum numbers that cannot be reached by $\bar qq$ states.} $J^{PC}=0^{--}$ in
this mass range.
The very same calculation also identifies $\psi(4230)$, $\psi(4360)$, and $\psi(4415)$ as
hadronic molecules with $J^{PC}=1^{--}$ and binding energies between
about 50 and 70 MeV in the channels $D_1\bar D$, $D_1\bar D^*$
and $D_2\bar D^*$, respectively.

A state that does not fit into the classification as $S$-wave hadronic molecules is the $Z_c(4430)$, since it sits
in the mass range of the $D^*\bar D_2$ threshold, but has positive parity. 
A possible explanation for a two-hadron structure of it could be a $P$-wave $D^*\bar D_1$
state~\cite{He:2017mbh}. This explanation calls for assigning the $Y(4390)$ as its $S$-wave companion.
If this explanation were correct, there should be a signature of the $Z_c(4430)$ in the $D^*D^*\pi$ subsystem 
of e.g. $B\to K D^*D^*\pi$ (so far the $Z_c(4430)$ showed up only in $B\to KZ_c(4430)$).
While the model study of Ref.~\cite{Wang:2023ivd} does not confirm the mentioned assignment, it finds a series of isoscalar $P$-wave states.
In Ref.~\cite{Lin:2024qcq} another  meson exchange model is employed to
connect a predicted $P$-wave $D\bar D^*$
resonance with the well established $X(3872)$ and $Z_c(3900)$ as well as the $T_{cc}(3875)$ all having a mass very close to the $D\bar D^*$ threshold.

\begin{figure}[t]
	\centering
	\parbox{7cm}{\includegraphics[width=8cm,height=5.cm]{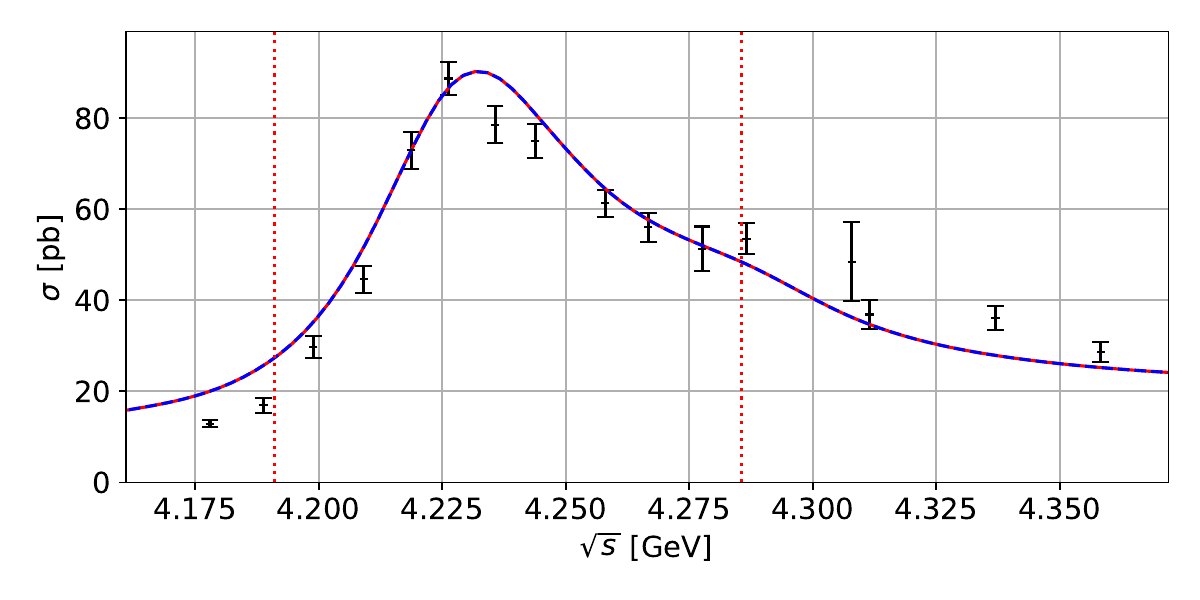}}
	\parbox{7cm}{\includegraphics[width=8cm,height=5.cm]{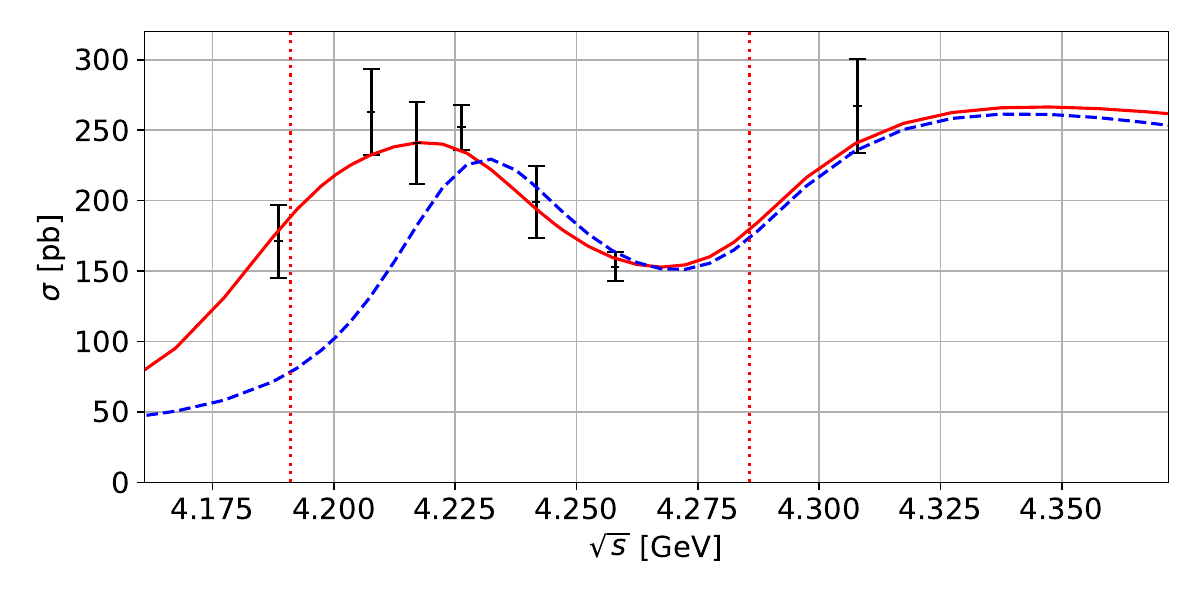}}
	\caption{The reactions $e^+e^-\to J/\psi \pi\pi$ (left panel) and  $e^+e^-\to D^*\bar D\pi$ (right panel)
	calculated in the molecular model. The red solid line shows the results of the full
	model of Ref.~\cite{vonDetten:2024eie}, while for the blue dashed line the effect of
	the $\psi(4160)$ was switched of. The data are from Refs.~\cite{BESIII:2022qal} and \cite{BESIII:2018iea} for the left
	and right panel, respectively. }
	\label{Fig:Y4230decays}
\end{figure}

The key characteristic of a hadronic molecule is that it has a strong coupling to its constituents.
Therefore, the nearby pole must lead to some imprint of the pertinent continuum threshold in observables.
This is illustrated in Fig.~\ref{Fig:Y4230decays} on the example of two decay channels of $Y(4230)$.
The lines are from a model study that includes the $D_1\bar D$ molecular dynamics---in particular 
the effect of the $D_1\bar D$ channel is included fully dynamically~\cite{vonDetten:2024eie}. 
Moreover, the data (also in other channels) called for the inclusion of not only the $Y(4230)$ but also
the $\psi(4160)$. With these ingredients it was possible to describe almost all available data
for the $Y(4230)$ with a consistent model. Especially, no extra exotic
state near 4320~MeV is necessary to describe the data, contrary to the analyses of e.g. Refs.~\cite{BESIII:2022qal,Nakamura:2023obk}, since the highly asymmetric line
shape of the $J/\psi \pi\pi$ channel is generated from the $D_1\bar D$ cut.
The line-shape that emerged for the $D^*\bar D\pi$ channel is characteristic for the decay for
a hadronic molecule with an unstable constituent having a width of the same order of magnitude
as the binding energy~\cite{Braaten:2007dw,Hanhart:2010wh}.

Typically the interaction of two heavy mesons is in principle modelled analogously to the nucleon-nucleon interaction, reviewed
in Ref.~\cite{Epelbaum:2008ga}. However, since the data situation is a lot worse than for the two-nucleon system,
there is no consensus yet on the most relevant contribution to the binding potential. While many follow Ref.~\cite{Voloshin:1976ap}
which suggest that vector meson exchanges should provide the bulk of the binding~\cite{Dong:2021juy,Gamermann:2006nm},
others follow Ref.~\cite{Tornqvist:1993ng} assuming the one-pion exchange as the prime contribution to the binding~\cite{Wang:2013kva}.
Other works include both mechanism together with others~\cite{Qiu:2023uno,Ke:2021rxd,Li:2012ss}---clearly the list
of references is not exhaustive.

In addition to model calculations there are effective field theory (EFT) studies for doubly heavy molecular states.
It should be clear that they suffer from
less predictive power than model predictions. For example 
\begin{itemize}
\item meson-meson and meson-antimeson scattering are not related in an non-relativistic EFT,
\item the short ranged operators for different total isospin channels are not related and thus the existence
of e.g. an isovector partner of a molecular $X(3872)$ can be deduced only from an analysis of high quality data~\cite{Zhang:2024fxy} and
\item it is not even possible to exploit heavy quark
flavor symmetry for doubly heavy systems~\cite{Baru:2018qkb}.
\end{itemize}
However,
 EFTs allow one to fully and systematically 
exploiting the implications of the symmetries of the underlying theory (here QCD) on the hadronic observables and thus
allow for model independent insights.
There are in general three different classes of approaches available in the literature, namely
pionless EFT~\cite{AlFiky:2005jd,Albaladejo:2015dsa}, EFT with perturbative pions~\cite{Fleming:2007rp,Valderrama:2012jv,Mehen:2011yh,Dai:2019hrf,Dai:2023mxm,Jansen:2013cba}, and
EFT with non-perturbative pions~\cite{Baru:2011rs,Baru:2013rta,Schmidt:2018vvl,Baru:2015tfa,Baru:2016iwj,Wang:2018jlv,Baru:2019xnh,Du:2021zzh}.
Again, also those find their analogs in studies of the nucleon-nucleon interaction~\cite{Epelbaum:2008ga}.

\begin{figure}[t]
	\centering
	\includegraphics[width=12cm,height=7.cm]{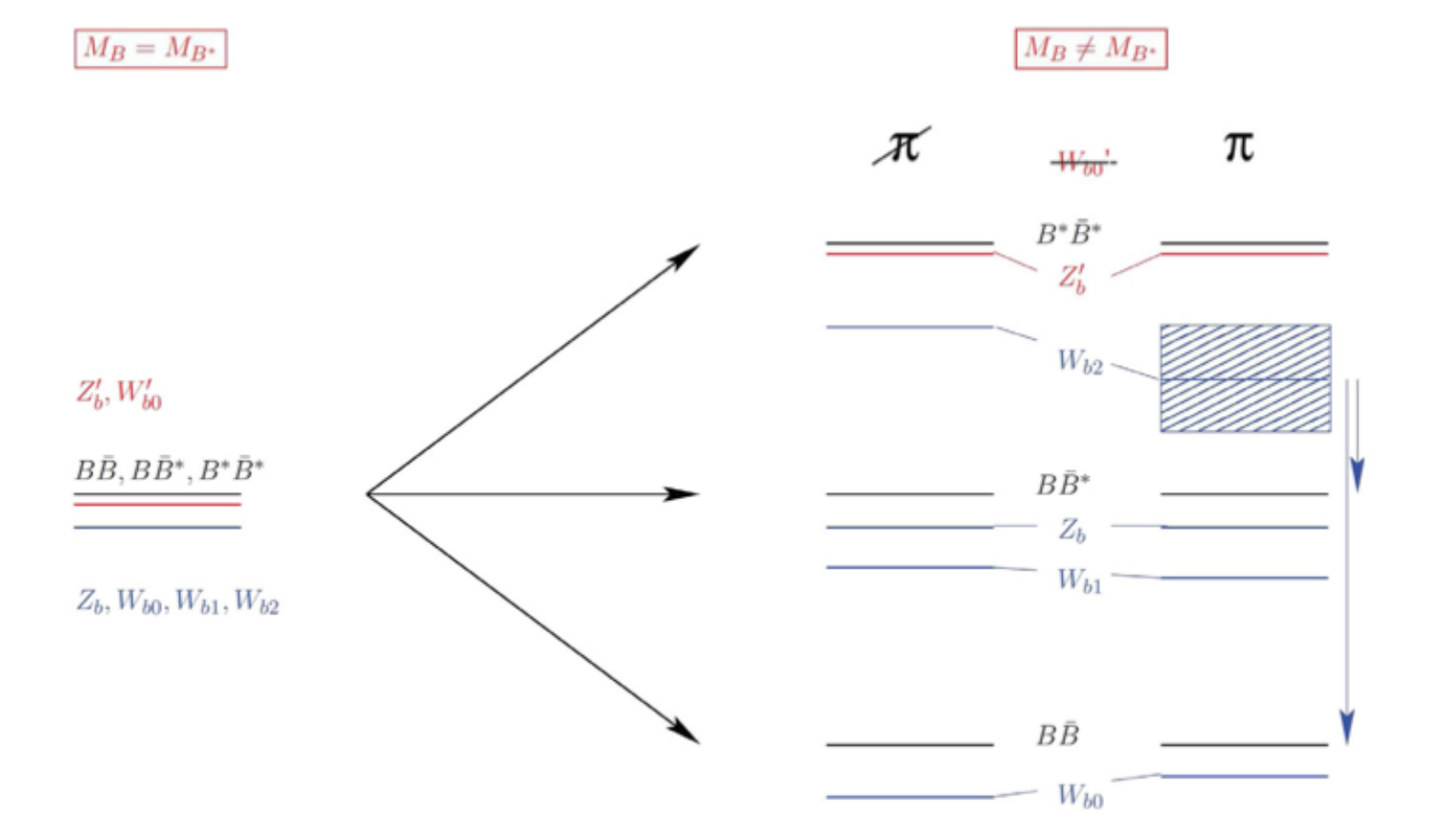}
	\caption{Effect of spin symmetry violation on the predicted spin multiplet structure for the 
	spin partner states of $Z_b(10650)$ and $Z_b(10610)$.
	Fig. from Ref.~\cite{Brambilla:2019esw}.}
	\label{Fig:Wspinmultiplets}
\end{figure}

One prediction of those effective field theories is that, if the bottomonium like exotic states $Z_b(10610)$ and $Z_b(10650)$
with $I(J^P)=1(1^+)$ are
both hadronic molecules with a decomposition $B\bar B^*$ and $B^*\bar B^*$, respectively, then there are necessarily
two multiplets of spin partner states at least in the heavy mass limit, one containing the $Z_b(10650)$ and a scalar
state, which disappears as soon as finite mass corrections are included for the $B$ mesons, 
and one containing the $Z_b(10650)$ together with 3 $W_{bJ}$ states, $J=0,1,2$~\cite{Mehen:2011yh,Voloshin:2011qa,Wang:2018jlv}. As spin symmetry violation is introduced into the
system, the degeneracy within the multiplets is lifted and the different states get attached to certain thresholds---see
Fig.~\ref{Fig:Wspinmultiplets} (while one of the scalar states gets unbound). What is also clear from the figure is
that for the spectrum the pionless and the pionfull theory give very similar results, however, the presence of pions
gives the $W_{b2}$ a significant width. It should be stressed that the tight connection between the spin
symmetry violation of the hadronic molecules and that of the constituents is a unique prediction of molecular models.

\section{Hidden-charm pentaquarks}

Multiquarks are expected not only in the meson sector
but also in the baryon sector, where the simplest structures are pentaquarks. The observation of a broad $\pc(4380)$ and a narrow $\pc(4450)$~\cite{Aaij:2015tga} in 2015
at LHCb was the first undebated observation of pentaquarks. They decay strongly into $\jp$. Since the $J/\psi$ cannot be produced in the process
 of the decay, the states contain 
at least five quarks, $c\bar{c}uud$. 
The data were analyses further in Refs.~\cite{Aaij:2016phn,Aaij:2016ymb} which supported
the existence of two states. 
A later analysis based on an order of magnitude larger data sample,
shows that the $\pc(4450)$ structure consists of two narrow overlapping peaks, $\pc(4440)$ and $\pc(4457)$, and a third narrow peak $\pc(4312)$ appears~\cite{Aaij:2019vzc}. However, there is no longer clear evidence for the broad $\pc(4380)$. 
As in the meson sector,
these discoveries were followed by numerous theoretical interpretations of the nature of the pentaquarks,
including hadronic molecules~\cite{Chen:2019bip,Chen:2019asm,Guo:2019fdo,Liu:2019tjn,He:2019ify,Guo:2019kdc,Shimizu:2019ptd,Xiao:2019mst,Xiao:2019aya,Wang:2019nwt,Meng:2019ilv,Wu:2019adv,Xiao:2019gjd,Voloshin:2019aut,Sakai:2019qph,Wang:2019hyc,Yamaguchi:2019seo,Liu:2019zvb,Lin:2019qiv,Wang:2019ato,Gutsche:2019mkg,Burns:2019iih,Du:2019pij,Wang:2019spc,Xu:2020gjl,Kuang:2020bnk,Peng:2020xrf,Peng:2020gwk,Xiao:2020frg,Dong:2021juy,Peng:2021hkr}, 
compact pentaquark states~\cite{Ali:2019npk,Zhu:2019iwm,Wang:2019got,Giron:2019bcs,Cheng:2019obk,Stancu:2019qga,Kuang:2020bnk}, hadro-charmonia~\cite{Eides:2015dtr,Eides:2019tgv,Anwar:2018bpu}, and cusp effects~\cite{Kuang:2020bnk}. 
 
The proximity of the $\Sigma_c\bar{D}^{(*)}$ thresholds
to these narrow pentaquark structures suggests that the corresponding
two-hadron states play an important role in the dynamics of the pentaquarks, suggesting an interpretation of their structure as hadronic molecules.
In the most common hadronic molecular picture, the $\pc(4312)$ is an $S$-wave $\Sigma_c\bar{D}$ bound state, while the $\pc(4440)$ and $\pc(4457)$ are bound states of $\Sigma_c\bar{D}^*$ with different spin structures~\cite{Liu:2019tjn,Xiao:2019aya,Sakai:2019qph,Du:2019pij,Xiao:2020frg}.  The origin of the peak from the $\pc(4312)$ is attributed to a virtual state of $\Sigma_c\bar{D}$ in Ref.~\cite{Fernandez-Ramirez:2019koa} based on amplitude analysis, which only fits data around the $\Sigma_c\bar{D}$ threshold. In Ref.~\cite{Kuang:2020bnk}, the final state interactions are constructed based on a $K$ matrix containing the $\jp$-$\Sigma_c\bar{D}$-$\Sigma_c\bar{D}^*$ channels.
The analysis suggests that $\pc(4312)$ is a $\Sigma_c\bar{D}$ molecule, while
the $\pc(4440)$ could be a compact pentaquark state, 
and the $\pc(4457)$ could be caused by the cusp effect.\footnote{A strong threshold cusp effect usually requires the existence of a near-threshold pole in an unphysical Riemann sheet (RS)~\cite{Guo:2014iya,Guo:2019twa,Dong:2020hxe}.}

While the proximity of the narrow $P_{c\bar c}$ peaks to the $\Sigma_c\bar{D}^{(*)}$ thresholds makes the 
molecular interpretation for the states natural, at least some peaks in the 
$\jp$ mass distributions can also be generated by triangle singularities~\cite{Guo:2015umn,Liu:2015fea,Mikhasenko:2015vca,Bayar:2016ftu,Aaij:2019vzc,Guo:2019twa}. 
A triangle singularity arises when all intermediate particles in a triangle loop are (almost) on the mass shell and the particles move collinearly. Therefore, its location is quite sensitive to the masses and widths of the particles involved~\cite{Guo:2019twa}. 
The potential triangle singularities have been discussed in Ref.~\cite{Aaij:2019vzc} for the 
three $P_{c\bar c}$ structures. Considering the realistic widths of the exchanged resonances, the $P_{c\bar c}(4312)$ 
and $\pc(4440)$ structures are unlikely to be caused by triangle singularities. However, the 
$\pc(4457)$ structure could in principle be generated by a triangle diagram with $D_{s1}^*(2860), \Lambda_c(2595)$ and $\bar{D}^{*0}$ in the intermediate state.

Under the assumption that the observed
pentaquarks are molecular in nature,
heavy quark spin symmetry allows one to predict spin partner
states.
Combining the light quark spins of the $\Sigma_c$, $j_{\rm light}^P=1^+$, and the ground state $D$ mesons,  $j_{\rm light}^P=(1/2)^+$, allows for two distinct total angular momenta of the light quarks, namely $1/2$ and $3/2$. Accordingly
the dynamics in the $\Sigma_c^{(*)}\bar D^{(*)}$ channels 
is controlled by two contact terms~\cite{Liu:2019tjn}.
It turns out that heavy-quark spin symmetry 
(HQSS)\footnote{It arises, since spin dependent couplings to a particle with mass $M$ vanish in the limit $M\to \infty$.} predicts seven $\pc$ states, divided into two heavy quark spin multiplets.  Three of these seven correspond to those reported by LHCb~\cite{Xiao:2013yca,Liu:2018zzu,Liu:2019tjn,Du:2019pij}:
While the $\pc(4312)$ is unambiguously assigned to the $J^P=\frac12^-$ $\Sigma_c\bar{D}$ bound state, 
there are two possible spin structures for the $\pc(4440)$ and $\pc(4457)$ identified as the $\Sigma_c\bar{D}^*$ bound states, namely $J^P=\frac12^-$ and $J^P=\frac32^-$~\cite{Guo:2019kdc,Liu:2019tjn,Du:2019pij}, 
and their spin assignment is not uniquely determined by HQSS alone~\cite{Chen:2019asm,Valderrama:2019chc,He:2019ify,Liu:2019zvb,Du:2019pij}. However, 
once the one-pion exchange (OPE) potentials were included, 
only one solution  could be found, suggesting that $\pc(4440)$ and $\pc(4457)$
couple dominantly to the $\Sigma_c\bar{D}^*$ with quantum numbers $J^P=\frac32^-$ and $\frac12^-$, respectively~\cite{Du:2019pij}.  In addition, there is evidence for an additional narrow state, also required by HQSS,
around $4.38$ GeV was found in the data with $J^P=\frac32^-$, which couples dominantly to the $\Sigma_c^*\bar{D}$ (see also Ref.~\cite{Xiao:2019mst,Xiao:2020frg}). 

While there is no evidence in the data yet for the three highlying states, they have to exist, if the pentaquarks
observed are indeed of molecular nature. In addition,
all the different states need to prominently decay
into the channels that form the molecular states
(that is $\Sigma_c \bar D$ for the $\pc(4312)$). The
pertinent predictions are provided in Ref.~\cite{Du:2021fmf}.
The relevant data are expected to be published in the coming years.

\section{All heavy tetraquarks}
\label{sec:multiquarks:allheavy}

While most of the experimental information is available for hadrons containing two heavy constituents, the LHCb measurements of di-$J/\psi$ production in proton-proton collisions at center-of-mass (c.m.) energies 7, 8 and 13~TeV revealed a new, potentially rich class of exotic states with four charmed (anti-)quarks~\cite{Aaij:2020fnh}. In fact, the measured line shape has a non-trivial shape that deviates significantly from the phase space distribution as well as from the exponential behaviour predicted from perturbative QCD for single- and double-parton scattering. The attention of both the experimental and theoretical communities was mainly drawn to the statistically significant peak structures observed in the energy range from 6.5~GeV to about 7.2~GeV. In particular, the most prominent structure was named $T_{cc\bar{c}\bar{c}}(6900)$ (also known as $X(6900)$)~\cite{ParticleDataGroup:2024cfk}. 
A fully-charmed compact tetraquark resonance is the most natural candidate to explain the structure. However, most of the theoretical studies give $\cccc$ ground states at a mass significantly lower than 6.9~GeV~\cite{Iwasaki:1975pv,Chao:1980dv,Badalian:1985es,Ader:1981db,Wu:2016vtq,Karliner:2016zzc,Wang:2017jtz,Liu:2019zuc,Bedolla:2019zwg,Chen:2020lgj}. Thus, one expects lower states, if there is a $\cccc$ resonance with a mass around 6.9~GeV. Due to a smaller phase space, such lighter states are expected to have smaller widths. However, there are no obvious narrower peaks in the reported double-$\jp$ spectrum.

When the data are analysed using a coupled-channel scattering formalism with vector charmonium pairs~\cite{Dong:2020nwy}, the number and locations of the poles in the 6900 MeV mass region appear to be rather poorly determined and strongly dependent on the model ingredients.
Nevertheless, if this approach does indeed capture the relevant dynamics, the data suggest the existence of a pole near the di-$J/\psi$ threshold in the double-$J/\psi$ production amplitude.  This state was named $X(6200)$ (or $T_{cc\bar{c}\bar{c}}(6200)$ according to the new naming scheme for exotic particles promoted by the Particle Data Group~\cite{ParticleDataGroup:2024cfk}). This finding was confirmed by the calculation in Ref.~\cite{Liang:2021fzr} and more recently in Ref.~\cite{Huang:2024jin}.  Due to the suppression of the signal near the threshold caused by the phase space factor, this pole cannot show up as a pronounced peak structure above the threshold in the double $J/\psi$ line shape. Instead, it can only be unambiguously identified by a comprehensive pole search in the coupled-channel scattering amplitudes. Meanwhile, a steep rise of the line shape just above the threshold provides evidence for the existence of such a near-threshold pole~\cite{Guo:2014iya}. As can be seen from the analysis in Ref.~\cite{Dong:2020nwy}, the behaviour of the signal just above the di-$J/\psi$ threshold does indeed call for the existence of the $X(6200)$ pole. 

Recently, data on the double $J/\psi$ production in $pp$ collisions have also arrived from two other LHC collaborations, namely CMS~\cite{CMS:2023owd} and ATLAS~\cite{ATLAS:2023bft}. Remarkably, in both cases the data description improved after the inclusion of an auxiliary Breit-Wigner resonance centred just above the production threshold. The need for its inclusion in the fitting function supports the existence of the $X(6200)$ pole in the amplitude~\cite{Song:2024ykq}.

\section{Single heavy tetraquarks}

Since their discovery in 2003 the lightest open
charm positive parity states containing strangeness remained largely a mystery---especially in light of the seemingly expected
properties of their non-strange partner states. In the past, attempts were made to explain the low lying $D_s$ states like $c\bar s$ mesons~\cite{Cahn:2003cw,Godfrey:2003kg,Colangelo:2003vg,Mehen:2005hc,Lakhina:2006fy}, chiral partners of the ground state $D_s$ and $D_s^*$ mesons~\cite{Bardeen:2003kt,Nowak:2003ra}, compact $[cq][\bar s\bar q]$ tetraquark states~\cite{Maiani:2004vq}, mixing of the $c\bar s$ and tetraquarks~\cite{Browder:2003fk}, a $D\pi$ atom for the $\dsz$~\cite{Szczepaniak:2003vy}, and $D^{(*)}K$
hadronic molecules~\cite{Barnes:2003dj,vanBeveren:2003kd,Chen:2004dy,Kolomeitsev:2003ac,Guo:2006fu,Guo:2006rp}. 
The experimental data show three features that need to be understood:
\begin{enumerate}\setlength\itemsep{-0.3em}
  \item[(1)] {\it The $D_s$ states are too light}: Both, $D_{s0}^*(2317)$ and $D_{s1}(2460)$ are much lighter than the quark model expectations for the lowest scalar and axial-vector $c\bar s$.
  \item[(2)] {\it Fine-tuning}: One has $M_{\dsone}-M_{\dsz}\simeq M_{D^{*\pm}}-M_{D^\pm}$ within 2~MeV. 
  \item[(3)] {\it Mass hierarchy}: One finds $M_{D_0^*(2300)}\sim M_{D_{s0}^*(2317)}$
   and $M_{D_1(2430)}\sim M_{D_{s1}(2460)}$, although usually adding a strange quark leads to an increase in mass of about 150-200 MeV.
\end{enumerate}
Items (1) and (2) could be understood in a compact tetraquark picture~\cite{Dmitrasinovic:2005gc}, however, leaving the third one unexplained.
On the other hand,
all these find a simultaneous natural explanation, if the lowest positive-parity charmed mesons are interpreted as hadronic mole\-cules.
In this case the flavor structure of this family of states is governed by the one that emerges
from the flavor decomposition representing the scattering of Goldstone bosons off $D$ mesons,
which may be expressed as~\cite{Albaladejo:2016lbb}
\begin{equation}
    [\mathbf{\bar 3}]\otimes [\mathbf{8}] = [\mathbf{\bar 3}] \oplus[\mathbf{6}]\oplus
[{\mathbf{\overline{15}}}] \ ,
\label{eq:mol_decomp}
\end{equation}
where the multiplets on the left refer to the $D$ mesons and the light pseudoscalars, respectively.
Note that we do not include scattering of the ninth pseudoscalar, the $\eta'$, here, since due to the action of the $U(1)_A$ anomaly it cannot be regarded as a Goldstone boson.
Non-strange isospin 1/2 multiplets appear in all three irreducible representations, however, chiral symmetry constraints dictate
that only the $[\mathbf{\bar 3}]$ and the $[\mathbf{6}]$ are attractive~\cite{Albaladejo:2016lbb}. 
In particular,
in this case the $D_0^*(2300)$ is interpreted as emerging from the interplay of two poles, one at 2105 MeV and one at 2451 MeV, with the lower one being a member of the same  SU(3) multiplet as the $D_{s0}^*(2317)$, the $[\mathbf{\bar 3}]$, where the attraction is the strongest.  The state at 2451 MeV is a member of the $[\mathbf{6}]$ representation of SU(3), where the interaction is weaker than in the $[\mathbf{\bar 3}]$, but still sufficiently strong
to generate a pole at physical quark masses
sufficiently close to the physical axis to show an impact on observables~\cite{Kolomeitsev:2003ac,Du:2017zvv,Du:2020pui,Guo:2018gyd}. 
It should be stressed that two-pole structures are occurring in 
various systems, since the compared to the naive quark model enlarged multiplet structure 
alluded to in Eq.~(\ref{eq:mol_decomp}) appears analogously
in various systems. For a recent review see Ref.~\cite{Meissner:2020khl}.
For the singly heavy, positive parity open flavor
states chiral symmetry 
constraints impose that the interaction of the particles in the $[{\mathbf{\overline{15}}}]$, is repulsive and thus no state should be found   in this channel. 

Thus, a crucial test of this interpretation is connected to the existence of the $\mathbf{[6]}$ state with a pole located at 2451~MeV
and the absence of a pole in
the $[{\mathbf{\overline{15}}}]$.
In contrast to this, quark model states with a quark composition $c\bar u$ or $c\bar d$ can appear only in the flavor $[\mathbf{\bar 3}]$
representation. 

One way 
to unravel the SU(3) structure 
underlying the spectrum of the lightest open charm scalar states
is to unambiguously establish the existence of the state in the 
$\mathbf{[6]}$, by determining whether it appears as a near threshold pole in the case when the Goldstone Boson mass (i.e. the pion mass) is even larger and near the SU(3) symmetric point, as predicted by unitarized chiral perturbation theory~\cite{Liu:2012zya}. This needs to be accompanied by the absence of
a state in the $[{\mathbf{\overline{15}}}]$.
First results indicated that, indeed, the $\mathbf{[\overline{15}]}$ state is \emph{repulsive} and the $\mathbf{[6]}$ state is \emph{attractive}, thus providing strong evidence for this state's molecular nature~\cite{Gregory:2021rgy}.
This finding was confirmed recently~\cite{Yeo:2024chk}
by a detailed  L\"uscher analysis imposing similar quark masses.

While these findings look like a strong support for a molecular structure
of the mentioned states, it remains to be studied, what e.g. the
compact tetraquark picture predicts for this system.
In particular, in Ref.~\cite{Maiani:2024quj} 
it is claimed that only scalar
diquarks should contribute to the formation
of the mentioned states. Then it is indeed a consequence
of the Pauli principle,
that 
also compact tetraquarks only appear in
the flavor $[\mathbf{\bar 3}]$ and $[\mathbf{6}]$ representations
and not in the $[{\mathbf{\overline{15}}}]$.
Further experimental and theoretical studies 
are necessary to reveal the nature of the lightest
positive parity open flavor states.

\section{Candidates for light multiquarks}
\label{sec:multiquarks:light}

The emerging multiquark states in the light quark sector like the light scalar mesons $f_0(500)$, $f_0(980)$, $K_0^*(700)$
and $a_0(980)$ and the $\Lambda(1405)$ where already discussed in some detail in other sections of this text. We therefore do not repeat those discussions here but refer the readers to the corresponding sections in this encyclopaedia. 

\begin{figure}[t]
	\begin{center}
	\includegraphics[width=14cm,height=3cm]{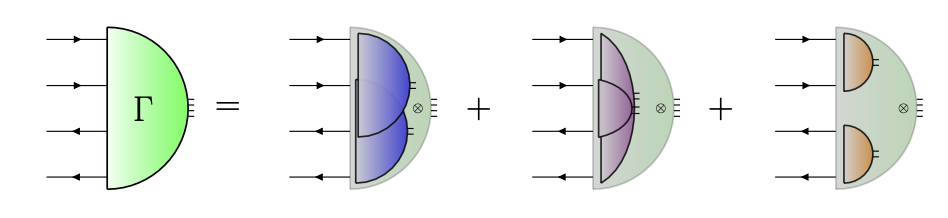}
	\caption{Illustration of the decomposition of tetraquark states into different components within the Dyson-Schwinger approach. The molecular component, the hadroquarkonium component and the compact tetraquark component are shown as diagram (a), (b) and (c), respectively. The figure is from Ref.~\cite{Hoffer:2024alv}.}
    \end{center}
	\label{fig:DS}
	\vspace{-4.7cm} 
    
\hspace{5.2cm}	(a) \hspace{3.1cm} (b) \hspace{3.1cm} (c)

		\vspace{4cm} 

\end{figure}

\section{Closing discussion on multiquark states}
\label{sec:multiquarks:subsec:outlook}

The sections above focused on the observable 
implications of various possible structures if they were
realised in an isolated form. These are the scenarios that most of the literature is so far dealing with. However, before closing this chapter, recent approaches that allow one to 
study the interplay of different scenarios should be mentioned.

Phenomenologically the interplay of conventional mesons
with multiquark states was studied in various works, however, the different works do not draw a clear picture.
While e.g. Refs.~\cite{Lu:2017yhl,Cincioglu:2016fkm}
call for a significant mixing of molecular and conventional
components, Refs.~\cite{Hammer:2016prh,Hanhart:2022qxq}
claim, that in a limit of large coupling of conventional states
to the continuum, the molecular structures that appear
as collective phenomena of the whole tower of
quark model states decouple from
the conventional state. 

One possible route with closer connection to QCD to access this field is to employ functional methods. Those were already introduced in other chapters of this encyclopedia, and applied to four-quark structures in Ref.~\cite{Wallbott:2019dng,Hoffer:2024alv,Hoffer:2024fgm}. 
The emerging ansatz for the four-body equations is shown in Fig.~\ref{fig:DS}.
In  principle the method allows one to investigate the relative importance of
the different components in the wave function of the exotic studied. The 
underlying interaction is already fixed from other studies. The technical problem
here is that the equations can be solved with well established techniques in the
space like regime, however, there are still some issues to overcome for a 
controlled continuation in the time-like regime, especially in the presence
of various continuum thresholds. 

Another very promising ansatz is the Born-Oppenheimer Effective Field Theory (BOEFT)~\cite{Brambilla:1999xf,Brambilla:2002nu,Braaten:2024tbm,Berwein:2024ztx}. This effective field theory inherited from molecular physics uses from static quark-quark or quark-antiquark potentials as input for tailor made coupled-channel Schr\"odinger equations. A first exploratory study of
the $X(3872)$ and the $T_{cc}(3875)$ can be found in Ref.~\cite{Brambilla:2024thx}. This effective field theory
holds the promise to combine hadronic and quark-gluon degrees of freedom in a field theoretically sound set up.
It still needs to be see how the subtle analytic structure of e.g. the one pion exchange amplitude discussed in the 
previous section can be embedded into the formalism.

\label{sec1:subsec1}

\section{Conclusions}
\label{sec:conclusions}
Given the fast developments both theoretically and experimentally, there is reason to believe that the nature of the recently discovered QCD exotics will be clarified within the next decade. Those insights will provide an additional important step to understand the inner workings of QCD in the non-perturbative regime.

\begin{ack}[Acknowledgments]%
 C.H. acknowledges the support from the CAS President's International Fellowship Initiative (PIFI) (Grant No. 2025PD0087). 
\end{ack}


\bibliographystyle{elsarticle-num} 
\bibliography{reference}

\begin{thebibliography}{100}
\expandafter\ifx\csname url\endcsname\relax
  \def\url#1{\texttt{#1}}\fi
\expandafter\ifx\csname urlprefix\endcsname\relax\def\urlprefix{URL }\fi
\expandafter\ifx\csname href\endcsname\relax
  \def\href#1#2{#2} \def\path#1{#1}\fi

\bibitem{Gell-Mann:1964ewy}
M.~Gell-Mann, A schematic model of baryons and mesons, Phys. Lett. 8 (1964)
  214--215.

\bibitem{Zweig:1964ruk}
G.~Zweig, {An SU(3) model for strong interaction symmetry and its breaking.
  Version 1} (1 1964).
\newblock \href {https://doi.org/10.17181/CERN-TH-401}
  {\path{doi:10.17181/CERN-TH-401}}.

\bibitem{Jaffe:1975fd}
R.~L. Jaffe, K.~Johnson, {Unconventional States of Confined Quarks and Gluons},
  Phys. Lett. B 60 (1976) 201--204.
\newblock \href {https://doi.org/10.1016/0370-2693(76)90423-8}
  {\path{doi:10.1016/0370-2693(76)90423-8}}.

\bibitem{Voloshin:1976ap}
M.~B. Voloshin, L.~B. Okun, {Hadron Molecules and Charmonium Atom}, JETP Lett.
  23 (1976) 333--336.

\bibitem{Tornqvist:1993ng}
N.~A. Tornqvist, {From the deuteron to deusons, an analysis of deuteron - like
  meson meson bound states}, Z. Phys. C 61 (1994) 525--537.
\newblock \href {http://arxiv.org/abs/hep-ph/9310247}
  {\path{arXiv:hep-ph/9310247}}, \href {https://doi.org/10.1007/BF01413192}
  {\path{doi:10.1007/BF01413192}}.

\bibitem{Hosaka:2016pey}
A.~Hosaka, T.~Iijima, K.~Miyabayashi, Y.~Sakai, S.~Yasui, Exotic hadrons with
  heavy flavors: {{X}}, {{Y}}, {{Z}}, and related states, PTEP 2016 (2016)
  062C01.
\newblock \href {http://arxiv.org/abs/1603.09229} {\path{arXiv:1603.09229}},
  \href {https://doi.org/10.1093/ptep/ptw045} {\path{doi:10.1093/ptep/ptw045}}.

\bibitem{Esposito:2016noz}
A.~Esposito, A.~Pilloni, A.~D. Polosa, {Multiquark Resonances}, Phys. Rept. 668
  (2017) 1--97.
\newblock \href {http://arxiv.org/abs/1611.07920} {\path{arXiv:1611.07920}},
  \href {https://doi.org/10.1016/j.physrep.2016.11.002}
  {\path{doi:10.1016/j.physrep.2016.11.002}}.

\bibitem{Guo:2017jvc}
F.-K. Guo, C.~Hanhart, U.-G. Mei{\ss}ner, Q.~Wang, Q.~Zhao, B.-S. Zou, Hadronic
  molecules, Rev. Mod. Phys. 90~(1) (2018) 015004.
\newblock \href {http://arxiv.org/abs/1705.00141} {\path{arXiv:1705.00141}},
  \href {https://doi.org/10.1103/RevModPhys.90.015004}
  {\path{doi:10.1103/RevModPhys.90.015004}}.

\bibitem{Olsen:2017bmm}
S.~L. Olsen, T.~Skwarnicki, D.~Zieminska, Nonstandard heavy mesons and baryons:
  {{Experimental}} evidence, Rev. Mod. Phys. 90~(1) (2018) 015003.
\newblock \href {http://arxiv.org/abs/1708.04012} {\path{arXiv:1708.04012}},
  \href {https://doi.org/10.1103/RevModPhys.90.015003}
  {\path{doi:10.1103/RevModPhys.90.015003}}.

\bibitem{Karliner:2017qhf}
M.~Karliner, J.~L. Rosner, T.~Skwarnicki, Multiquark states, Ann. Rev. Nucl.
  Part. Sci. 68 (2018) 17--44.
\newblock \href {http://arxiv.org/abs/1711.10626} {\path{arXiv:1711.10626}},
  \href {https://doi.org/10.1146/annurev-nucl-101917-020902}
  {\path{doi:10.1146/annurev-nucl-101917-020902}}.

\bibitem{Brambilla:2019esw}
N.~Brambilla, S.~Eidelman, C.~Hanhart, A.~Nefediev, C.-P. Shen, C.~E. Thomas,
  A.~Vairo, C.-Z. Yuan, The {{{\emph{XYZ}}}} states: {{Experimental}} and
  theoretical status and perspectives, Phys. Rept. 873~(TUM-EFT 125/19) (2020)
  1--154.
\newblock \href {http://arxiv.org/abs/1907.07583} {\path{arXiv:1907.07583}},
  \href {https://doi.org/10.1016/j.physrep.2020.05.001}
  {\path{doi:10.1016/j.physrep.2020.05.001}}.

\bibitem{Yang:2020atz}
G.~Yang, J.~Ping, J.~Segovia, Tetra- and penta-quark structures in the
  constituent quark model, Symmetry 12~(11) (2020) 1869.
\newblock \href {http://arxiv.org/abs/2009.00238} {\path{arXiv:2009.00238}},
  \href {https://doi.org/10.3390/sym12111869} {\path{doi:10.3390/sym12111869}}.

\bibitem{Chen:2022asf}
H.-X. Chen, W.~Chen, X.~Liu, Y.-R. Liu, S.-L. Zhu, An updated review of the new
  hadron states, Rept. Prog. Phys. 86~(2) (2022) 026201.
\newblock \href {http://arxiv.org/abs/2204.02649} {\path{arXiv:2204.02649}},
  \href {https://doi.org/10.1088/1361-6633/aca3b6}
  {\path{doi:10.1088/1361-6633/aca3b6}}.

\bibitem{Meng:2022ozq}
L.~Meng, B.~Wang, G.-J. Wang, S.-L. Zhu, Chiral perturbation theory for heavy
  hadrons and chiral effective field theory for heavy hadronic molecules, Phys.
  Rept. 1019 (2023) 2266.
\newblock \href {http://arxiv.org/abs/2204.08716} {\path{arXiv:2204.08716}},
  \href {https://doi.org/10.1016/j.physrep.2023.04.003}
  {\path{doi:10.1016/j.physrep.2023.04.003}}.

\bibitem{Cleven:2015era}
M.~Cleven, F.-K. Guo, C.~Hanhart, Q.~Wang, Q.~Zhao, {Employing spin symmetry to
  disentangle different models for the XYZ states}, Phys. Rev. D 92~(1) (2015)
  014005.
\newblock \href {http://arxiv.org/abs/1505.01771} {\path{arXiv:1505.01771}},
  \href {https://doi.org/10.1103/PhysRevD.92.014005}
  {\path{doi:10.1103/PhysRevD.92.014005}}.

\bibitem{Dubynskiy:2008mq}
S.~Dubynskiy, M.~B. Voloshin, {Hadro-Charmonium}, Phys. Lett. B 666 (2008)
  344--346.
\newblock \href {http://arxiv.org/abs/0803.2224} {\path{arXiv:0803.2224}},
  \href {https://doi.org/10.1016/j.physletb.2008.07.086}
  {\path{doi:10.1016/j.physletb.2008.07.086}}.

\bibitem{Dong:2021lkh}
X.-K. Dong, V.~Baru, F.-K. Guo, C.~Hanhart, A.~Nefediev, B.-S. Zou, {Is the
  existence of a J/\ensuremath{\psi}J/\ensuremath{\psi} bound state
  plausible?}, Sci. Bull. 66~(24) (2021) 2462--2470.
\newblock \href {http://arxiv.org/abs/2107.03946} {\path{arXiv:2107.03946}},
  \href {https://doi.org/10.1016/j.scib.2021.09.009}
  {\path{doi:10.1016/j.scib.2021.09.009}}.

\bibitem{BESIII:2022qal}
M.~Ablikim, et~al., {Study of the resonance structures in the process $e^+e^+
  \to \pi^+\pi^- J/ \psi$}, Phys. Rev. D 106~(7) (2022) 072001.
\newblock \href {http://arxiv.org/abs/2206.08554} {\path{arXiv:2206.08554}},
  \href {https://doi.org/10.1103/PhysRevD.106.072001}
  {\path{doi:10.1103/PhysRevD.106.072001}}.

\bibitem{BESIII:2018iea}
M.~Ablikim, et~al., {Evidence of a resonant structure in the $e^+e^-\to
  \pi^+D^0D^{*-}$ cross section between 4.05 and 4.60 GeV}, Phys. Rev. Lett.
  122~(10) (2019) 102002.
\newblock \href {http://arxiv.org/abs/1808.02847} {\path{arXiv:1808.02847}},
  \href {https://doi.org/10.1103/PhysRevLett.122.102002}
  {\path{doi:10.1103/PhysRevLett.122.102002}}.

\bibitem{Wang:2013kra}
Q.~Wang, M.~Cleven, F.-K. Guo, C.~Hanhart, U.-G. Mei\ss{}ner, X.-G. Wu,
  Q.~Zhao, {Y(4260): hadronic molecule versus hadro-charmonium interpretation},
  Phys. Rev. D 89~(3) (2014) 034001.
\newblock \href {http://arxiv.org/abs/1309.4303} {\path{arXiv:1309.4303}},
  \href {https://doi.org/10.1103/PhysRevD.89.034001}
  {\path{doi:10.1103/PhysRevD.89.034001}}.

\bibitem{Li:2013ssa}
X.~Li, M.~B. Voloshin, {$Y$(4260) and $Y$(4360) as mixed hadrocharmonium}, Mod.
  Phys. Lett. A 29~(12) (2014) 1450060.
\newblock \href {http://arxiv.org/abs/1309.1681} {\path{arXiv:1309.1681}},
  \href {https://doi.org/10.1142/S0217732314500606}
  {\path{doi:10.1142/S0217732314500606}}.

\bibitem{Voloshin:2018vym}
M.~B. Voloshin, {$Z_c(4100)$ and $Z_c(4200)$ as hadrocharmonium}, Phys. Rev. D
  98~(9) (2018) 094028.
\newblock \href {http://arxiv.org/abs/1810.08146} {\path{arXiv:1810.08146}},
  \href {https://doi.org/10.1103/PhysRevD.98.094028}
  {\path{doi:10.1103/PhysRevD.98.094028}}.

\bibitem{Guo:2008zg}
F.-K. Guo, C.~Hanhart, U.-G. Meissner, {Evidence that the Y(4660) is a
  f(0)(980)psi-prime bound state}, Phys. Lett. B 665 (2008) 26--29.
\newblock \href {http://arxiv.org/abs/0803.1392} {\path{arXiv:0803.1392}},
  \href {https://doi.org/10.1016/j.physletb.2008.05.057}
  {\path{doi:10.1016/j.physletb.2008.05.057}}.

\bibitem{Guo:2009id}
F.-K. Guo, C.~Hanhart, U.-G. Meissner, {Implications of heavy quark spin
  symmetry on heavy meson hadronic molecules}, Phys. Rev. Lett. 102 (2009)
  242004.
\newblock \href {http://arxiv.org/abs/0904.3338} {\path{arXiv:0904.3338}},
  \href {https://doi.org/10.1103/PhysRevLett.102.242004}
  {\path{doi:10.1103/PhysRevLett.102.242004}}.

\bibitem{Barabanov:2020jvn}
M.~Y. Barabanov, et~al., {Diquark correlations in hadron physics: Origin,
  impact and evidence}, Prog. Part. Nucl. Phys. 116 (2021) 103835.
\newblock \href {http://arxiv.org/abs/2008.07630} {\path{arXiv:2008.07630}},
  \href {https://doi.org/10.1016/j.ppnp.2020.103835}
  {\path{doi:10.1016/j.ppnp.2020.103835}}.

\bibitem{Maiani:2004vq}
L.~Maiani, F.~Piccinini, A.~D. Polosa, V.~Riquer, {Diquark-antidiquarks with
  hidden or open charm and the nature of X(3872)}, Phys. Rev. D 71 (2005)
  014028.
\newblock \href {http://arxiv.org/abs/hep-ph/0412098}
  {\path{arXiv:hep-ph/0412098}}, \href
  {https://doi.org/10.1103/PhysRevD.71.014028}
  {\path{doi:10.1103/PhysRevD.71.014028}}.

\bibitem{Ali:2017wsf}
A.~Ali, L.~Maiani, A.~V. Borisov, I.~Ahmed, M.~Jamil~Aslam, A.~Y. Parkhomenko,
  A.~D. Polosa, A.~Rehman, {A new look at the Y tetraquarks and $\Omega _c$
  baryons in the diquark model}, Eur. Phys. J. C 78~(1) (2018) 29.
\newblock \href {http://arxiv.org/abs/1708.04650} {\path{arXiv:1708.04650}},
  \href {https://doi.org/10.1140/epjc/s10052-017-5501-6}
  {\path{doi:10.1140/epjc/s10052-017-5501-6}}.

\bibitem{Maiani:2014aja}
L.~Maiani, F.~Piccinini, A.~D. Polosa, V.~Riquer, {The Z(4430) and a New
  Paradigm for Spin Interactions in Tetraquarks}, Phys. Rev. D 89 (2014)
  114010.
\newblock \href {http://arxiv.org/abs/1405.1551} {\path{arXiv:1405.1551}},
  \href {https://doi.org/10.1103/PhysRevD.89.114010}
  {\path{doi:10.1103/PhysRevD.89.114010}}.

\bibitem{Giron:2019cfc}
J.~F. Giron, R.~F. Lebed, C.~T. Peterson, {The Dynamical Diquark Model: Fine
  Structure and Isospin}, JHEP 01 (2020) 124.
\newblock \href {http://arxiv.org/abs/1907.08546} {\path{arXiv:1907.08546}},
  \href {https://doi.org/10.1007/JHEP01(2020)124}
  {\path{doi:10.1007/JHEP01(2020)124}}.

\bibitem{Maiani:2017kyi}
L.~Maiani, A.~D. Polosa, V.~Riquer, {A Theory of X and Z Multiquark
  Resonances}, Phys. Lett. B 778 (2018) 247--251.
\newblock \href {http://arxiv.org/abs/1712.05296} {\path{arXiv:1712.05296}},
  \href {https://doi.org/10.1016/j.physletb.2018.01.039}
  {\path{doi:10.1016/j.physletb.2018.01.039}}.

\bibitem{Brodsky:2014xia}
S.~J. Brodsky, D.~S. Hwang, R.~F. Lebed, {Dynamical Picture for the Formation
  and Decay of the Exotic XYZ Mesons}, Phys. Rev. Lett. 113~(11) (2014) 112001.
\newblock \href {http://arxiv.org/abs/1406.7281} {\path{arXiv:1406.7281}},
  \href {https://doi.org/10.1103/PhysRevLett.113.112001}
  {\path{doi:10.1103/PhysRevLett.113.112001}}.

\bibitem{Maiani:2021tri}
L.~Maiani, A.~D. Polosa, V.~Riquer, {The new resonances Zcs(3985) and Zcs(4003)
  (almost) fill two tetraquark nonets of broken SU(3)f}, Sci. Bull. 66 (2021)
  1616--1619.
\newblock \href {http://arxiv.org/abs/2103.08331} {\path{arXiv:2103.08331}},
  \href {https://doi.org/10.1016/j.scib.2021.04.040}
  {\path{doi:10.1016/j.scib.2021.04.040}}.

\bibitem{Esposito:2013fma}
A.~Esposito, M.~Papinutto, A.~Pilloni, A.~D. Polosa, N.~Tantalo, {Doubly
  charmed tetraquarks in $B_c$ and $\Xi_{bc}$ decays}, Phys. Rev. D 88~(5)
  (2013) 054029.
\newblock \href {http://arxiv.org/abs/1307.2873} {\path{arXiv:1307.2873}},
  \href {https://doi.org/10.1103/PhysRevD.88.054029}
  {\path{doi:10.1103/PhysRevD.88.054029}}.

\bibitem{Karliner:2017qjm}
M.~Karliner, J.~L. Rosner, {Discovery of doubly-charmed $\Xi_{cc}$ baryon
  implies a stable ($b b \bar{u} \bar{d}$) tetraquark}, Phys. Rev. Lett.
  119~(20) (2017) 202001.
\newblock \href {http://arxiv.org/abs/1707.07666} {\path{arXiv:1707.07666}},
  \href {https://doi.org/10.1103/PhysRevLett.119.202001}
  {\path{doi:10.1103/PhysRevLett.119.202001}}.

\bibitem{Eichten:2017ffp}
E.~J. Eichten, C.~Quigg, {Heavy-quark symmetry implies stable heavy tetraquark
  mesons $Q_iQ_j \bar q_k \bar q_l$}, Phys. Rev. Lett. 119~(20) (2017) 202002.
\newblock \href {http://arxiv.org/abs/1707.09575} {\path{arXiv:1707.09575}},
  \href {https://doi.org/10.1103/PhysRevLett.119.202002}
  {\path{doi:10.1103/PhysRevLett.119.202002}}.

\bibitem{Ballot:1983iv}
J.~l. Ballot, J.~M. Richard, {FOUR QUARK STATES IN ADDITIVE POTENTIALS}, Phys.
  Lett. B 123 (1983) 449--451.
\newblock \href {https://doi.org/10.1016/0370-2693(83)90991-7}
  {\path{doi:10.1016/0370-2693(83)90991-7}}.

\bibitem{Zouzou:1986qh}
S.~Zouzou, B.~Silvestre-Brac, C.~Gignoux, J.~M. Richard, {FOUR QUARK BOUND
  STATES}, Z. Phys. C 30 (1986) 457.
\newblock \href {https://doi.org/10.1007/BF01557611}
  {\path{doi:10.1007/BF01557611}}.

\bibitem{Brink:1998as}
D.~M. Brink, F.~Stancu, {Tetraquarks with heavy flavors}, Phys. Rev. D 57
  (1998) 6778--6787.
\newblock \href {https://doi.org/10.1103/PhysRevD.57.6778}
  {\path{doi:10.1103/PhysRevD.57.6778}}.

\bibitem{Park:2018wjk}
W.~Park, S.~Noh, S.~H. Lee, {Masses of the doubly heavy tetraquarks in a
  constituent quark model}, Nucl. Phys. A 983 (2019) 1--19.
\newblock \href {http://arxiv.org/abs/1809.05257} {\path{arXiv:1809.05257}},
  \href {https://doi.org/10.1016/j.nuclphysa.2018.12.019}
  {\path{doi:10.1016/j.nuclphysa.2018.12.019}}.

\bibitem{Noh:2021lqs}
S.~Noh, W.~Park, S.~H. Lee, {The Doubly-heavy Tetraquarks
  ($qq'\bar{Q}\bar{Q'}$) in a Constituent Quark Model with a Complete Set of
  Harmonic Oscillator Bases}, Phys. Rev. D 103 (2021) 114009.
\newblock \href {http://arxiv.org/abs/2102.09614} {\path{arXiv:2102.09614}},
  \href {https://doi.org/10.1103/PhysRevD.103.114009}
  {\path{doi:10.1103/PhysRevD.103.114009}}.

\bibitem{Ma:2023int}
Y.~Ma, L.~Meng, Y.-K. Chen, S.-L. Zhu, {Doubly heavy tetraquark states in the
  constituent quark model using diffusion Monte~Carlo method}, Phys. Rev. D
  109~(7) (2024) 074001.
\newblock \href {http://arxiv.org/abs/2309.17068} {\path{arXiv:2309.17068}},
  \href {https://doi.org/10.1103/PhysRevD.109.074001}
  {\path{doi:10.1103/PhysRevD.109.074001}}.

\bibitem{Bicudo:2012qt}
P.~Bicudo, M.~Wagner, {Lattice QCD signal for a bottom-bottom tetraquark},
  Phys. Rev. D 87~(11) (2013) 114511.
\newblock \href {http://arxiv.org/abs/1209.6274} {\path{arXiv:1209.6274}},
  \href {https://doi.org/10.1103/PhysRevD.87.114511}
  {\path{doi:10.1103/PhysRevD.87.114511}}.

\bibitem{Brown:2012tm}
Z.~S. Brown, K.~Orginos, {Tetraquark bound states in the heavy-light
  heavy-light system}, Phys. Rev. D 86 (2012) 114506.
\newblock \href {http://arxiv.org/abs/1210.1953} {\path{arXiv:1210.1953}},
  \href {https://doi.org/10.1103/PhysRevD.86.114506}
  {\path{doi:10.1103/PhysRevD.86.114506}}.

\bibitem{Bicudo:2015kna}
P.~Bicudo, K.~Cichy, A.~Peters, M.~Wagner, {BB interactions with static bottom
  quarks from Lattice QCD}, Phys. Rev. D 93~(3) (2016) 034501.
\newblock \href {http://arxiv.org/abs/1510.03441} {\path{arXiv:1510.03441}},
  \href {https://doi.org/10.1103/PhysRevD.93.034501}
  {\path{doi:10.1103/PhysRevD.93.034501}}.

\bibitem{Bicudo:2016ooe}
P.~Bicudo, J.~Scheunert, M.~Wagner, {Including heavy spin effects in the
  prediction of a $\bar{b} \bar{b} u d$ tetraquark with lattice QCD
  potentials}, Phys. Rev. D 95~(3) (2017) 034502.
\newblock \href {http://arxiv.org/abs/1612.02758} {\path{arXiv:1612.02758}},
  \href {https://doi.org/10.1103/PhysRevD.95.034502}
  {\path{doi:10.1103/PhysRevD.95.034502}}.

\bibitem{Francis:2016hui}
A.~Francis, R.~J. Hudspith, R.~Lewis, K.~Maltman, {Lattice Prediction for
  Deeply Bound Doubly Heavy Tetraquarks}, Phys. Rev. Lett. 118~(14) (2017)
  142001.
\newblock \href {http://arxiv.org/abs/1607.05214} {\path{arXiv:1607.05214}},
  \href {https://doi.org/10.1103/PhysRevLett.118.142001}
  {\path{doi:10.1103/PhysRevLett.118.142001}}.

\bibitem{Junnarkar:2018twb}
P.~Junnarkar, N.~Mathur, M.~Padmanath, {Study of doubly heavy tetraquarks in
  Lattice QCD}, Phys. Rev. D 99~(3) (2019) 034507.
\newblock \href {http://arxiv.org/abs/1810.12285} {\path{arXiv:1810.12285}},
  \href {https://doi.org/10.1103/PhysRevD.99.034507}
  {\path{doi:10.1103/PhysRevD.99.034507}}.

\bibitem{Meinel:2022lzo}
S.~Meinel, M.~Pflaumer, M.~Wagner, {Search for
  b\textasciimacron{}b\textasciimacron{}us and
  b\textasciimacron{}c\textasciimacron{}ud tetraquark bound states using
  lattice QCD}, Phys. Rev. D 106~(3) (2022) 034507.
\newblock \href {http://arxiv.org/abs/2205.13982} {\path{arXiv:2205.13982}},
  \href {https://doi.org/10.1103/PhysRevD.106.034507}
  {\path{doi:10.1103/PhysRevD.106.034507}}.

\bibitem{Mohanta:2020eed}
P.~Mohanta, S.~Basak, {Construction of $bb\bar{u}\bar{d}$ tetraquark states on
  lattice with NRQCD bottom and HISQ up and down quarks}, Phys. Rev. D 102~(9)
  (2020) 094516.
\newblock \href {http://arxiv.org/abs/2008.11146} {\path{arXiv:2008.11146}},
  \href {https://doi.org/10.1103/PhysRevD.102.094516}
  {\path{doi:10.1103/PhysRevD.102.094516}}.

\bibitem{Hudspith:2023loy}
R.~J. Hudspith, D.~Mohler, {Exotic tetraquark states with two
  b\textasciimacron{} quarks and JP=0+ and 1+ Bs states in a nonperturbatively
  tuned lattice NRQCD setup}, Phys. Rev. D 107~(11) (2023) 114510.
\newblock \href {http://arxiv.org/abs/2303.17295} {\path{arXiv:2303.17295}},
  \href {https://doi.org/10.1103/PhysRevD.107.114510}
  {\path{doi:10.1103/PhysRevD.107.114510}}.

\bibitem{Alexandrou:2024iwi}
C.~Alexandrou, J.~Finkenrath, T.~Leontiou, S.~Meinel, M.~Pflaumer, M.~Wagner,
  {$\bar b \bar b u d$ and $\bar b \bar b u s$ tetraquarks from lattice QCD
  using symmetric correlation matrices with both local and scattering
  interpolating operators} (4 2024).
\newblock \href {http://arxiv.org/abs/2404.03588} {\path{arXiv:2404.03588}}.

\bibitem{Navarra:2007yw}
F.~S. Navarra, M.~Nielsen, S.~H. Lee, {QCD sum rules study of QQ - anti-u
  anti-d mesons}, Phys. Lett. B 649 (2007) 166--172.
\newblock \href {http://arxiv.org/abs/hep-ph/0703071}
  {\path{arXiv:hep-ph/0703071}}, \href
  {https://doi.org/10.1016/j.physletb.2007.04.010}
  {\path{doi:10.1016/j.physletb.2007.04.010}}.

\bibitem{Du:2012wp}
M.-L. Du, W.~Chen, X.-L. Chen, S.-L. Zhu, {Exotic $QQ\bar{q}\bar{q}$,
  $QQ\bar{q}\bar{s}$ and $QQ\bar{s}\bar{s}$ states}, Phys. Rev. D 87~(1) (2013)
  014003.
\newblock \href {http://arxiv.org/abs/1209.5134} {\path{arXiv:1209.5134}},
  \href {https://doi.org/10.1103/PhysRevD.87.014003}
  {\path{doi:10.1103/PhysRevD.87.014003}}.

\bibitem{Wang:2017uld}
Z.-G. Wang, {Analysis of the axialvector doubly heavy tetraquark states with
  QCD sum rules}, Acta Phys. Polon. B 49 (2018) 1781.
\newblock \href {http://arxiv.org/abs/1708.04545} {\path{arXiv:1708.04545}},
  \href {https://doi.org/10.5506/APhysPolB.49.1781}
  {\path{doi:10.5506/APhysPolB.49.1781}}.

\bibitem{Kaplan:1998tg}
D.~B. Kaplan, M.~J. Savage, M.~B. Wise, {A New expansion for nucleon-nucleon
  interactions}, Phys. Lett. B 424 (1998) 390--396.
\newblock \href {http://arxiv.org/abs/nucl-th/9801034}
  {\path{arXiv:nucl-th/9801034}}, \href
  {https://doi.org/10.1016/S0370-2693(98)00210-X}
  {\path{doi:10.1016/S0370-2693(98)00210-X}}.

\bibitem{Epelbaum:2008ga}
E.~Epelbaum, H.-W. Hammer, U.-G. Meissner, {Modern Theory of Nuclear Forces},
  Rev. Mod. Phys. 81 (2009) 1773--1825.
\newblock \href {http://arxiv.org/abs/0811.1338} {\path{arXiv:0811.1338}},
  \href {https://doi.org/10.1103/RevModPhys.81.1773}
  {\path{doi:10.1103/RevModPhys.81.1773}}.

\bibitem{Braaten:2004rn}
E.~Braaten, H.~W. Hammer, {Universality in few-body systems with large
  scattering length}, Phys. Rept. 428 (2006) 259--390.
\newblock \href {http://arxiv.org/abs/cond-mat/0410417}
  {\path{arXiv:cond-mat/0410417}}, \href
  {https://doi.org/10.1016/j.physrep.2006.03.001}
  {\path{doi:10.1016/j.physrep.2006.03.001}}.

\bibitem{Weinberg:1965zz}
S.~Weinberg, {Evidence That the Deuteron Is Not an Elementary Particle}, Phys.
  Rev. 137 (1965) B672--B678.
\newblock \href {https://doi.org/10.1103/PhysRev.137.B672}
  {\path{doi:10.1103/PhysRev.137.B672}}.

\bibitem{Baru:2003qq}
V.~Baru, J.~Haidenbauer, C.~Hanhart, Y.~Kalashnikova, A.~E. Kudryavtsev,
  {Evidence that the a(0)(980) and f(0)(980) are not elementary particles},
  Phys. Lett. B 586 (2004) 53--61.
\newblock \href {http://arxiv.org/abs/hep-ph/0308129}
  {\path{arXiv:hep-ph/0308129}}, \href
  {https://doi.org/10.1016/j.physletb.2004.01.088}
  {\path{doi:10.1016/j.physletb.2004.01.088}}.

\bibitem{Baru:2010ww}
V.~Baru, C.~Hanhart, Y.~S. Kalashnikova, A.~E. Kudryavtsev, A.~V. Nefediev,
  {Interplay of quark and meson degrees of freedom in a near-threshold
  resonance}, Eur. Phys. J. A 44 (2010) 93--103.
\newblock \href {http://arxiv.org/abs/1001.0369} {\path{arXiv:1001.0369}},
  \href {https://doi.org/10.1140/epja/i2010-10929-7}
  {\path{doi:10.1140/epja/i2010-10929-7}}.

\bibitem{Hanhart:2011jz}
C.~Hanhart, Y.~S. Kalashnikova, A.~V. Nefediev, {Interplay of quark and meson
  degrees of freedom in a near-threshold resonance: multi-channel case}, Eur.
  Phys. J. A 47 (2011) 101--110.
\newblock \href {http://arxiv.org/abs/1106.1185} {\path{arXiv:1106.1185}},
  \href {https://doi.org/10.1140/epja/i2011-11101-9}
  {\path{doi:10.1140/epja/i2011-11101-9}}.

\bibitem{Guo:2016bjq}
F.~K. Guo, C.~Hanhart, Y.~S. Kalashnikova, P.~Matuschek, R.~V. Mizuk, A.~V.
  Nefediev, Q.~Wang, J.~L. Wynen, {Interplay of quark and meson degrees of
  freedom in near-threshold states: A practical parametrization for line
  shapes}, Phys. Rev. D 93~(7) (2016) 074031.
\newblock \href {http://arxiv.org/abs/1602.00940} {\path{arXiv:1602.00940}},
  \href {https://doi.org/10.1103/PhysRevD.93.074031}
  {\path{doi:10.1103/PhysRevD.93.074031}}.

\bibitem{Weinberg:1962hj}
S.~Weinberg, {Elementary particle theory of composite particles}, Phys. Rev.
  130 (1963) 776--783.
\newblock \href {https://doi.org/10.1103/PhysRev.130.776}
  {\path{doi:10.1103/PhysRev.130.776}}.

\bibitem{Weinberg:1963zza}
S.~Weinberg, {Quasiparticles and the Born Series}, Phys. Rev. 131 (1963)
  440--460.
\newblock \href {https://doi.org/10.1103/PhysRev.131.440}
  {\path{doi:10.1103/PhysRev.131.440}}.

\bibitem{Song:2022yvz}
J.~Song, L.~R. Dai, E.~Oset, {How much is the compositeness of a bound state
  constrained by a and $r_0$? The role of the interaction range}, Eur. Phys. J.
  A 58~(7) (2022) 133.
\newblock \href {http://arxiv.org/abs/2201.04414} {\path{arXiv:2201.04414}},
  \href {https://doi.org/10.1140/epja/s10050-022-00753-3}
  {\path{doi:10.1140/epja/s10050-022-00753-3}}.

\bibitem{Matuschek:2020gqe}
I.~Matuschek, V.~Baru, F.-K. Guo, C.~Hanhart, {On the nature of near-threshold
  bound and virtual states}, Eur. Phys. J. A 57~(3) (2021) 101.
\newblock \href {http://arxiv.org/abs/2007.05329} {\path{arXiv:2007.05329}},
  \href {https://doi.org/10.1140/epja/s10050-021-00413-y}
  {\path{doi:10.1140/epja/s10050-021-00413-y}}.

\bibitem{Hanhart:2008mx}
C.~Hanhart, J.~R. Pelaez, G.~Rios, {Quark mass dependence of the rho and sigma
  from dispersion relations and Chiral Perturbation Theory}, Phys. Rev. Lett.
  100 (2008) 152001.
\newblock \href {http://arxiv.org/abs/0801.2871} {\path{arXiv:0801.2871}},
  \href {https://doi.org/10.1103/PhysRevLett.100.152001}
  {\path{doi:10.1103/PhysRevLett.100.152001}}.

\bibitem{Hanhart:2014ssa}
C.~Hanhart, J.~R. Pelaez, G.~Rios, {Remarks on pole trajectories for
  resonances}, Phys. Lett. B 739 (2014) 375--382.
\newblock \href {http://arxiv.org/abs/1407.7452} {\path{arXiv:1407.7452}},
  \href {https://doi.org/10.1016/j.physletb.2014.11.011}
  {\path{doi:10.1016/j.physletb.2014.11.011}}.

\bibitem{Guo:2009ct}
F.-K. Guo, C.~Hanhart, U.-G. Meissner, {Interactions between heavy mesons and
  Goldstone bosons from chiral dynamics}, Eur. Phys. J. A 40 (2009) 171--179.
\newblock \href {http://arxiv.org/abs/0901.1597} {\path{arXiv:0901.1597}},
  \href {https://doi.org/10.1140/epja/i2009-10762-1}
  {\path{doi:10.1140/epja/i2009-10762-1}}.

\bibitem{Bruns:2019xgo}
P.~C. Bruns, {Spatial interpretation of ''compositeness'' for finite-range
  potentials} (5 2019).
\newblock \href {http://arxiv.org/abs/1905.09196} {\path{arXiv:1905.09196}}.

\bibitem{Oller:2017alp}
J.~A. Oller, {New results from a number operator interpretation of the
  compositeness of bound and resonant states}, Annals Phys. 396 (2018)
  429--458.
\newblock \href {http://arxiv.org/abs/1710.00991} {\path{arXiv:1710.00991}},
  \href {https://doi.org/10.1016/j.aop.2018.07.023}
  {\path{doi:10.1016/j.aop.2018.07.023}}.

\bibitem{Kang:2016jxw}
X.-W. Kang, J.~A. Oller, {Different pole structures in line shapes of the
  $X(3872)$}, Eur. Phys. J. C 77~(6) (2017) 399.
\newblock \href {http://arxiv.org/abs/1612.08420} {\path{arXiv:1612.08420}},
  \href {https://doi.org/10.1140/epjc/s10052-017-4961-z}
  {\path{doi:10.1140/epjc/s10052-017-4961-z}}.

\bibitem{Sekihara:2016xnq}
T.~Sekihara, {Two-body wave functions and compositeness from scattering
  amplitudes. I. General properties with schematic models}, Phys. Rev. C 95~(2)
  (2017) 025206.
\newblock \href {http://arxiv.org/abs/1609.09496} {\path{arXiv:1609.09496}},
  \href {https://doi.org/10.1103/PhysRevC.95.025206}
  {\path{doi:10.1103/PhysRevC.95.025206}}.

\bibitem{Kamiya:2016oao}
Y.~Kamiya, T.~Hyodo, {Generalized weak-binding relations of compositeness in
  effective field theory}, PTEP 2017~(2) (2017) 023D02.
\newblock \href {http://arxiv.org/abs/1607.01899} {\path{arXiv:1607.01899}},
  \href {https://doi.org/10.1093/ptep/ptw188} {\path{doi:10.1093/ptep/ptw188}}.

\bibitem{Guo:2015daa}
Z.-H. Guo, J.~A. Oller, {Probabilistic interpretation of compositeness relation
  for resonances}, Phys. Rev. D 93~(9) (2016) 096001.
\newblock \href {http://arxiv.org/abs/1508.06400} {\path{arXiv:1508.06400}},
  \href {https://doi.org/10.1103/PhysRevD.93.096001}
  {\path{doi:10.1103/PhysRevD.93.096001}}.

\bibitem{Sekihara:2014kya}
T.~Sekihara, T.~Hyodo, D.~Jido, {Comprehensive analysis of the wave function of
  a hadronic resonance and its compositeness}, PTEP 2015 (2015) 063D04.
\newblock \href {http://arxiv.org/abs/1411.2308} {\path{arXiv:1411.2308}},
  \href {https://doi.org/10.1093/ptep/ptv081} {\path{doi:10.1093/ptep/ptv081}}.

\bibitem{Aceti:2012dd}
F.~Aceti, E.~Oset, {Wave functions of composite hadron states and relationship
  to couplings of scattering amplitudes for general partial waves}, Phys. Rev.
  D 86 (2012) 014012.
\newblock \href {http://arxiv.org/abs/1202.4607} {\path{arXiv:1202.4607}},
  \href {https://doi.org/10.1103/PhysRevD.86.014012}
  {\path{doi:10.1103/PhysRevD.86.014012}}.

\bibitem{Gamermann:2009uq}
D.~Gamermann, J.~Nieves, E.~Oset, E.~Ruiz~Arriola, {Couplings in coupled
  channels versus wave functions: application to the $X(3872)$ resonance},
  Phys. Rev. D 81 (2010) 014029.
\newblock \href {http://arxiv.org/abs/0911.4407} {\path{arXiv:0911.4407}},
  \href {https://doi.org/10.1103/PhysRevD.81.014029}
  {\path{doi:10.1103/PhysRevD.81.014029}}.

\bibitem{Kinugawa:2024kwb}
T.~Kinugawa, T.~Hyodo, {Compositeness of near-threshold $s$-wave resonances} (3
  2024).
\newblock \href {http://arxiv.org/abs/2403.12635} {\path{arXiv:2403.12635}}.

\bibitem{Esposito:2021vhu}
A.~Esposito, L.~Maiani, A.~Pilloni, A.~D. Polosa, V.~Riquer, {From the line
  shape of the X(3872) to its structure}, Phys. Rev. D 105~(3) (2022) L031503.
\newblock \href {http://arxiv.org/abs/2108.11413} {\path{arXiv:2108.11413}},
  \href {https://doi.org/10.1103/PhysRevD.105.L031503}
  {\path{doi:10.1103/PhysRevD.105.L031503}}.

\bibitem{Baru:2021ldu}
V.~Baru, X.-K. Dong, M.-L. Du, A.~Filin, F.-K. Guo, C.~Hanhart, A.~Nefediev,
  J.~Nieves, Q.~Wang, {Effective range expansion for narrow near-threshold
  resonances}, Phys. Lett. B 833 (2022) 137290.
\newblock \href {http://arxiv.org/abs/2110.07484} {\path{arXiv:2110.07484}},
  \href {https://doi.org/10.1016/j.physletb.2022.137290}
  {\path{doi:10.1016/j.physletb.2022.137290}}.

\bibitem{Li:2021cue}
Y.~Li, F.-K. Guo, J.-Y. Pang, J.-J. Wu, {Generalization of
  Weinberg\textquoteright{}s compositeness relations}, Phys. Rev. D 105~(7)
  (2022) L071502.
\newblock \href {http://arxiv.org/abs/2110.02766} {\path{arXiv:2110.02766}},
  \href {https://doi.org/10.1103/PhysRevD.105.L071502}
  {\path{doi:10.1103/PhysRevD.105.L071502}}.

\bibitem{Albaladejo:2022sux}
M.~Albaladejo, J.~Nieves, {Compositeness of S-wave weakly-bound states from
  next-to-leading order Weinberg\textquoteright{}s relations}, Eur. Phys. J. C
  82~(8) (2022) 724.
\newblock \href {http://arxiv.org/abs/2203.04864} {\path{arXiv:2203.04864}},
  \href {https://doi.org/10.1140/epjc/s10052-022-10695-1}
  {\path{doi:10.1140/epjc/s10052-022-10695-1}}.

\bibitem{Bethe:1949yr}
H.~A. Bethe, {Theory of the Effective Range in Nuclear Scattering}, Phys. Rev.
  76 (1949) 38--50.
\newblock \href {https://doi.org/10.1103/PhysRev.76.38}
  {\path{doi:10.1103/PhysRev.76.38}}.

\bibitem{Hanhart:2007yq}
C.~Hanhart, Y.~S. Kalashnikova, A.~E. Kudryavtsev, A.~V. Nefediev, {Reconciling
  the X(3872) with the near-threshold enhancement in the D0 anti-D*0 final
  state}, Phys. Rev. D 76 (2007) 034007.
\newblock \href {http://arxiv.org/abs/0704.0605} {\path{arXiv:0704.0605}},
  \href {https://doi.org/10.1103/PhysRevD.76.034007}
  {\path{doi:10.1103/PhysRevD.76.034007}}.

\bibitem{LHCb:2020xds}
R.~Aaij, et~al., {Study of the lineshape of the $\chi_{c1}(3872)$ state}, Phys.
  Rev. D 102~(9) (2020) 092005.
\newblock \href {http://arxiv.org/abs/2005.13419} {\path{arXiv:2005.13419}},
  \href {https://doi.org/10.1103/PhysRevD.102.092005}
  {\path{doi:10.1103/PhysRevD.102.092005}}.

\bibitem{Du:2023hlu}
M.-L. Du, A.~Filin, V.~Baru, X.-K. Dong, E.~Epelbaum, F.-K. Guo, C.~Hanhart,
  A.~Nefediev, J.~Nieves, Q.~Wang, {Role of Left-Hand Cut Contributions on Pole
  Extractions from Lattice Data: Case Study for Tcc(3875)+}, Phys. Rev. Lett.
  131~(13) (2023) 131903.
\newblock \href {http://arxiv.org/abs/2303.09441} {\path{arXiv:2303.09441}},
  \href {https://doi.org/10.1103/PhysRevLett.131.131903}
  {\path{doi:10.1103/PhysRevLett.131.131903}}.

\bibitem{Padmanath:2022cvl}
M.~Padmanath, S.~Prelovsek, {Signature of a Doubly Charm Tetraquark Pole in DD*
  Scattering on the Lattice}, Phys. Rev. Lett. 129~(3) (2022) 032002.
\newblock \href {http://arxiv.org/abs/2202.10110} {\path{arXiv:2202.10110}},
  \href {https://doi.org/10.1103/PhysRevLett.129.032002}
  {\path{doi:10.1103/PhysRevLett.129.032002}}.

\bibitem{Du:2024snq}
M.-L. Du, F.-K. Guo, B.~Wu, {Effective range expansion with the left-hand cut}
  (8 2024).
\newblock \href {http://arxiv.org/abs/2408.09375} {\path{arXiv:2408.09375}}.

\bibitem{Meng:2023bmz}
L.~Meng, V.~Baru, E.~Epelbaum, A.~A. Filin, A.~M. Gasparyan, {Solving the
  left-hand cut problem in lattice QCD: Tcc(3875)+ from finite volume energy
  levels}, Phys. Rev. D 109~(7) (2024) L071506.
\newblock \href {http://arxiv.org/abs/2312.01930} {\path{arXiv:2312.01930}},
  \href {https://doi.org/10.1103/PhysRevD.109.L071506}
  {\path{doi:10.1103/PhysRevD.109.L071506}}.

\bibitem{Hansen:2024ffk}
M.~T. Hansen, F.~Romero-L\'opez, S.~R. Sharpe, {Incorporating
  DD\ensuremath{\pi} effects and left-hand cuts in lattice QCD studies of the
  T$_{cc}$(3875)$^{+}$}, JHEP 06 (2024) 051.
\newblock \href {http://arxiv.org/abs/2401.06609} {\path{arXiv:2401.06609}},
  \href {https://doi.org/10.1007/JHEP06(2024)051}
  {\path{doi:10.1007/JHEP06(2024)051}}.

\bibitem{Bubna:2024izx}
R.~Bubna, H.-W. Hammer, F.~M\"uller, J.-Y. Pang, A.~Rusetsky, J.-J. Wu,
  {L\"uscher equation with long-range forces}, JHEP 05 (2024) 168.
\newblock \href {http://arxiv.org/abs/2402.12985} {\path{arXiv:2402.12985}},
  \href {https://doi.org/10.1007/JHEP05(2024)168}
  {\path{doi:10.1007/JHEP05(2024)168}}.

\bibitem{Abolnikov:2024key}
M.~Abolnikov, V.~Baru, E.~Epelbaum, A.~A. Filin, C.~Hanhart, L.~Meng, {Internal
  structure of the $T_{cc}(3875)^+$ from its light-quark mass dependence} (7
  2024).
\newblock \href {http://arxiv.org/abs/2407.04649} {\path{arXiv:2407.04649}}.

\bibitem{Collins:2024sfi}
S.~Collins, A.~Nefediev, M.~Padmanath, S.~Prelovsek, {Toward the quark mass
  dependence of Tcc+ from lattice QCD}, Phys. Rev. D 109~(9) (2024) 094509.
\newblock \href {http://arxiv.org/abs/2402.14715} {\path{arXiv:2402.14715}},
  \href {https://doi.org/10.1103/PhysRevD.109.094509}
  {\path{doi:10.1103/PhysRevD.109.094509}}.

\bibitem{Guo:2014taa}
F.-K. Guo, C.~Hanhart, Y.~S. Kalashnikova, U.-G. Mei\ss{}ner, A.~V. Nefediev,
  {What can radiative decays of the X(3872) teach us about its nature?}, Phys.
  Lett. B 742 (2015) 394--398.
\newblock \href {http://arxiv.org/abs/1410.6712} {\path{arXiv:1410.6712}},
  \href {https://doi.org/10.1016/j.physletb.2015.02.013}
  {\path{doi:10.1016/j.physletb.2015.02.013}}.

\bibitem{Swanson:2004pp}
E.~S. Swanson, {Diagnostic decays of the X(3872)}, Phys. Lett. B 598 (2004)
  197--202.
\newblock \href {http://arxiv.org/abs/hep-ph/0406080}
  {\path{arXiv:hep-ph/0406080}}, \href
  {https://doi.org/10.1016/j.physletb.2004.07.059}
  {\path{doi:10.1016/j.physletb.2004.07.059}}.

\bibitem{Grinstein:2024rcu}
B.~Grinstein, L.~Maiani, A.~D. Polosa, {Radiative decays of X(3872)
  discriminate between the molecular and compact interpretations}, Phys. Rev. D
  109~(7) (2024) 074009.
\newblock \href {http://arxiv.org/abs/2401.11623} {\path{arXiv:2401.11623}},
  \href {https://doi.org/10.1103/PhysRevD.109.074009}
  {\path{doi:10.1103/PhysRevD.109.074009}}.

\bibitem{Barnes:2003vb}
T.~Barnes, S.~Godfrey, {Charmonium options for the X(3872)}, Phys. Rev. D 69
  (2004) 054008.
\newblock \href {http://arxiv.org/abs/hep-ph/0311162}
  {\path{arXiv:hep-ph/0311162}}, \href
  {https://doi.org/10.1103/PhysRevD.69.054008}
  {\path{doi:10.1103/PhysRevD.69.054008}}.

\bibitem{Dong:2008gb}
Y.-b. Dong, A.~Faessler, T.~Gutsche, V.~E. Lyubovitskij, {Estimate for the
  X(3872) ---\ensuremath{>} gamma J/psi decay width}, Phys. Rev. D 77 (2008)
  094013.
\newblock \href {http://arxiv.org/abs/0802.3610} {\path{arXiv:0802.3610}},
  \href {https://doi.org/10.1103/PhysRevD.77.094013}
  {\path{doi:10.1103/PhysRevD.77.094013}}.

\bibitem{Dong:2009uf}
Y.~Dong, A.~Faessler, T.~Gutsche, V.~E. Lyubovitskij, {J/psi gamma and psi(2S)
  gamma decay modes of the X(3872)}, J. Phys. G 38 (2011) 015001.
\newblock \href {http://arxiv.org/abs/0909.0380} {\path{arXiv:0909.0380}},
  \href {https://doi.org/10.1088/0954-3899/38/1/015001}
  {\path{doi:10.1088/0954-3899/38/1/015001}}.

\bibitem{Giacosa:2019zxw}
F.~Giacosa, M.~Piotrowska, S.~Coito, {$X(3872)$ as virtual companion pole of
  the charm\textendash{}anticharm state $\chi_{c1}(2P)$}, Int. J. Mod. Phys. A
  34~(29) (2019) 1950173.
\newblock \href {http://arxiv.org/abs/1903.06926} {\path{arXiv:1903.06926}},
  \href {https://doi.org/10.1142/S0217751X19501732}
  {\path{doi:10.1142/S0217751X19501732}}.

\bibitem{Lebed:2022vks}
R.~F. Lebed, S.~R. Martinez, {Diabatic representation of exotic hadrons in the
  dynamical diquark model}, Phys. Rev. D 106~(7) (2022) 074007.
\newblock \href {http://arxiv.org/abs/2207.01101} {\path{arXiv:2207.01101}},
  \href {https://doi.org/10.1103/PhysRevD.106.074007}
  {\path{doi:10.1103/PhysRevD.106.074007}}.

\bibitem{LHCb:2024tpv}
R.~Aaij, et~al., {Probing the nature of the $\chi_{c1}(3872)$ state using
  radiative decays} (6 2024).
\newblock \href {http://arxiv.org/abs/2406.17006} {\path{arXiv:2406.17006}}.

\bibitem{Wang:2013cya}
Q.~Wang, C.~Hanhart, Q.~Zhao, {Decoding the riddle of $Y(4260)$ and
  $Z_c(3900)$}, Phys. Rev. Lett. 111~(13) (2013) 132003.
\newblock \href {http://arxiv.org/abs/1303.6355} {\path{arXiv:1303.6355}},
  \href {https://doi.org/10.1103/PhysRevLett.111.132003}
  {\path{doi:10.1103/PhysRevLett.111.132003}}.

\bibitem{Wang:2013hga}
Q.~Wang, C.~Hanhart, Q.~Zhao, {Systematic study of the singularity mechanism in
  heavy quarkonium decays}, Phys. Lett. B 725~(1-3) (2013) 106--110.
\newblock \href {http://arxiv.org/abs/1305.1997} {\path{arXiv:1305.1997}},
  \href {https://doi.org/10.1016/j.physletb.2013.06.049}
  {\path{doi:10.1016/j.physletb.2013.06.049}}.

\bibitem{Guo:2013zbw}
F.-K. Guo, C.~Hanhart, U.-G. Mei\ss{}ner, Q.~Wang, Q.~Zhao, {Production of the
  X(3872) in charmonia radiative decays}, Phys. Lett. B 725 (2013) 127--133.
\newblock \href {http://arxiv.org/abs/1306.3096} {\path{arXiv:1306.3096}},
  \href {https://doi.org/10.1016/j.physletb.2013.06.053}
  {\path{doi:10.1016/j.physletb.2013.06.053}}.

\bibitem{BESIII:2013fnz}
M.~Ablikim, et~al., {Observation of $e^+e^? ? ?X$(3872) at BESIII}, Phys. Rev.
  Lett. 112~(9) (2014) 092001.
\newblock \href {http://arxiv.org/abs/1310.4101} {\path{arXiv:1310.4101}},
  \href {https://doi.org/10.1103/PhysRevLett.112.092001}
  {\path{doi:10.1103/PhysRevLett.112.092001}}.

\bibitem{Godfrey:1985xj}
S.~Godfrey, N.~Isgur, {Mesons in a Relativized Quark Model with
  Chromodynamics}, Phys. Rev. D32 (1985) 189--231.
\newblock \href {https://doi.org/10.1103/PhysRevD.32.189}
  {\path{doi:10.1103/PhysRevD.32.189}}.

\bibitem{Filin:2010se}
A.~A. Filin, A.~Romanov, V.~Baru, C.~Hanhart, Y.~S. Kalashnikova, A.~E.
  Kudryavtsev, U.~G. Meissner, A.~V. Nefediev, {Comment on `Possibility of
  Deeply Bound Hadronic Molecules from Single Pion Exchange'}, Phys. Rev. Lett.
  105 (2010) 019101.
\newblock \href {http://arxiv.org/abs/1004.4789} {\path{arXiv:1004.4789}},
  \href {https://doi.org/10.1103/PhysRevLett.105.019101}
  {\path{doi:10.1103/PhysRevLett.105.019101}}.

\bibitem{Guo:2011dd}
F.-K. Guo, U.-G. Meissner, {More kaonic bound states and a comprehensive
  interpretation of the $D_{sJ}$ states}, Phys. Rev. D 84 (2011) 014013.
\newblock \href {http://arxiv.org/abs/1102.3536} {\path{arXiv:1102.3536}},
  \href {https://doi.org/10.1103/PhysRevD.84.014013}
  {\path{doi:10.1103/PhysRevD.84.014013}}.

\bibitem{Ding:2008gr}
G.-J. Ding, {Are Y(4260) and Z+(2) are D(1) D or D(0) D* Hadronic Molecules?},
  Phys. Rev. D 79 (2009) 014001.
\newblock \href {http://arxiv.org/abs/0809.4818} {\path{arXiv:0809.4818}},
  \href {https://doi.org/10.1103/PhysRevD.79.014001}
  {\path{doi:10.1103/PhysRevD.79.014001}}.

\bibitem{Cleven:2013mka}
M.~Cleven, Q.~Wang, F.-K. Guo, C.~Hanhart, U.-G. Mei\ss{}ner, Q.~Zhao,
  {$Y(4260)$ as the first $S$-wave open charm vector molecular state?}, Phys.
  Rev. D 90~(7) (2014) 074039.
\newblock \href {http://arxiv.org/abs/1310.2190} {\path{arXiv:1310.2190}},
  \href {https://doi.org/10.1103/PhysRevD.90.074039}
  {\path{doi:10.1103/PhysRevD.90.074039}}.

\bibitem{Ji:2022blw}
T.~Ji, X.-K. Dong, F.-K. Guo, B.-S. Zou, {Prediction of a Narrow Exotic
  Hadronic State with Quantum Numbers $J^{PC}=0^{--}$}, Phys. Rev. Lett.
  129~(10) (2022) 102002.
\newblock \href {http://arxiv.org/abs/2205.10994} {\path{arXiv:2205.10994}},
  \href {https://doi.org/10.1103/PhysRevLett.129.102002}
  {\path{doi:10.1103/PhysRevLett.129.102002}}.

\bibitem{He:2017mbh}
J.~He, D.-Y. Chen, {Interpretation of $Y(4390)$ as an isoscalar partner of
  $Z(4430)$ from $D^*(2010)\bar{D}_1(2420)$ interaction}, Eur. Phys. J. C
  77~(6) (2017) 398.
\newblock \href {http://arxiv.org/abs/1704.08776} {\path{arXiv:1704.08776}},
  \href {https://doi.org/10.1140/epjc/s10052-017-4973-8}
  {\path{doi:10.1140/epjc/s10052-017-4973-8}}.

\bibitem{Wang:2023ivd}
Z.-P. Wang, F.-L. Wang, G.-J. Wang, X.~Liu, {Exploring Charmonium-like
  Molecular Resonances from Deeply Bound $D \bar D_1$, $D^* \bar D_1$, and $D^*
  \bar D_2^*$ Molecules} (12 2023).
\newblock \href {http://arxiv.org/abs/2312.03512} {\path{arXiv:2312.03512}}.

\bibitem{Lin:2024qcq}
Z.-Y. Lin, J.-Z. Wang, J.-B. Cheng, L.~Meng, S.-L. Zhu, {Identify the new state
  $Y(3872)$ as the P-wave $D\bar{D}^*/\bar{D}D^*$ resonance} (3 2024).
\newblock \href {http://arxiv.org/abs/2403.01727} {\path{arXiv:2403.01727}}.

\bibitem{vonDetten:2024eie}
L.~von Detten, V.~Baru, C.~Hanhart, Q.~Wang, D.~Winney, Q.~Zhao, {How many
  vector charmoniumlike states lie in the mass range
  4.2\textendash{}4.35~GeV?}, Phys. Rev. D 109~(11) (2024) 116002.
\newblock \href {http://arxiv.org/abs/2402.03057} {\path{arXiv:2402.03057}},
  \href {https://doi.org/10.1103/PhysRevD.109.116002}
  {\path{doi:10.1103/PhysRevD.109.116002}}.

\bibitem{Nakamura:2023obk}
S.~X. Nakamura, X.~H. Li, H.~P. Peng, Z.~T. Sun, X.~R. Zhou, {Global
  coupled-channel analysis of $e^+e^-\to c\bar{c}$ processes in
  $\sqrt{s}=3.75-4.7$ GeV} (12 2023).
\newblock \href {http://arxiv.org/abs/2312.17658} {\path{arXiv:2312.17658}}.

\bibitem{Braaten:2007dw}
E.~Braaten, M.~Lu, {Line shapes of the X(3872)}, Phys. Rev. D 76 (2007) 094028.
\newblock \href {http://arxiv.org/abs/0709.2697} {\path{arXiv:0709.2697}},
  \href {https://doi.org/10.1103/PhysRevD.76.094028}
  {\path{doi:10.1103/PhysRevD.76.094028}}.

\bibitem{Hanhart:2010wh}
C.~Hanhart, Y.~S. Kalashnikova, A.~V. Nefediev, {Lineshapes for composite
  particles with unstable constituents}, Phys. Rev. D 81 (2010) 094028.
\newblock \href {http://arxiv.org/abs/1002.4097} {\path{arXiv:1002.4097}},
  \href {https://doi.org/10.1103/PhysRevD.81.094028}
  {\path{doi:10.1103/PhysRevD.81.094028}}.

\bibitem{Dong:2021juy}
X.-K. Dong, F.-K. Guo, B.-S. Zou, {A survey of heavy-antiheavy hadronic
  molecules}, Progr. Phys. 41 (2021) 65--93.
\newblock \href {http://arxiv.org/abs/2101.01021} {\path{arXiv:2101.01021}},
  \href {https://doi.org/10.13725/j.cnki.pip.2021.02.001}
  {\path{doi:10.13725/j.cnki.pip.2021.02.001}}.

\bibitem{Gamermann:2006nm}
D.~Gamermann, E.~Oset, D.~Strottman, M.~J. Vicente~Vacas, {Dynamically
  generated open and hidden charm meson systems}, Phys. Rev. D 76 (2007)
  074016.
\newblock \href {http://arxiv.org/abs/hep-ph/0612179}
  {\path{arXiv:hep-ph/0612179}}, \href
  {https://doi.org/10.1103/PhysRevD.76.074016}
  {\path{doi:10.1103/PhysRevD.76.074016}}.

\bibitem{Wang:2013kva}
P.~Wang, X.~G. Wang, {Study on X(3872) from effective field theory with pion
  exchange interaction}, Phys. Rev. Lett. 111~(4) (2013) 042002.
\newblock \href {http://arxiv.org/abs/1304.0846} {\path{arXiv:1304.0846}},
  \href {https://doi.org/10.1103/PhysRevLett.111.042002}
  {\path{doi:10.1103/PhysRevLett.111.042002}}.

\bibitem{Qiu:2023uno}
L.~Qiu, C.~Gong, Q.~Zhao, {Coupled-channel description of charmed heavy
  hadronic molecules within the meson-exchange model and its implication},
  Phys. Rev. D 109~(7) (2024) 076016.
\newblock \href {http://arxiv.org/abs/2311.10067} {\path{arXiv:2311.10067}},
  \href {https://doi.org/10.1103/PhysRevD.109.076016}
  {\path{doi:10.1103/PhysRevD.109.076016}}.

\bibitem{Ke:2021rxd}
H.-W. Ke, X.-H. Liu, X.-Q. Li, {Possible molecular states of $D^{(*)}D^{(*)}$
  and $B^{(*)}B^{(*)}$ within the Bethe\textendash{}Salpeter framework}, Eur.
  Phys. J. C 82~(2) (2022) 144.
\newblock \href {http://arxiv.org/abs/2112.14142} {\path{arXiv:2112.14142}},
  \href {https://doi.org/10.1140/epjc/s10052-022-10092-8}
  {\path{doi:10.1140/epjc/s10052-022-10092-8}}.

\bibitem{Li:2012ss}
N.~Li, Z.-F. Sun, X.~Liu, S.-L. Zhu, {Coupled-channel analysis of the possible
  $D^{(*)}D^{(*)}, \overline{B}^{(*)}\overline{B}^{(*)}$ and
  $D^{(*)}\overline{B}^{(*)}$ molecular states}, Phys. Rev. D 88~(11) (2013)
  114008.
\newblock \href {http://arxiv.org/abs/1211.5007} {\path{arXiv:1211.5007}},
  \href {https://doi.org/10.1103/PhysRevD.88.114008}
  {\path{doi:10.1103/PhysRevD.88.114008}}.

\bibitem{Zhang:2024fxy}
Z.-H. Zhang, T.~Ji, X.-K. Dong, F.-K. Guo, C.~Hanhart, U.-G. Mei\ss{}ner,
  A.~Rusetsky, {Predicting isovector charmonium-like states from $X(3872)$
  properties}, JHEP 08 (2024) 130.
\newblock \href {http://arxiv.org/abs/2404.11215} {\path{arXiv:2404.11215}},
  \href {https://doi.org/https://doi.org/10.1007/JHEP08(2024)130}
  {\path{doi:https://doi.org/10.1007/JHEP08(2024)130}}.

\bibitem{Baru:2018qkb}
V.~Baru, E.~Epelbaum, J.~Gegelia, C.~Hanhart, U.~G. Mei\ss{}ner, A.~V.
  Nefediev, {Remarks on the Heavy-Quark Flavour Symmetry for doubly heavy
  hadronic molecules}, Eur. Phys. J. C 79~(1) (2019) 46.
\newblock \href {http://arxiv.org/abs/1810.06921} {\path{arXiv:1810.06921}},
  \href {https://doi.org/10.1140/epjc/s10052-019-6560-7}
  {\path{doi:10.1140/epjc/s10052-019-6560-7}}.

\bibitem{AlFiky:2005jd}
M.~T. AlFiky, F.~Gabbiani, A.~A. Petrov, {X(3872): Hadronic molecules in
  effective field theory}, Phys. Lett. B 640 (2006) 238--245.
\newblock \href {http://arxiv.org/abs/hep-ph/0506141}
  {\path{arXiv:hep-ph/0506141}}, \href
  {https://doi.org/10.1016/j.physletb.2006.07.069}
  {\path{doi:10.1016/j.physletb.2006.07.069}}.

\bibitem{Albaladejo:2015dsa}
M.~Albaladejo, F.~K. Guo, C.~Hidalgo-Duque, J.~Nieves, M.~P. Valderrama, {Decay
  widths of the spin-2 partners of the X(3872)}, Eur. Phys. J. C 75~(11) (2015)
  547.
\newblock \href {http://arxiv.org/abs/1504.00861} {\path{arXiv:1504.00861}},
  \href {https://doi.org/10.1140/epjc/s10052-015-3753-6}
  {\path{doi:10.1140/epjc/s10052-015-3753-6}}.

\bibitem{Fleming:2007rp}
S.~Fleming, M.~Kusunoki, T.~Mehen, U.~van Kolck, {Pion interactions in the
  $X(3872)$}, Phys. Rev. D 76 (2007) 034006.
\newblock \href {http://arxiv.org/abs/hep-ph/0703168}
  {\path{arXiv:hep-ph/0703168}}, \href
  {https://doi.org/10.1103/PhysRevD.76.034006}
  {\path{doi:10.1103/PhysRevD.76.034006}}.

\bibitem{Valderrama:2012jv}
M.~P. Valderrama, {Power Counting and Perturbative One Pion Exchange in Heavy
  Meson Molecules}, Phys. Rev. D 85 (2012) 114037.
\newblock \href {http://arxiv.org/abs/1204.2400} {\path{arXiv:1204.2400}},
  \href {https://doi.org/10.1103/PhysRevD.85.114037}
  {\path{doi:10.1103/PhysRevD.85.114037}}.

\bibitem{Mehen:2011yh}
T.~Mehen, J.~W. Powell, {Heavy Quark Symmetry Predictions for Weakly Bound
  B-Meson Molecules}, Phys. Rev. D 84 (2011) 114013.
\newblock \href {http://arxiv.org/abs/1109.3479} {\path{arXiv:1109.3479}},
  \href {https://doi.org/10.1103/PhysRevD.84.114013}
  {\path{doi:10.1103/PhysRevD.84.114013}}.

\bibitem{Dai:2019hrf}
L.~Dai, F.-K. Guo, T.~Mehen, {Revisiting $X(3872)\to D^0 \bar{D}^0 \pi^0$ in an
  effective field theory for the $X$(3872)}, Phys. Rev. D 101~(5) (2020)
  054024.
\newblock \href {http://arxiv.org/abs/1912.04317} {\path{arXiv:1912.04317}},
  \href {https://doi.org/10.1103/PhysRevD.101.054024}
  {\path{doi:10.1103/PhysRevD.101.054024}}.

\bibitem{Dai:2023mxm}
L.~Dai, S.~Fleming, R.~Hodges, T.~Mehen, {Strong decays of Tcc+ at NLO in an
  effective field theory}, Phys. Rev. D 107~(7) (2023) 076001.
\newblock \href {http://arxiv.org/abs/2301.11950} {\path{arXiv:2301.11950}},
  \href {https://doi.org/10.1103/PhysRevD.107.076001}
  {\path{doi:10.1103/PhysRevD.107.076001}}.

\bibitem{Jansen:2013cba}
M.~Jansen, H.~W. Hammer, Y.~Jia, {Light quark mass dependence of the X(3872) in
  an effective field theory}, Phys. Rev. D 89~(1) (2014) 014033.
\newblock \href {http://arxiv.org/abs/1310.6937} {\path{arXiv:1310.6937}},
  \href {https://doi.org/10.1103/PhysRevD.89.014033}
  {\path{doi:10.1103/PhysRevD.89.014033}}.

\bibitem{Baru:2011rs}
V.~Baru, A.~A. Filin, C.~Hanhart, Y.~S. Kalashnikova, A.~E. Kudryavtsev, A.~V.
  Nefediev, {Three-body $D\bar{D}\pi$ dynamics for the X(3872)}, Phys. Rev. D
  84 (2011) 074029.
\newblock \href {http://arxiv.org/abs/1108.5644} {\path{arXiv:1108.5644}},
  \href {https://doi.org/10.1103/PhysRevD.84.074029}
  {\path{doi:10.1103/PhysRevD.84.074029}}.

\bibitem{Baru:2013rta}
V.~Baru, E.~Epelbaum, A.~A. Filin, C.~Hanhart, U.~G. Meissner, A.~V. Nefediev,
  {Quark mass dependence of the X(3872) binding energy}, Phys. Lett. B 726
  (2013) 537--543.
\newblock \href {http://arxiv.org/abs/1306.4108} {\path{arXiv:1306.4108}},
  \href {https://doi.org/10.1016/j.physletb.2013.08.073}
  {\path{doi:10.1016/j.physletb.2013.08.073}}.

\bibitem{Schmidt:2018vvl}
M.~Schmidt, M.~Jansen, H.~W. Hammer, {Threshold Effects and the Line Shape of
  the X(3872) in Effective Field Theory}, Phys. Rev. D 98~(1) (2018) 014032.
\newblock \href {http://arxiv.org/abs/1804.00375} {\path{arXiv:1804.00375}},
  \href {https://doi.org/10.1103/PhysRevD.98.014032}
  {\path{doi:10.1103/PhysRevD.98.014032}}.

\bibitem{Baru:2015tfa}
V.~Baru, E.~Epelbaum, A.~A. Filin, J.~Gegelia, A.~V. Nefediev, {Binding energy
  of the $X(3872)$ at unphysical pion masses}, Phys. Rev. D 92~(11) (2015)
  114016.
\newblock \href {http://arxiv.org/abs/1509.01789} {\path{arXiv:1509.01789}},
  \href {https://doi.org/10.1103/PhysRevD.92.114016}
  {\path{doi:10.1103/PhysRevD.92.114016}}.

\bibitem{Baru:2016iwj}
V.~Baru, E.~Epelbaum, A.~A. Filin, C.~Hanhart, U.-G. Mei\ss{}ner, A.~V.
  Nefediev, {Heavy-quark spin symmetry partners of the X (3872) revisited},
  Phys. Lett. B 763 (2016) 20--28.
\newblock \href {http://arxiv.org/abs/1605.09649} {\path{arXiv:1605.09649}},
  \href {https://doi.org/10.1016/j.physletb.2016.10.008}
  {\path{doi:10.1016/j.physletb.2016.10.008}}.

\bibitem{Wang:2018jlv}
Q.~Wang, V.~Baru, A.~A. Filin, C.~Hanhart, A.~V. Nefediev, J.~L. Wynen, {Line
  shapes of the $Z_b(10610)$ and $Z_b(10650)$ in the elastic and inelastic
  channels revisited}, Phys. Rev. D 98~(7) (2018) 074023.
\newblock \href {http://arxiv.org/abs/1805.07453} {\path{arXiv:1805.07453}},
  \href {https://doi.org/10.1103/PhysRevD.98.074023}
  {\path{doi:10.1103/PhysRevD.98.074023}}.

\bibitem{Baru:2019xnh}
V.~Baru, E.~Epelbaum, A.~A. Filin, C.~Hanhart, A.~V. Nefediev, Q.~Wang, {Spin
  partners $W_{bJ}$ from the line shapes of the $Z_b(10610)$ and $Z_b(10650)$},
  Phys. Rev. D 99~(9) (2019) 094013.
\newblock \href {http://arxiv.org/abs/1901.10319} {\path{arXiv:1901.10319}},
  \href {https://doi.org/10.1103/PhysRevD.99.094013}
  {\path{doi:10.1103/PhysRevD.99.094013}}.

\bibitem{Du:2021zzh}
M.-L. Du, V.~Baru, X.-K. Dong, A.~Filin, F.-K. Guo, C.~Hanhart, A.~Nefediev,
  J.~Nieves, Q.~Wang, {Coupled-channel approach to Tcc+ including three-body
  effects}, Phys. Rev. D 105~(1) (2022) 014024.
\newblock \href {http://arxiv.org/abs/2110.13765} {\path{arXiv:2110.13765}},
  \href {https://doi.org/10.1103/PhysRevD.105.014024}
  {\path{doi:10.1103/PhysRevD.105.014024}}.

\bibitem{Voloshin:2011qa}
M.~B. Voloshin, {Radiative transitions from Upsilon(5S) to molecular
  bottomonium}, Phys. Rev. D 84 (2011) 031502.
\newblock \href {http://arxiv.org/abs/1105.5829} {\path{arXiv:1105.5829}},
  \href {https://doi.org/10.1103/PhysRevD.84.031502}
  {\path{doi:10.1103/PhysRevD.84.031502}}.

\bibitem{Aaij:2015tga}
R.~Aaij, et~al., {Observation of $J/\psi p$ Resonances Consistent with
  Pentaquark States in $\Lambda_b^0 \to J/\psi K^- p$ Decays}, Phys. Rev. Lett.
  115 (2015) 072001.
\newblock \href {http://arxiv.org/abs/1507.03414} {\path{arXiv:1507.03414}},
  \href {https://doi.org/10.1103/PhysRevLett.115.072001}
  {\path{doi:10.1103/PhysRevLett.115.072001}}.

\bibitem{Aaij:2016phn}
R.~Aaij, et~al., {Model-independent evidence for $J/\psi p$ contributions to
  $\Lambda_b^0\to J/\psi p K^-$ decays}, Phys. Rev. Lett. 117~(8) (2016)
  082002.
\newblock \href {http://arxiv.org/abs/1604.05708} {\path{arXiv:1604.05708}},
  \href {https://doi.org/10.1103/PhysRevLett.117.082002}
  {\path{doi:10.1103/PhysRevLett.117.082002}}.

\bibitem{Aaij:2016ymb}
R.~Aaij, et~al., {Evidence for exotic hadron contributions to $\Lambda_b^0 \to
  J/\psi p \pi^-$ decays}, Phys. Rev. Lett. 117~(8) (2016) 082003, [Addendum:
  Phys.Rev.Lett. 117, 109902 (2016), Addendum: Phys.Rev.Lett. 118, 119901
  (2017)].
\newblock \href {http://arxiv.org/abs/1606.06999} {\path{arXiv:1606.06999}},
  \href {https://doi.org/10.1103/PhysRevLett.117.082003}
  {\path{doi:10.1103/PhysRevLett.117.082003}}.

\bibitem{Aaij:2019vzc}
R.~Aaij, et~al., {Observation of a narrow pentaquark state, $P_c(4312)^+$, and
  of two-peak structure of the $P_c(4450)^+$}, Phys. Rev. Lett. 122~(22) (2019)
  222001.
\newblock \href {http://arxiv.org/abs/1904.03947} {\path{arXiv:1904.03947}},
  \href {https://doi.org/10.1103/PhysRevLett.122.222001}
  {\path{doi:10.1103/PhysRevLett.122.222001}}.

\bibitem{Chen:2019bip}
H.-X. Chen, W.~Chen, S.-L. Zhu, {Possible interpretations of the $P_c(4312)$,
  $P_c(4440)$, and $P_c(4457)$}, Phys. Rev. D 100~(5) (2019) 051501.
\newblock \href {http://arxiv.org/abs/1903.11001} {\path{arXiv:1903.11001}},
  \href {https://doi.org/10.1103/PhysRevD.100.051501}
  {\path{doi:10.1103/PhysRevD.100.051501}}.

\bibitem{Chen:2019asm}
R.~Chen, Z.-F. Sun, X.~Liu, S.-L. Zhu, {Strong LHCb evidence supporting the
  existence of the hidden-charm molecular pentaquarks}, Phys. Rev. D 100~(1)
  (2019) 011502.
\newblock \href {http://arxiv.org/abs/1903.11013} {\path{arXiv:1903.11013}},
  \href {https://doi.org/10.1103/PhysRevD.100.011502}
  {\path{doi:10.1103/PhysRevD.100.011502}}.

\bibitem{Guo:2019fdo}
F.-K. Guo, H.-J. Jing, U.-G. Mei\ss{}ner, S.~Sakai, {Isospin breaking decays as
  a diagnosis of the hadronic molecular structure of the $P_c(4457)$}, Phys.
  Rev. D 99~(9) (2019) 091501.
\newblock \href {http://arxiv.org/abs/1903.11503} {\path{arXiv:1903.11503}},
  \href {https://doi.org/10.1103/PhysRevD.99.091501}
  {\path{doi:10.1103/PhysRevD.99.091501}}.

\bibitem{Liu:2019tjn}
M.-Z. Liu, Y.-W. Pan, F.-Z. Peng, M.~S\'anchez~S\'anchez, L.-S. Geng,
  A.~Hosaka, M.~Pavon~Valderrama, {Emergence of a complete heavy-quark spin
  symmetry multiplet: seven molecular pentaquarks in light of the latest LHCb
  analysis}, Phys. Rev. Lett. 122~(24) (2019) 242001.
\newblock \href {http://arxiv.org/abs/1903.11560} {\path{arXiv:1903.11560}},
  \href {https://doi.org/10.1103/PhysRevLett.122.242001}
  {\path{doi:10.1103/PhysRevLett.122.242001}}.

\bibitem{He:2019ify}
J.~He, {Study of $P_c(4457)$, $P_c(4440)$, and $P_c(4312)$ in a quasipotential
  Bethe-Salpeter equation approach}, Eur. Phys. J. C 79~(5) (2019) 393.
\newblock \href {http://arxiv.org/abs/1903.11872} {\path{arXiv:1903.11872}},
  \href {https://doi.org/10.1140/epjc/s10052-019-6906-1}
  {\path{doi:10.1140/epjc/s10052-019-6906-1}}.

\bibitem{Guo:2019kdc}
Z.-H. Guo, J.~A. Oller, {Anatomy of the newly observed hidden-charm pentaquark
  states: $P_c(4312)$, $P_c(4440)$ and $P_c(4457)$}, Phys. Lett. B 793 (2019)
  144--149.
\newblock \href {http://arxiv.org/abs/1904.00851} {\path{arXiv:1904.00851}},
  \href {https://doi.org/10.1016/j.physletb.2019.04.053}
  {\path{doi:10.1016/j.physletb.2019.04.053}}.

\bibitem{Shimizu:2019ptd}
Y.~Shimizu, Y.~Yamaguchi, M.~Harada, {Heavy quark spin multiplet structure of
  $P_c(4312)$, $P_c(4440)$, and $P_c(4457)$} (4 2019).
\newblock \href {http://arxiv.org/abs/1904.00587} {\path{arXiv:1904.00587}}.

\bibitem{Xiao:2019mst}
C.-J. Xiao, Y.~Huang, Y.-B. Dong, L.-S. Geng, D.-Y. Chen, {Exploring the
  molecular scenario of Pc(4312) , Pc(4440) , and Pc(4457)}, Phys. Rev. D
  100~(1) (2019) 014022.
\newblock \href {http://arxiv.org/abs/1904.00872} {\path{arXiv:1904.00872}},
  \href {https://doi.org/10.1103/PhysRevD.100.014022}
  {\path{doi:10.1103/PhysRevD.100.014022}}.

\bibitem{Xiao:2019aya}
C.~W. Xiao, J.~Nieves, E.~Oset, {Heavy quark spin symmetric molecular states
  from ${\bar D}^{(*)}\Sigma_c^{(*)}$ and other coupled channels in the light
  of the recent LHCb pentaquarks}, Phys. Rev. D 100~(1) (2019) 014021.
\newblock \href {http://arxiv.org/abs/1904.01296} {\path{arXiv:1904.01296}},
  \href {https://doi.org/10.1103/PhysRevD.100.014021}
  {\path{doi:10.1103/PhysRevD.100.014021}}.

\bibitem{Wang:2019nwt}
F.-L. Wang, R.~Chen, Z.-W. Liu, X.~Liu, {Probing new types of $P_c$ states
  inspired by the interaction between $S$-wave charmed baryon and anti-charmed
  meson in a $\bar T$ doublet}, Phys. Rev. C 101~(2) (2020) 025201.
\newblock \href {http://arxiv.org/abs/1905.03636} {\path{arXiv:1905.03636}},
  \href {https://doi.org/10.1103/PhysRevC.101.025201}
  {\path{doi:10.1103/PhysRevC.101.025201}}.

\bibitem{Meng:2019ilv}
L.~Meng, B.~Wang, G.-J. Wang, S.-L. Zhu, {The hidden charm pentaquark states
  and $\Sigma_c\bar{D}^{(*)}$ interaction in chiral perturbation theory}, Phys.
  Rev. D 100~(1) (2019) 014031.
\newblock \href {http://arxiv.org/abs/1905.04113} {\path{arXiv:1905.04113}},
  \href {https://doi.org/10.1103/PhysRevD.100.014031}
  {\path{doi:10.1103/PhysRevD.100.014031}}.

\bibitem{Wu:2019adv}
J.-J. Wu, T.~S.~H. Lee, B.-S. Zou, {Nucleon resonances with hidden charm in
  \ensuremath{\gamma}p reactions}, Phys. Rev. C 100~(3) (2019) 035206.
\newblock \href {http://arxiv.org/abs/1906.05375} {\path{arXiv:1906.05375}},
  \href {https://doi.org/10.1103/PhysRevC.100.035206}
  {\path{doi:10.1103/PhysRevC.100.035206}}.

\bibitem{Xiao:2019gjd}
C.~W. Xiao, J.~Nieves, E.~Oset, {Prediction of hidden charm strange molecular
  baryon states with heavy quark spin symmetry}, Phys. Lett. B 799 (2019)
  135051.
\newblock \href {http://arxiv.org/abs/1906.09010} {\path{arXiv:1906.09010}},
  \href {https://doi.org/10.1016/j.physletb.2019.135051}
  {\path{doi:10.1016/j.physletb.2019.135051}}.

\bibitem{Voloshin:2019aut}
M.~B. Voloshin, {Some decay properties of hidden-charm pentaquarks as
  baryon-meson molecules}, Phys. Rev. D 100~(3) (2019) 034020.
\newblock \href {http://arxiv.org/abs/1907.01476} {\path{arXiv:1907.01476}},
  \href {https://doi.org/10.1103/PhysRevD.100.034020}
  {\path{doi:10.1103/PhysRevD.100.034020}}.

\bibitem{Sakai:2019qph}
S.~Sakai, H.-J. Jing, F.-K. Guo, {Decays of $P_c$ into $J/\psi N$ and $\eta_cN$
  with heavy quark spin symmetry}, Phys. Rev. D 100~(7) (2019) 074007.
\newblock \href {http://arxiv.org/abs/1907.03414} {\path{arXiv:1907.03414}},
  \href {https://doi.org/10.1103/PhysRevD.100.074007}
  {\path{doi:10.1103/PhysRevD.100.074007}}.

\bibitem{Wang:2019hyc}
Z.-G. Wang, X.~Wang, {Analysis of the strong decays of the $P_c(4312)$ as a
  pentaquark molecular state with QCD sum rules}, Chin. Phys. C 44 (2020)
  103102.
\newblock \href {http://arxiv.org/abs/1907.04582} {\path{arXiv:1907.04582}},
  \href {https://doi.org/10.1088/1674-1137/ababf7}
  {\path{doi:10.1088/1674-1137/ababf7}}.

\bibitem{Yamaguchi:2019seo}
Y.~Yamaguchi, H.~Garc\'\i{}a-Tecocoatzi, A.~Giachino, A.~Hosaka, E.~Santopinto,
  S.~Takeuchi, M.~Takizawa, {$P_c$ pentaquarks with chiral tensor and quark
  dynamics}, Phys. Rev. D 101~(9) (2020) 091502.
\newblock \href {http://arxiv.org/abs/1907.04684} {\path{arXiv:1907.04684}},
  \href {https://doi.org/10.1103/PhysRevD.101.091502}
  {\path{doi:10.1103/PhysRevD.101.091502}}.

\bibitem{Liu:2019zvb}
M.-Z. Liu, T.-W. Wu, M.~S\'anchez~S\'anchez, M.~P. Valderrama, L.-S. Geng,
  J.-J. Xie, {Spin-parities of the $P_c(4440)$ and $P_c(4457)$ in the
  one-boson-exchange model}, Phys. Rev. D 103~(5) (2021) 054004.
\newblock \href {http://arxiv.org/abs/1907.06093} {\path{arXiv:1907.06093}},
  \href {https://doi.org/10.1103/PhysRevD.103.054004}
  {\path{doi:10.1103/PhysRevD.103.054004}}.

\bibitem{Lin:2019qiv}
Y.-H. Lin, B.-S. Zou, {Strong decays of the latest LHCb pentaquark candidates
  in hadronic molecule pictures}, Phys. Rev. D 100~(5) (2019) 056005.
\newblock \href {http://arxiv.org/abs/1908.05309} {\path{arXiv:1908.05309}},
  \href {https://doi.org/10.1103/PhysRevD.100.056005}
  {\path{doi:10.1103/PhysRevD.100.056005}}.

\bibitem{Wang:2019ato}
B.~Wang, L.~Meng, S.-L. Zhu, {Hidden-charm and hidden-bottom molecular
  pentaquarks in chiral effective field theory}, JHEP 11 (2019) 108.
\newblock \href {http://arxiv.org/abs/1909.13054} {\path{arXiv:1909.13054}},
  \href {https://doi.org/10.1007/JHEP11(2019)108}
  {\path{doi:10.1007/JHEP11(2019)108}}.

\bibitem{Gutsche:2019mkg}
T.~Gutsche, V.~E. Lyubovitskij, {Structure and decays of hidden heavy
  pentaquarks}, Phys. Rev. D 100~(9) (2019) 094031.
\newblock \href {http://arxiv.org/abs/1910.03984} {\path{arXiv:1910.03984}},
  \href {https://doi.org/10.1103/PhysRevD.100.094031}
  {\path{doi:10.1103/PhysRevD.100.094031}}.

\bibitem{Burns:2019iih}
T.~J. Burns, E.~S. Swanson, {Molecular interpretation of the $P_c$(4440) and
  $P_c$(4457) states}, Phys. Rev. D 100~(11) (2019) 114033.
\newblock \href {http://arxiv.org/abs/1908.03528} {\path{arXiv:1908.03528}},
  \href {https://doi.org/10.1103/PhysRevD.100.114033}
  {\path{doi:10.1103/PhysRevD.100.114033}}.

\bibitem{Du:2019pij}
M.-L. Du, V.~Baru, F.-K. Guo, C.~Hanhart, U.-G. Mei\ss{}ner, J.~A. Oller,
  Q.~Wang, {Interpretation of the LHCb $P_c$ States as Hadronic Molecules and
  Hints of a Narrow $P_c(4380)$}, Phys. Rev. Lett. 124~(7) (2020) 072001.
\newblock \href {http://arxiv.org/abs/1910.11846} {\path{arXiv:1910.11846}},
  \href {https://doi.org/10.1103/PhysRevLett.124.072001}
  {\path{doi:10.1103/PhysRevLett.124.072001}}.

\bibitem{Wang:2019spc}
G.-J. Wang, L.-Y. Xiao, R.~Chen, X.-H. Liu, X.~Liu, S.-L. Zhu, {Probing
  hidden-charm decay properties of $P_c$ states in a molecular scenario}, Phys.
  Rev. D 102~(3) (2020) 036012.
\newblock \href {http://arxiv.org/abs/1911.09613} {\path{arXiv:1911.09613}},
  \href {https://doi.org/10.1103/PhysRevD.102.036012}
  {\path{doi:10.1103/PhysRevD.102.036012}}.

\bibitem{Xu:2020gjl}
H.~Xu, Q.~Li, C.-H. Chang, G.-L. Wang, {Recently observed $P_c$ as molecular
  states and possible mixture of $P_c(4457)$}, Phys. Rev. D 101~(5) (2020)
  054037.
\newblock \href {http://arxiv.org/abs/2001.02980} {\path{arXiv:2001.02980}},
  \href {https://doi.org/10.1103/PhysRevD.101.054037}
  {\path{doi:10.1103/PhysRevD.101.054037}}.

\bibitem{Kuang:2020bnk}
S.-Q. Kuang, L.-Y. Dai, X.-W. Kang, D.-L. Yao, {Pole analysis on the hadron
  spectroscopy of $\Lambda_b\to J/\Psi p K^-$}, Eur. Phys. J. C 80~(5) (2020)
  433.
\newblock \href {http://arxiv.org/abs/2002.11959} {\path{arXiv:2002.11959}},
  \href {https://doi.org/10.1140/epjc/s10052-020-8008-5}
  {\path{doi:10.1140/epjc/s10052-020-8008-5}}.

\bibitem{Peng:2020xrf}
F.-Z. Peng, M.-Z. Liu, M.~S\'anchez~S\'anchez, M.~Pavon~Valderrama,
  {Heavy-hadron molecules from light-meson-exchange saturation}, Phys. Rev. D
  102 (2020) 114020.
\newblock \href {http://arxiv.org/abs/2004.05658} {\path{arXiv:2004.05658}},
  \href {https://doi.org/10.1103/PhysRevD.102.114020}
  {\path{doi:10.1103/PhysRevD.102.114020}}.

\bibitem{Peng:2020gwk}
F.-Z. Peng, J.-X. Lu, M.~S\'anchez~S\'anchez, M.-J. Yan, M.~Pavon~Valderrama,
  {Peaks within peaks and the possible two-peak structure of the Pc(4457) : The
  effective field theory perspective}, Phys. Rev. D 103~(1) (2021) 014023.
\newblock \href {http://arxiv.org/abs/2007.01198} {\path{arXiv:2007.01198}},
  \href {https://doi.org/10.1103/PhysRevD.103.014023}
  {\path{doi:10.1103/PhysRevD.103.014023}}.

\bibitem{Xiao:2020frg}
C.~W. Xiao, J.~X. Lu, J.~J. Wu, L.~S. Geng, {How to reveal the nature of three
  or more pentaquark states}, Phys. Rev. D 102~(5) (2020) 056018.
\newblock \href {http://arxiv.org/abs/2007.12106} {\path{arXiv:2007.12106}},
  \href {https://doi.org/10.1103/PhysRevD.102.056018}
  {\path{doi:10.1103/PhysRevD.102.056018}}.

\bibitem{Peng:2021hkr}
F.-Z. Peng, M.~S\'anchez~S\'anchez, M.-J. Yan, M.~Pavon~Valderrama,
  {Heavy-hadron molecular spectrum from light-meson exchange saturation} (1
  2021).
\newblock \href {http://arxiv.org/abs/2101.07213} {\path{arXiv:2101.07213}}.

\bibitem{Ali:2019npk}
A.~Ali, A.~Y. Parkhomenko, {Interpretation of the narrow $J/\psi p$ Peaks in
  $\Lambda_b \to J/\psi p K^-$ decay in the compact diquark model}, Phys. Lett.
  B 793 (2019) 365--371.
\newblock \href {http://arxiv.org/abs/1904.00446} {\path{arXiv:1904.00446}},
  \href {https://doi.org/10.1016/j.physletb.2019.05.002}
  {\path{doi:10.1016/j.physletb.2019.05.002}}.

\bibitem{Zhu:2019iwm}
R.~Zhu, X.~Liu, H.~Huang, C.-F. Qiao, {Analyzing doubly heavy tetra- and
  penta-quark states by variational method}, Phys. Lett. B 797 (2019) 134869.
\newblock \href {http://arxiv.org/abs/1904.10285} {\path{arXiv:1904.10285}},
  \href {https://doi.org/10.1016/j.physletb.2019.134869}
  {\path{doi:10.1016/j.physletb.2019.134869}}.

\bibitem{Wang:2019got}
Z.-G. Wang, {Analysis of the $P_c(4312)$, $P_c(4440)$, $P_c(4457)$ and related
  hidden-charm pentaquark states with QCD sum rules}, Int. J. Mod. Phys. A
  35~(01) (2020) 2050003.
\newblock \href {http://arxiv.org/abs/1905.02892} {\path{arXiv:1905.02892}},
  \href {https://doi.org/10.1142/S0217751X20500037}
  {\path{doi:10.1142/S0217751X20500037}}.

\bibitem{Giron:2019bcs}
J.~F. Giron, R.~F. Lebed, C.~T. Peterson, {The Dynamical Diquark Model: First
  Numerical Results}, JHEP 05 (2019) 061.
\newblock \href {http://arxiv.org/abs/1903.04551} {\path{arXiv:1903.04551}},
  \href {https://doi.org/10.1007/JHEP05(2019)061}
  {\path{doi:10.1007/JHEP05(2019)061}}.

\bibitem{Cheng:2019obk}
J.-B. Cheng, Y.-R. Liu, {$P_c(4457)^+$, $P_c(4440)^+$, and $P_c(4312)^+$:
  molecules or compact pentaquarks?}, Phys. Rev. D 100~(5) (2019) 054002.
\newblock \href {http://arxiv.org/abs/1905.08605} {\path{arXiv:1905.08605}},
  \href {https://doi.org/10.1103/PhysRevD.100.054002}
  {\path{doi:10.1103/PhysRevD.100.054002}}.

\bibitem{Stancu:2019qga}
F.~Stancu, {Spectrum of the $uudc \bar{c}$ hidden charm pentaquark with an
  SU(4) flavor-spin hyperfine interaction}, Eur. Phys. J. C 79~(11) (2019) 957.
\newblock \href {http://arxiv.org/abs/1902.07101} {\path{arXiv:1902.07101}},
  \href {https://doi.org/10.1140/epjc/s10052-019-7474-0}
  {\path{doi:10.1140/epjc/s10052-019-7474-0}}.

\bibitem{Eides:2015dtr}
M.~I. Eides, V.~Y. Petrov, M.~V. Polyakov, {Narrow Nucleon-$\psi(2S)$ Bound
  State and LHCb Pentaquarks}, Phys. Rev. D 93~(5) (2016) 054039.
\newblock \href {http://arxiv.org/abs/1512.00426} {\path{arXiv:1512.00426}},
  \href {https://doi.org/10.1103/PhysRevD.93.054039}
  {\path{doi:10.1103/PhysRevD.93.054039}}.

\bibitem{Eides:2019tgv}
M.~I. Eides, V.~Y. Petrov, M.~V. Polyakov, {New LHCb pentaquarks as
  hadrocharmonium states}, Mod. Phys. Lett. A 35~(18) (2020) 2050151.
\newblock \href {http://arxiv.org/abs/1904.11616} {\path{arXiv:1904.11616}},
  \href {https://doi.org/10.1142/S0217732320501515}
  {\path{doi:10.1142/S0217732320501515}}.

\bibitem{Anwar:2018bpu}
J.~Ferretti, E.~Santopinto, M.~Naeem~Anwar, M.~A. Bedolla, {The
  baryo-quarkonium picture for hidden-charm and bottom pentaquarks and LHCb
  $P_{\rm c}(4380)$ and $P_{\rm c}(4450)$ states}, Phys. Lett. B 789 (2019)
  562--567.
\newblock \href {http://arxiv.org/abs/1807.01207} {\path{arXiv:1807.01207}},
  \href {https://doi.org/10.1016/j.physletb.2018.09.047}
  {\path{doi:10.1016/j.physletb.2018.09.047}}.

\bibitem{Fernandez-Ramirez:2019koa}
C.~Fern\'andez-Ram\'\i{}rez, A.~Pilloni, M.~Albaladejo, A.~Jackura, V.~Mathieu,
  M.~Mikhasenko, J.~A. Silva-Castro, A.~P. Szczepaniak, {Interpretation of the
  LHCb $P_c$(4312)$^+$ Signal}, Phys. Rev. Lett. 123~(9) (2019) 092001.
\newblock \href {http://arxiv.org/abs/1904.10021} {\path{arXiv:1904.10021}},
  \href {https://doi.org/10.1103/PhysRevLett.123.092001}
  {\path{doi:10.1103/PhysRevLett.123.092001}}.

\bibitem{Guo:2014iya}
F.-K. Guo, C.~Hanhart, Q.~Wang, Q.~Zhao, {Could the near-threshold $XYZ$ states
  be simply kinematic effects?}, Phys. Rev. D 91~(5) (2015) 051504.
\newblock \href {http://arxiv.org/abs/1411.5584} {\path{arXiv:1411.5584}},
  \href {https://doi.org/10.1103/PhysRevD.91.051504}
  {\path{doi:10.1103/PhysRevD.91.051504}}.

\bibitem{Guo:2019twa}
F.-K. Guo, X.-H. Liu, S.~Sakai, {Threshold cusps and triangle singularities in
  hadronic reactions}, Prog. Part. Nucl. Phys. 112 (2020) 103757.
\newblock \href {http://arxiv.org/abs/1912.07030} {\path{arXiv:1912.07030}},
  \href {https://doi.org/10.1016/j.ppnp.2020.103757}
  {\path{doi:10.1016/j.ppnp.2020.103757}}.

\bibitem{Dong:2020hxe}
X.-K. Dong, F.-K. Guo, B.-S. Zou, {Explaining the Many Threshold Structures in
  the Heavy-Quark Hadron Spectrum}, Phys. Rev. Lett. 126~(15) (2021) 152001.
\newblock \href {http://arxiv.org/abs/2011.14517} {\path{arXiv:2011.14517}},
  \href {https://doi.org/10.1103/PhysRevLett.126.152001}
  {\path{doi:10.1103/PhysRevLett.126.152001}}.

\bibitem{Guo:2015umn}
F.-K. Guo, U.-G. Mei\ss{}ner, W.~Wang, Z.~Yang, {How to reveal the exotic
  nature of the P$_c$(4450)}, Phys. Rev. D 92~(7) (2015) 071502.
\newblock \href {http://arxiv.org/abs/1507.04950} {\path{arXiv:1507.04950}},
  \href {https://doi.org/10.1103/PhysRevD.92.071502}
  {\path{doi:10.1103/PhysRevD.92.071502}}.

\bibitem{Liu:2015fea}
X.-H. Liu, Q.~Wang, Q.~Zhao, {Understanding the newly observed heavy pentaquark
  candidates}, Phys. Lett. B 757 (2016) 231--236.
\newblock \href {http://arxiv.org/abs/1507.05359} {\path{arXiv:1507.05359}},
  \href {https://doi.org/10.1016/j.physletb.2016.03.089}
  {\path{doi:10.1016/j.physletb.2016.03.089}}.

\bibitem{Mikhasenko:2015vca}
M.~Mikhasenko, {A triangle singularity and the LHCb pentaquarks} (7 2015).
\newblock \href {http://arxiv.org/abs/1507.06552} {\path{arXiv:1507.06552}}.

\bibitem{Bayar:2016ftu}
M.~Bayar, F.~Aceti, F.-K. Guo, E.~Oset, {A Discussion on Triangle Singularities
  in the $\Lambda_b \to J/\psi K^{-} p$ Reaction}, Phys. Rev. D 94~(7) (2016)
  074039.
\newblock \href {http://arxiv.org/abs/1609.04133} {\path{arXiv:1609.04133}},
  \href {https://doi.org/10.1103/PhysRevD.94.074039}
  {\path{doi:10.1103/PhysRevD.94.074039}}.

\bibitem{Xiao:2013yca}
C.~W. Xiao, J.~Nieves, E.~Oset, {Combining heavy quark spin and local hidden
  gauge symmetries in the dynamical generation of hidden charm baryons}, Phys.
  Rev. D 88 (2013) 056012.
\newblock \href {http://arxiv.org/abs/1304.5368} {\path{arXiv:1304.5368}},
  \href {https://doi.org/10.1103/PhysRevD.88.056012}
  {\path{doi:10.1103/PhysRevD.88.056012}}.

\bibitem{Liu:2018zzu}
M.-Z. Liu, F.-Z. Peng, M.~S\'anchez~S\'anchez, M.~P. Valderrama, {Heavy-quark
  symmetry partners of the $P_c(4450)$ pentaquark}, Phys. Rev. D 98~(11) (2018)
  114030.
\newblock \href {http://arxiv.org/abs/1811.03992} {\path{arXiv:1811.03992}},
  \href {https://doi.org/10.1103/PhysRevD.98.114030}
  {\path{doi:10.1103/PhysRevD.98.114030}}.

\bibitem{Valderrama:2019chc}
M.~Pavon~Valderrama, {One pion exchange and the quantum numbers of the
  P$_c$(4440) and P$_c$(4457) pentaquarks}, Phys. Rev. D 100~(9) (2019) 094028.
\newblock \href {http://arxiv.org/abs/1907.05294} {\path{arXiv:1907.05294}},
  \href {https://doi.org/10.1103/PhysRevD.100.094028}
  {\path{doi:10.1103/PhysRevD.100.094028}}.

\bibitem{Du:2021fmf}
M.-L. Du, V.~Baru, F.-K. Guo, C.~Hanhart, U.-G. Mei\ss{}ner, J.~A. Oller,
  Q.~Wang, {Revisiting the nature of the P$_{c}$ pentaquarks}, JHEP 08 (2021)
  157.
\newblock \href {http://arxiv.org/abs/2102.07159} {\path{arXiv:2102.07159}},
  \href {https://doi.org/10.1007/JHEP08(2021)157}
  {\path{doi:10.1007/JHEP08(2021)157}}.

\bibitem{Aaij:2020fnh}
R.~Aaij, et~al., {Observation of structure in the $J /\psi$ -pair mass
  spectrum}, Sci. Bull. 65~(23) (2020) 1983--1993.
\newblock \href {http://arxiv.org/abs/2006.16957} {\path{arXiv:2006.16957}},
  \href {https://doi.org/10.1016/j.scib.2020.08.032}
  {\path{doi:10.1016/j.scib.2020.08.032}}.

\bibitem{ParticleDataGroup:2024cfk}
S.~Navas, {Others}, Review of {{Particle Physics}}, Phys. Rev. D 110~(3) (2024)
  030001.
\newblock \href {https://doi.org/10.1103/PhysRevD.110.030001}
  {\path{doi:10.1103/PhysRevD.110.030001}}.

\bibitem{Iwasaki:1975pv}
Y.~Iwasaki, {A Possible Model for New Resonances-Exotics and Hidden Charm},
  Prog. Theor. Phys. 54 (1975) 492.
\newblock \href {https://doi.org/10.1143/PTP.54.492}
  {\path{doi:10.1143/PTP.54.492}}.

\bibitem{Chao:1980dv}
K.-T. Chao, {The (cc) - ($\bar{cc}$) (Diquark - Anti-Diquark) States in $e^+
  e^-$ Annihilation}, Z. Phys. C 7 (1981) 317.
\newblock \href {https://doi.org/10.1007/BF01431564}
  {\path{doi:10.1007/BF01431564}}.

\bibitem{Badalian:1985es}
A.~M. Badalian, B.~L. Ioffe, A.~V. Smilga, {FOUR QUARK STATES IN THE HEAVY
  QUARK SYSTEM}, Nucl. Phys. B 281 (1987) 85.
\newblock \href {https://doi.org/10.1016/0550-3213(87)90248-3}
  {\path{doi:10.1016/0550-3213(87)90248-3}}.

\bibitem{Ader:1981db}
J.~P. Ader, J.~M. Richard, P.~Taxil, {DO NARROW HEAVY MULTI - QUARK STATES
  EXIST?}, Phys. Rev. D 25 (1982) 2370.
\newblock \href {https://doi.org/10.1103/PhysRevD.25.2370}
  {\path{doi:10.1103/PhysRevD.25.2370}}.

\bibitem{Wu:2016vtq}
J.~Wu, Y.-R. Liu, K.~Chen, X.~Liu, S.-L. Zhu, {Heavy-flavored tetraquark states
  with the $QQ\bar{Q}\bar{Q}$ configuration}, Phys. Rev. D 97~(9) (2018)
  094015.
\newblock \href {http://arxiv.org/abs/1605.01134} {\path{arXiv:1605.01134}},
  \href {https://doi.org/10.1103/PhysRevD.97.094015}
  {\path{doi:10.1103/PhysRevD.97.094015}}.

\bibitem{Karliner:2016zzc}
M.~Karliner, S.~Nussinov, J.~L. Rosner, {$Q Q \bar Q \bar Q$ states: masses,
  production, and decays}, Phys. Rev. D 95~(3) (2017) 034011.
\newblock \href {http://arxiv.org/abs/1611.00348} {\path{arXiv:1611.00348}},
  \href {https://doi.org/10.1103/PhysRevD.95.034011}
  {\path{doi:10.1103/PhysRevD.95.034011}}.

\bibitem{Wang:2017jtz}
Z.-G. Wang, {Analysis of the $QQ\bar{Q}\bar{Q}$ tetraquark states with QCD sum
  rules}, Eur. Phys. J. C 77~(7) (2017) 432.
\newblock \href {http://arxiv.org/abs/1701.04285} {\path{arXiv:1701.04285}},
  \href {https://doi.org/10.1140/epjc/s10052-017-4997-0}
  {\path{doi:10.1140/epjc/s10052-017-4997-0}}.

\bibitem{Liu:2019zuc}
M.-S. Liu, Q.-F. L\"u, X.-H. Zhong, Q.~Zhao, {All-heavy tetraquarks}, Phys.
  Rev. D 100~(1) (2019) 016006.
\newblock \href {http://arxiv.org/abs/1901.02564} {\path{arXiv:1901.02564}},
  \href {https://doi.org/10.1103/PhysRevD.100.016006}
  {\path{doi:10.1103/PhysRevD.100.016006}}.

\bibitem{Bedolla:2019zwg}
M.~A. Bedolla, J.~Ferretti, C.~D. Roberts, E.~Santopinto, {Spectrum of
  fully-heavy tetraquarks from a diquark+antidiquark perspective}, Eur. Phys.
  J. C 80~(11) (2020) 1004.
\newblock \href {http://arxiv.org/abs/1911.00960} {\path{arXiv:1911.00960}},
  \href {https://doi.org/10.1140/epjc/s10052-020-08579-3}
  {\path{doi:10.1140/epjc/s10052-020-08579-3}}.

\bibitem{Chen:2020lgj}
X.~Chen, {Fully-charm tetraquarks: $cc\bar{c}\bar{c}$} (1 2020).
\newblock \href {http://arxiv.org/abs/2001.06755} {\path{arXiv:2001.06755}}.

\bibitem{Dong:2020nwy}
X.-K. Dong, V.~Baru, F.-K. Guo, C.~Hanhart, A.~Nefediev, {Coupled-Channel
  Interpretation of the LHCb Double-~$J/\psi$~Spectrum and Hints of a New State
  Near the~ $J/\psi J/\psi$~~Threshold}, Phys. Rev. Lett. 126~(13) (2021)
  132001, [Erratum: Phys.Rev.Lett. 127, 119901 (2021)].
\newblock \href {http://arxiv.org/abs/2009.07795} {\path{arXiv:2009.07795}},
  \href {https://doi.org/10.1103/PhysRevLett.127.119901}
  {\path{doi:10.1103/PhysRevLett.127.119901}}.

\bibitem{Liang:2021fzr}
Z.-R. Liang, X.-Y. Wu, D.-L. Yao, Hunting for states in the recent {{LHCb}}
  di-{{J}}/{$\psi$} invariant mass spectrum, Phys. Rev. D 104~(3) (2021)
  034034.
\newblock \href {http://arxiv.org/abs/2104.08589} {\path{arXiv:2104.08589}},
  \href {https://doi.org/10.1103/PhysRevD.104.034034}
  {\path{doi:10.1103/PhysRevD.104.034034}}.

\bibitem{Huang:2024jin}
Q.~Huang, R.~Chen, J.~He, X.~Liu, {Discovering a Novel Dynamics Mechanism for
  Charmonium Scattering} (7 2024).
\newblock \href {http://arxiv.org/abs/2407.16316} {\path{arXiv:2407.16316}}.

\bibitem{CMS:2023owd}
A.~Hayrapetyan, et~al., {New Structures in the
  J/\ensuremath{\psi}J/\ensuremath{\psi} Mass Spectrum in Proton-Proton
  Collisions at s=13\,\,TeV}, Phys. Rev. Lett. 132~(11) (2024) 111901.
\newblock \href {http://arxiv.org/abs/2306.07164} {\path{arXiv:2306.07164}},
  \href {https://doi.org/10.1103/PhysRevLett.132.111901}
  {\path{doi:10.1103/PhysRevLett.132.111901}}.

\bibitem{ATLAS:2023bft}
G.~Aad, et~al., {Observation of an Excess of Dicharmonium Events in the
  Four-Muon Final State with the ATLAS Detector}, Phys. Rev. Lett. 131~(15)
  (2023) 151902.
\newblock \href {http://arxiv.org/abs/2304.08962} {\path{arXiv:2304.08962}},
  \href {https://doi.org/10.1103/PhysRevLett.131.151902}
  {\path{doi:10.1103/PhysRevLett.131.151902}}.

\bibitem{Song:2024ykq}
Y.-L. Song, Y.~Zhang, V.~Baru, F.-K. Guo, C.~Hanhart, A.~Nefediev, {Towards a
  precision determination of the $X(6200)$ parameters from data} (11 2024).
\newblock \href {http://arxiv.org/abs/2411.12062} {\path{arXiv:2411.12062}}.

\bibitem{Cahn:2003cw}
R.~N. Cahn, J.~D. Jackson, {Spin orbit and tensor forces in heavy quark light
  quark mesons: Implications of the new D(s) state at 2.32-GeV}, Phys. Rev. D68
  (2003) 037502.
\newblock \href {http://arxiv.org/abs/hep-ph/0305012}
  {\path{arXiv:hep-ph/0305012}}, \href
  {https://doi.org/10.1103/PhysRevD.68.037502}
  {\path{doi:10.1103/PhysRevD.68.037502}}.

\bibitem{Godfrey:2003kg}
S.~Godfrey, {Testing the nature of the D(sJ)*(2317)+ and D(sJ)(2463)+ states
  using radiative transitions}, Phys. Lett. B568 (2003) 254--260.
\newblock \href {http://arxiv.org/abs/hep-ph/0305122}
  {\path{arXiv:hep-ph/0305122}}, \href
  {https://doi.org/10.1016/j.physletb.2003.06.049}
  {\path{doi:10.1016/j.physletb.2003.06.049}}.

\bibitem{Colangelo:2003vg}
P.~Colangelo, F.~De~Fazio, {Understanding D(sJ)(2317)}, Phys. Lett. B570 (2003)
  180--184.
\newblock \href {http://arxiv.org/abs/hep-ph/0305140}
  {\path{arXiv:hep-ph/0305140}}, \href
  {https://doi.org/10.1016/j.physletb.2003.08.003}
  {\path{doi:10.1016/j.physletb.2003.08.003}}.

\bibitem{Mehen:2005hc}
T.~Mehen, R.~P. Springer, {Even- and odd-parity charmed meson masses in heavy
  hadron chiral perturbation theory}, Phys. Rev. D72 (2005) 034006.
\newblock \href {http://arxiv.org/abs/hep-ph/0503134}
  {\path{arXiv:hep-ph/0503134}}, \href
  {https://doi.org/10.1103/PhysRevD.72.034006}
  {\path{doi:10.1103/PhysRevD.72.034006}}.

\bibitem{Lakhina:2006fy}
O.~Lakhina, E.~S. Swanson, {A Canonical Ds(2317)?}, Phys. Lett. B650 (2007)
  159--165.
\newblock \href {http://arxiv.org/abs/hep-ph/0608011}
  {\path{arXiv:hep-ph/0608011}}, \href
  {https://doi.org/10.1016/j.physletb.2007.01.075}
  {\path{doi:10.1016/j.physletb.2007.01.075}}.

\bibitem{Bardeen:2003kt}
W.~A. Bardeen, E.~J. Eichten, C.~T. Hill, {Chiral multiplets of heavy - light
  mesons}, Phys. Rev. D68 (2003) 054024.
\newblock \href {http://arxiv.org/abs/hep-ph/0305049}
  {\path{arXiv:hep-ph/0305049}}, \href
  {https://doi.org/10.1103/PhysRevD.68.054024}
  {\path{doi:10.1103/PhysRevD.68.054024}}.

\bibitem{Nowak:2003ra}
M.~A. Nowak, M.~Rho, I.~Zahed, {Chiral doubling of heavy light hadrons: BABAR
  2317-MeV/c**2 and CLEO 2463-MeV/c**2 discoveries}, Acta Phys. Polon. B35
  (2004) 2377--2392.
\newblock \href {http://arxiv.org/abs/hep-ph/0307102}
  {\path{arXiv:hep-ph/0307102}}.

\bibitem{Browder:2003fk}
T.~E. Browder, S.~Pakvasa, A.~A. Petrov, {Comment on the new D(s)(*)+ pi0
  resonances}, Phys. Lett. B578 (2004) 365--368.
\newblock \href {http://arxiv.org/abs/hep-ph/0307054}
  {\path{arXiv:hep-ph/0307054}}, \href
  {https://doi.org/10.1016/j.physletb.2003.10.067}
  {\path{doi:10.1016/j.physletb.2003.10.067}}.

\bibitem{Szczepaniak:2003vy}
A.~P. Szczepaniak, {Description of the D*(s)(2320) resonance as the D pi atom},
  Phys. Lett. B567 (2003) 23--26.
\newblock \href {http://arxiv.org/abs/hep-ph/0305060}
  {\path{arXiv:hep-ph/0305060}}, \href
  {https://doi.org/10.1016/S0370-2693(03)00865-7}
  {\path{doi:10.1016/S0370-2693(03)00865-7}}.

\bibitem{Barnes:2003dj}
T.~Barnes, F.~E. Close, H.~J. Lipkin, {Implications of a DK molecule at
  2.32-GeV}, Phys. Rev. D68 (2003) 054006.
\newblock \href {http://arxiv.org/abs/hep-ph/0305025}
  {\path{arXiv:hep-ph/0305025}}, \href
  {https://doi.org/10.1103/PhysRevD.68.054006}
  {\path{doi:10.1103/PhysRevD.68.054006}}.

\bibitem{vanBeveren:2003kd}
E.~van Beveren, G.~Rupp, {Observed D(s)(2317) and tentative D(2030) as the
  charmed cousins of the light scalar nonet}, Phys. Rev. Lett. 91 (2003)
  012003.
\newblock \href {http://arxiv.org/abs/hep-ph/0305035}
  {\path{arXiv:hep-ph/0305035}}, \href
  {https://doi.org/10.1103/PhysRevLett.91.012003}
  {\path{doi:10.1103/PhysRevLett.91.012003}}.

\bibitem{Chen:2004dy}
Y.-Q. Chen, X.-Q. Li, {A Comprehensive four-quark interpretation of D(s)(2317),
  D(s)(2457) and D(s)(2632)}, Phys. Rev. Lett. 93 (2004) 232001.
\newblock \href {http://arxiv.org/abs/hep-ph/0407062}
  {\path{arXiv:hep-ph/0407062}}, \href
  {https://doi.org/10.1103/PhysRevLett.93.232001}
  {\path{doi:10.1103/PhysRevLett.93.232001}}.

\bibitem{Kolomeitsev:2003ac}
E.~E. Kolomeitsev, M.~F.~M. Lutz, {On Heavy light meson resonances and chiral
  symmetry}, Phys. Lett. B582 (2004) 39--48.
\newblock \href {http://arxiv.org/abs/hep-ph/0307133}
  {\path{arXiv:hep-ph/0307133}}, \href
  {https://doi.org/10.1016/j.physletb.2003.10.118}
  {\path{doi:10.1016/j.physletb.2003.10.118}}.

\bibitem{Guo:2006fu}
F.-K. Guo, P.-N. Shen, H.-C. Chiang, R.-G. Ping, B.-S. Zou, {Dynamically
  generated 0+ heavy mesons in a heavy chiral unitary approach}, Phys. Lett.
  B641 (2006) 278--285.
\newblock \href {http://arxiv.org/abs/hep-ph/0603072}
  {\path{arXiv:hep-ph/0603072}}, \href
  {https://doi.org/10.1016/j.physletb.2006.08.064}
  {\path{doi:10.1016/j.physletb.2006.08.064}}.

\bibitem{Guo:2006rp}
F.-K. Guo, P.-N. Shen, H.-C. Chiang, {Dynamically generated 1+ heavy mesons},
  Phys. Lett. B647 (2007) 133--139.
\newblock \href {http://arxiv.org/abs/hep-ph/0610008}
  {\path{arXiv:hep-ph/0610008}}, \href
  {https://doi.org/10.1016/j.physletb.2007.01.050}
  {\path{doi:10.1016/j.physletb.2007.01.050}}.

\bibitem{Dmitrasinovic:2005gc}
V.~Dmitrasinovic, {D(s0)+(2317) - D(0)(2308) mass difference as evidence for
  tetraquarks}, Phys. Rev. Lett. 94 (2005) 162002.
\newblock \href {https://doi.org/10.1103/PhysRevLett.94.162002}
  {\path{doi:10.1103/PhysRevLett.94.162002}}.

\bibitem{Albaladejo:2016lbb}
M.~Albaladejo, P.~Fernandez-Soler, F.-K. Guo, J.~Nieves, {Two-pole structure of
  the $D^\ast_0(2400)$}, Phys. Lett. B 767 (2017) 465--469.
\newblock \href {http://arxiv.org/abs/1610.06727} {\path{arXiv:1610.06727}},
  \href {https://doi.org/10.1016/j.physletb.2017.02.036}
  {\path{doi:10.1016/j.physletb.2017.02.036}}.

\bibitem{Du:2017zvv}
M.-L. Du, M.~Albaladejo, P.~Fernández-Soler, F.-K. Guo, C.~Hanhart, U.-G.
  Meißner, J.~Nieves, D.-L. Yao, {Towards a new paradigm for heavy-light meson
  spectroscopy}, Phys. Rev. D98~(9) (2018) 094018.
\newblock \href {http://arxiv.org/abs/1712.07957} {\path{arXiv:1712.07957}},
  \href {https://doi.org/10.1103/PhysRevD.98.094018}
  {\path{doi:10.1103/PhysRevD.98.094018}}.

\bibitem{Du:2020pui}
M.-L. Du, F.-K. Guo, C.~Hanhart, B.~Kubis, U.-G. Mei\ss{}ner, {Where is the
  lightest charmed scalar meson?}, Phys. Rev. Lett. 126~(19) (2021) 192001.
\newblock \href {http://arxiv.org/abs/2012.04599} {\path{arXiv:2012.04599}},
  \href {https://doi.org/10.1103/PhysRevLett.126.192001}
  {\path{doi:10.1103/PhysRevLett.126.192001}}.

\bibitem{Guo:2018gyd}
X.-Y. Guo, Y.~Heo, M.~F.~M. Lutz, {On chiral excitations with exotic quantum
  numbers}, Phys. Lett. B 791 (2019) 86--91.
\newblock \href {http://arxiv.org/abs/1809.01311} {\path{arXiv:1809.01311}},
  \href {https://doi.org/10.1016/j.physletb.2019.02.022}
  {\path{doi:10.1016/j.physletb.2019.02.022}}.

\bibitem{Meissner:2020khl}
U.-G. Mei\ss{}ner, {Two-pole structures in QCD: Facts, not fantasy!}, Symmetry
  12~(6) (2020) 981.
\newblock \href {http://arxiv.org/abs/2005.06909} {\path{arXiv:2005.06909}},
  \href {https://doi.org/10.3390/sym12060981} {\path{doi:10.3390/sym12060981}}.

\bibitem{Liu:2012zya}
L.~Liu, K.~Orginos, F.-K. Guo, C.~Hanhart, U.-G. Meissner, {Interactions of
  charmed mesons with light pseudoscalar mesons from lattice QCD and
  implications on the nature of the $D_{s0}^*(2317)$}, Phys. Rev. D87~(1)
  (2013) 014508.
\newblock \href {http://arxiv.org/abs/1208.4535} {\path{arXiv:1208.4535}},
  \href {https://doi.org/10.1103/PhysRevD.87.014508}
  {\path{doi:10.1103/PhysRevD.87.014508}}.

\bibitem{Gregory:2021rgy}
E.~B. Gregory, F.-K. Guo, C.~Hanhart, S.~Krieg, T.~Luu, {Confirmation of the
  existence of an exotic state in the $\pi D$ system} (6 2021).
\newblock \href {http://arxiv.org/abs/2106.15391} {\path{arXiv:2106.15391}}.

\bibitem{Yeo:2024chk}
J.~D.~E. Yeo, C.~E. Thomas, D.~J. Wilson, {DK/D\ensuremath{\pi} scattering and
  an exotic virtual bound state at the SU(3) flavour symmetric point from
  lattice QCD}, JHEP 07 (2024) 012.
\newblock \href {http://arxiv.org/abs/2403.10498} {\path{arXiv:2403.10498}},
  \href {https://doi.org/10.1007/JHEP07(2024)012}
  {\path{doi:10.1007/JHEP07(2024)012}}.

\bibitem{Maiani:2024quj}
L.~Maiani, A.~D. Polosa, V.~Riquer, {Open charm tetraquarks in broken SU(3)F
  symmetry}, Phys. Rev. D 110~(3) (2024) 034014.
\newblock \href {http://arxiv.org/abs/2405.08545} {\path{arXiv:2405.08545}},
  \href {https://doi.org/10.1103/PhysRevD.110.034014}
  {\path{doi:10.1103/PhysRevD.110.034014}}.

\bibitem{Hoffer:2024alv}
J.~Hoffer, G.~Eichmann, C.~S. Fischer, {Hidden-flavor four-quark states in the
  charm and bottom region}, Phys. Rev. D 109~(7) (2024) 074025.
\newblock \href {http://arxiv.org/abs/2402.12830} {\path{arXiv:2402.12830}},
  \href {https://doi.org/10.1103/PhysRevD.109.074025}
  {\path{doi:10.1103/PhysRevD.109.074025}}.

\bibitem{Lu:2017yhl}
Y.~Lu, M.~N. Anwar, B.-S. Zou, {$X(4260)$ Revisited: A Coupled Channel
  Perspective}, Phys. Rev. D 96~(11) (2017) 114022.
\newblock \href {http://arxiv.org/abs/1705.00449} {\path{arXiv:1705.00449}},
  \href {https://doi.org/10.1103/PhysRevD.96.114022}
  {\path{doi:10.1103/PhysRevD.96.114022}}.

\bibitem{Cincioglu:2016fkm}
E.~Cincioglu, J.~Nieves, A.~Ozpineci, A.~U. Yilmazer, {Quarkonium Contribution
  to Meson Molecules}, Eur. Phys. J. C 76~(10) (2016) 576.
\newblock \href {http://arxiv.org/abs/1606.03239} {\path{arXiv:1606.03239}},
  \href {https://doi.org/10.1140/epjc/s10052-016-4413-1}
  {\path{doi:10.1140/epjc/s10052-016-4413-1}}.

\bibitem{Hammer:2016prh}
I.~K. Hammer, C.~Hanhart, A.~V. Nefediev, {Remarks on meson loop effects on
  quark models}, Eur. Phys. J. A 52~(11) (2016) 330.
\newblock \href {http://arxiv.org/abs/1607.06971} {\path{arXiv:1607.06971}},
  \href {https://doi.org/10.1140/epja/i2016-16330-8}
  {\path{doi:10.1140/epja/i2016-16330-8}}.

\bibitem{Hanhart:2022qxq}
C.~Hanhart, A.~Nefediev, {Do near-threshold molecular states mix with
  neighboring Q\textasciimacron{}Q states?}, Phys. Rev. D 106~(11) (2022)
  114003.
\newblock \href {http://arxiv.org/abs/2209.10165} {\path{arXiv:2209.10165}},
  \href {https://doi.org/10.1103/PhysRevD.106.114003}
  {\path{doi:10.1103/PhysRevD.106.114003}}.

\bibitem{Wallbott:2019dng}
P.~C. Wallbott, G.~Eichmann, C.~S. Fischer, {$X(3872)$ as a four-quark state in
  a Dyson-Schwinger/Bethe-Salpeter approach}, Phys. Rev. D 100~(1) (2019)
  014033.
\newblock \href {http://arxiv.org/abs/1905.02615} {\path{arXiv:1905.02615}},
  \href {https://doi.org/10.1103/PhysRevD.100.014033}
  {\path{doi:10.1103/PhysRevD.100.014033}}.

\bibitem{Hoffer:2024fgm}
J.~Hoffer, G.~Eichmann, C.~S. Fischer, {The structure of open-flavour
  four-quark states in the charm and bottom region} (9 2024).
\newblock \href {http://arxiv.org/abs/2409.05779} {\path{arXiv:2409.05779}}.

\bibitem{Brambilla:1999xf}
N.~Brambilla, A.~Pineda, J.~Soto, A.~Vairo, {Potential NRQCD: An Effective
  theory for heavy quarkonium}, Nucl. Phys. B 566 (2000) 275.
\newblock \href {http://arxiv.org/abs/hep-ph/9907240}
  {\path{arXiv:hep-ph/9907240}}, \href
  {https://doi.org/10.1016/S0550-3213(99)00693-8}
  {\path{doi:10.1016/S0550-3213(99)00693-8}}.

\bibitem{Brambilla:2002nu}
N.~Brambilla, D.~Eiras, A.~Pineda, J.~Soto, A.~Vairo, {Inclusive decays of
  heavy quarkonium to light particles}, Phys. Rev. D 67 (2003) 034018.
\newblock \href {http://arxiv.org/abs/hep-ph/0208019}
  {\path{arXiv:hep-ph/0208019}}, \href
  {https://doi.org/10.1103/PhysRevD.67.034018}
  {\path{doi:10.1103/PhysRevD.67.034018}}.

\bibitem{Braaten:2024tbm}
E.~Braaten, R.~Bruschini, {Exotic Hidden-heavy Hadrons and Where to Find Them}
  (9 2024).
\newblock \href {http://arxiv.org/abs/2409.08002} {\path{arXiv:2409.08002}}.

\bibitem{Berwein:2024ztx}
M.~Berwein, N.~Brambilla, A.~Mohapatra, A.~Vairo, {Hybrids, tetraquarks,
  pentaquarks, doubly heavy baryons, and quarkonia in Born-Oppenheimer
  effective theory}, Phys. Rev. D 110~(9) (2024) 094040.
\newblock \href {http://arxiv.org/abs/2408.04719} {\path{arXiv:2408.04719}},
  \href {https://doi.org/10.1103/PhysRevD.110.094040}
  {\path{doi:10.1103/PhysRevD.110.094040}}.

\bibitem{Brambilla:2024thx}
N.~Brambilla, A.~Mohapatra, T.~Scirpa, A.~Vairo, {The nature of
  $\chi_{c1}\left(3872\right)$ and $T_{cc}^+\left(3875\right)$} (11 2024).
\newblock \href {http://arxiv.org/abs/2411.14306} {\path{arXiv:2411.14306}}.

\end{thebibliography}

\end{document}